\documentclass{aa}
\usepackage{graphicx}
\usepackage{txfonts}
\usepackage{afterpage}

\def\ltsima{$\; \buildrel < \over \sim \;$}
\def\simlt{\lower.5ex\hbox{\ltsima}}
\def\gtsima{$\; \buildrel > \over \sim \;$}
\def\simgt{\lower.5ex\hbox{\gtsima}}
\def\cgs{{erg cm$^{-2}$ s$^{-1}$}}
\def\ergs{{erg s$^{-1}$}}
\def\cm2{{cm$^{-2}$}}

\def\lum{{$L_{2-10}$}}

\def\p1{{Paper I}}

\def\xmm{{\em XMM--Newton}}
\def\chandra~{{\em Chandra}}

\def\chandra{{\em Chandra}}

\def\xmm{{\em XMM--Newton}}

\def\nh{{N$_{\rm H}$}}

\def\f14{{10$^{-14}$}}
\def\f13{{10$^{-13}$}}
\def\f12{{10$^{-12}$}}
\def\f11{{10$^{-11}$}}
\def\e22{{10$^{22}$}}

\def\feka{{Fe K$\alpha$}}

\def\3c{{3C 234}}
\def\l58{{$L_{5.8 \mu m}$}}

\def\tor{{\it Tor}}
\def\plcabs{{\it Pl}}
\def\xray{X-ray}

\begin{document}

   \title{Compton Thick AGN in the XMM-COSMOS survey}

   \author{G. Lanzuisi\inst{1,2}
          \and
          P. Ranalli\inst{1}
          \and
          I. Georgantopoulos\inst{1}
          \and
          A. Georgakakis\inst{1,3}
          \and
	  I. Delvecchio\inst{4}
          \and
          T. Akylas\inst{1}
          \and
          S. Berta\inst{3}
          \and
          A. Bongiorno\inst{5}
          \and 
          M. Brusa\inst{4}
          \and 
          N. Cappelluti\inst{2}
          \and
          F. Civano\inst{6}
          \and
          A. Comastri\inst{2}
          \and
          R. Gilli\inst{2}
          \and
          C. Gruppioni\inst{2}
          \and
          G. Hasinger\inst{7}
          \and
	  K. Iwasawa\inst{8,9}
	  \and
	  A. Koekemoer\inst{10}
          \and
          E. Lusso\inst{11}
          \and
          S. Marchesi\inst{4,6}
          \and
          V. Mainieri\inst{12}
          \and
          A. Merloni\inst{3}
          \and
          M. Mignoli\inst{2}
          \and
          E. Piconcelli\inst{5}
          \and
          F. Pozzi\inst{4}
          \and
          D. J. Rosario\inst{3}
          \and
          M. Salvato\inst{3}
          \and
          J. Silverman\inst{13}          
          \and
          B. Trakhtenbrot\inst{14} 
          \and
          C. Vignali\inst{4}
          \and
          G. Zamorani\inst{2}
          }

   \institute{Institute of Astronomy \& Astrophysics, National Observatory of Athens, Palaia Penteli, 15236, Athens, Greece
             \and
             INAF - Osservatorio Astronomico di Bologna, Via Ranzani 1, 40127, Bologna, Italy  
             \and
             Max-Planck-Institut f\"ur extraterrestrische Physik,  Giessenbachstrasse, 85748 Garching, Germany
             \and 
             Dipartimento di Fisica e di Astronomia, Universit\'a di Bologna, Via Ranzani 1, 40127, Bologna, Italy  
             \and
             INAF - Osservatorio Astronomico di Roma, via Frascati 33, 00040, Monte Porzio Catone (RM), Italy 
             \and
             Yale Center for Astronomy and Astrophysics, 260 Whitney Avenue, New Haven, CT 06520, USA
             \and
             Institute for Astronomy, 2680 Woodlawn Drive, Honolulu, HI 96822-1839, USA
             \and 
             ICREA and Institut de Ci\`encies del Cosmos, Universitat de Barcelona, IEEC-UB, Mart\'i i Franqu\`es, 1, 08028 Barcelona, Spain
             \and
             Institut de Ci\'encies del Cosmos (ICCUB), Universitat de Barcelona, Mart\'i i Franqu\'es, 1 E-08028 Barcelona, SPAIN
             \and 
             Space Telescope Science Institute, 3700 San Martin Drive, Baltimore, MD 21218, USA
             \and
             Max Planck Institut fur Astronomie, K\"onigstuhl 17, D-69117 Heidelberg, Germany
             \and 
             European Southern Observatory, Karl-Schwarzschild-Strasse 2, D-85748 Garching, Germany
             \and 
             Kavli Institute for the Physics and Mathematics of the Universe (IPMU) 5-1-5 Kashiwanoha Kashiwa, Chiba 277-8583, Japan
             \and
             Department of Physics, Institute for Astronomy, ETH Zurich, Wolfgang-Pauli-Strasse 27, CH-8093 Zurich, Switzerland (Zwicky postdoctoral fellow)
             }

   \date{}

\abstract{
Heavily obscured, Compton Thick (CT, \nh$>$10$^{24}$ cm$^{-2}$) Active Galactic Nuclei (AGN) may represent an important phase in 
AGN/galaxy co-evolution and are expected to provide a significant contribution to the cosmic X-ray background at its peak. 
However, unambiguously identifying CT AGN beyond the local Universe is a challenging task even in the deepest X-ray surveys,
and given the expected low spatial density of these sources in the 2-10 keV band, large area surveys are needed to collect sizable samples.
Through direct X-ray spectra analysis, we selected 39 heavily 
obscured AGN (\nh$>$3$\times10^{23}$ cm$^{-2}$) at bright \xray\ fluxes (F$_{2-10}$ $\simgt 10^{-14}$ erg s$^{-1}$ cm$^{-2}$) 
in the 2 deg$^2$ XMM-COSMOS survey.
After selecting CT AGN based on the fit of a simple absorbed two power law model to the shallow \xmm\ data,
the presence of {\it bona-fide} CT AGN was confirmed in 80\% 
of the sources using deeper \chandra\ data and more complex models.
The final sample comprises 10 CT AGN (6 of them also have a detected \feka\ line with EW$\sim1$ keV), 
spanning a large range of redshift (z$\sim0.1-2.5$) and luminosity (L$_{2-10}\sim10^{43.5}-10^{45}$ \ergs)
and is complemented by 29 heavily obscured AGN spanning the same redshift and luminosity range.
We collected the rich multi-wavelength information available for all these sources, 
in order to study the distribution of SMBH and host properties, such as BH mass (M$_{BH}$), Eddington ratio ($\lambda_{Edd}$), stellar mass (M$_*$), 
specific star formation rate (sSFR)
in comparison with a sample of unobscured AGN. We find that highly obscured sources tend to have significantly smaller M$_{BH}$ 
and higher $\lambda_{Edd}$ with respect to unobscured sources, while a weaker evolution in M$_*$ is observed. 
The sSFR of highly obscured sources is consistent with the one observed in the main sequence of star forming galaxies,
at all redshift.
We also present and briefly discuss optical spectra, broad band spectral energy distribution (SED) and morphology for the sample of 10 CT AGN.
Both the optical spectra and SED agree with the classification as highly obscured sources: all the available optical spectra are dominated by the 
stellar component of the host galaxy,
and to reproduce the broad band SED, an highly obscured torus component is needed for all the CT sources.
Exploiting the high resolution {\it Hubble}-ACS images available, we are able to show that these highly obscured sources have a 
significantly larger merger fraction with respect to other \xray\ selected samples of AGN.
Finally we discuss the implications of our findings in the context of AGN/galaxy co-evolutionary models, and 
compare our results with the predictions of \xray\ background synthesis models.  
}


   \maketitle
%

\section{Introduction}

Observational  and theoretical  arguments suggest  that  the obscured
phase  of  super massive  black  hole  (SMBH)  growth  holds  important
information  on both  the accretion  history of  the Universe  and the
interplay  between  Active   Galactic  Nuclei (AGN) and  their  host
galaxies.
Obscured  AGN for  example,  are essential  for  reconciling the  mass
function of black holes in  the local Universe with that expected from
AGN relics,  i.e. inferred by  integrating the luminosity  function of
AGN    via   the   continuity    equation   (e.g. Soltan 1982; Marconi 
et al. 2004).  The spectral shape of the diffuse X-ray background also
requires a  large number of mildly  obscured AGN (Gilli et al. 2007)
and even possibly  deeply buried ones with column  densities in excess
of   the  Compton  thick   limit  (CT AGN, $\rm   N_H  \ga   10^{24}$ cm$^{-2}$; 
Treister et al. 2009;  Akylas et al. 2012).

X-ray surveys provide a relatively unbiased census of the accretion history in the Universe,
as they can penetrate large amounts of dust and gas, especially in the hard, 2-10 keV band.
X-ray surveys  indeed find  that  the  accretion  density of  the
Universe is dominated by black holes that grow their mass behind large
columns  of dust  and gas  clouds  (e.g. Mainieri et al. 2002; Ueda et
al. 2003; Tozzi et al. 2006;
Akylas et al. 2006; Buchner et al. 2014; Ueda et al. 2014).

However, the exact {\it intrinsic} fraction of CT AGN remains highly uncertain, 
ranging from about 10\% of the total AGN 
population up to 35\%. 
Owing to ultra-hard X-ray surveys above 10 keV performed with SWIFT and INTEGRAL,
CT AGN are commonly observed in the local Universe, 
representing  up to 20\% of local active galaxies at energies 15-200 
keV down to a flux limit of $10^{-11}$ \cgs (see Burlon et al. 2011 and references therein). 
Still, the identification of CT AGN beyond the local universe is challenging.
Recent results on preliminary NuSTAR (Harrison et al. 2013) data were 
able to put only an upper limit ( $<33\%$)
to the fraction of CT AGN between $z=0.5-1$ (Alexander et al. 2013).

Within the AGN/galaxy  co-evolution perspective, dust and gas enshrouded AGN
represent an evolutionary point that is critical for understanding how
the  growth of  black holes  relates to  the build-up  of  the stellar
populations of their hosts.  The generic picture proposed involves gas
inflows     triggered      by     internal (Hopkins \& Hernquist 2006;
Ciotti \& Ostriker 1997; Ciotti \& Ostriker 2007;  Bournaud et al. 2012;
Gabor \& Bournaud 2013) or external (Sanders et al. 1988; Di Matteo et al. 2005;
Hopkins et al. 2006) processes. 
These  result in a period  of rapid black  hole growth that
takes place within dense dust and gas cocoons.  It is then followed by
a blow-out  stage during which some  form of AGN  feedback depletes the
gas reservoirs  thereby regulating  both the star-formation  and black
hole growth.  The  study of obscured AGN has  the potential to provide
important pieces of  the puzzle, such as the  nature of the triggering
mechanism, the  physics of AGN outflows  and their impact  on the host
galaxy.

The  evidence above motivated  numerous studies  to identify  the most
heavily  obscured  AGN,  determine  their space  density  relative  to
unobscured   sources   and   study   their  host   galaxy   properties
(e.g. Hickox et al. 2009; Brusa et al. 2009; Mainieri et al. 2011; 
Donley et al. 2012; Rovilos et al. 2014;
Delvecchio et al. 2014).   At X-rays  in particular,
there has  been an explosion recently  in the quality  and quantity of
data available,  a development that  led to direct constraints  on the
fraction of even the most deeply shrouded Compton Thick sources over a
large range of redshifts (e.g. Tozzi et al. 2006; Burlon et al. 2011;
Comastri et al. 2011; Brightman \& Ueda 2012; Georgantopoulos et al. 2013;
Buchner et al. 2014; Brightman et al. 2014).
Despite   significant  progress,  there   are  still
differences in the heavily  obscured AGN samples compiled by different
groups using the  same data.  This is primarily  related to variations
in the  spectral analysis  methods, e.g. the  complexity of  the X-ray
spectral models  adopted or how uncertainties due to photometric
redshifts or the Poisson nature of X-ray spectra are propagated in the
analysis.  It  has been  shown for example,  that a common  feature of
obscured  AGN spectra is emission  at soft  energies in excess to the primary continuum, 
possibly due to  scattered   radiation    into   the   line   of   sight
(Brightman \& Nandra 2012; Buchner et al. 2014).  Taking this component
into account is clearly important for identifying the most obscured AGN
in X-ray  surveys.  Recently, Buchner et al (2014) also demonstrated the
importance of  advanced statistical methods that  properly account for
various sources of uncertainty,  allow robust X-ray spectral parameter
estimation and  hence, the  identification of secure  heavily obscured
AGN samples.

In  this paper  we search  for the  most heavily  obscured AGN in the
COSMOS    field    (Scoville et al. 2007)  starting from the XMM-COSMOS catalog
(Hasinger et al. 2007; Cappelluti et al 2009) and using deeper \chandra\ data (Elvis et al. 2009)
to test the efficiency of our selection method for CT AGN.
Moreover, using the wealth of multi-wavelength data available in the COSMOS field,
we explore the accretion and
host galaxy  properties of the obscured  AGN sample, in order to place them in the context of AGN-Galaxy co-evolution scenarios.
This stage should be characterized by small BH masses, high accretion and star formation (SF) rates  
(e.g., Fabian 1999; Page et al. 2004; Draper \& Ballantyne 2010) ,
and could be possibly merger-driven (Hopkins et al. 2008). Our multi-wavelength analysis attempts to provide constraints on such models.
Finally we present an atlas of the  multi-wavelength properties (i.e. broad band SED,
optical spectra, morphology) of the sample of 10 {\it bona fide} CT AGN. 
Throughout the paper, a standard $\Lambda-$CDM cosmology with $H_0=70$ km s$^{-1}$ Mpc$^{-1}$, $\Omega_\Lambda=0.7$ and $\Omega_M=0.3$ is used.
Errors are given at 90\% confidence level.

\section{The data set}

\subsection{Multi-wavelength coverage in the COSMOS field} 

One of the main goals of the Cosmic Evolution Survey (COSMOS;
Scoville et al. 2007) is to trace star formation and nuclear activity
along with the mass assembly history of galaxies as a function of
redshift. 
The 2 deg$^2$ area of the {\it HST} COSMOS Treasury program 
is bounded by $9^h57.5^m<R.A.<10^h03.5^m;1^\circ27.50<$ DEC $< 2^\circ57.50$.
The field has a unique deep and wide multi-wavelength coverage, 
from the optical band ({\it Hubble, Subaru, VLT}
and other ground-based telescopes), to the infrared ({\it Spitzer, Herschel}),
\xray\ (\xmm\ and \chandra) and radio (Very Large
Array (VLA) and future Jansky-VLA, P.I. V. Smolcic) bands.
Large dedicated ground-based spectroscopy programs in the optical 
with Magellan/IMACS (Trump et al. 2009), VLT/VIMOS (Lilly et al. 2009), Subaru-FMOS (P.I. J. Silverman), and DEIMOS-Keck (P.I. N. Scoville) 
have been completed or are well under way.
Very accurate photometric redshifts are available for both the galaxy population ($\Delta z/(1 + z) <1\%$; Ilbert et al. 2009) 
and the AGN population ($\Delta z/(1 + z)\sim1.5\%$ Salvato et al. 2009, 2011).

The COSMOS field has been observed with \xmm\ for a total of $\sim1.5$ Ms at a rather homogeneous depth of
$\sim$60 ks (Hasinger et al. 2007, Cappelluti et al. 2009). 
The \xmm\ catalog used in this work includes $\sim1800$ point-like sources,
detected in one or more of the 3 adopted bands (0.5-2, 2-10 and 5-10 keV),
and having a hard (2-10 keV) band limiting flux of $2.5 \times 10^{-15}$ \cgs (and $5.6 (9.3) \times 10^{-15}$ \cgs on 50\% (90\%) of the area).
The adopted likelihood threshold corresponds to a probability of $\sim4.5\times10^{-5}$ 
that a catalog source is a spurious background fluctuation. 
197 sources are classified as stars or unclassified (Brusa et al. 2010),
so they are excluded from the following X-ray spectral analysis.

At the hard band flux limit of the XMM-COSMOS survey, current CXB models predict a fraction of CT-AGN  between 1 and 3\%
(Gilli et al. 2007; Treister et al. 2009, Ueda et al. 2014), while at a flux limit of F$_{2-10}$ $\sim10^{-14}$ \cgs, above which a basic spectral analysis is viable, 
this fraction is even smaller ($\simlt1\%$, see Sec.~8).
Therefore, the XMM-COSMOS catalog may seem not the best place where to look for, in order to select a large sample of CT AGN, 
even taking into account the large area covered.
However, the great advantage of the XMM-COSMOS data set 
is that there are deeper X-ray observations available in the same area: the \chandra-COSMOS and its new extension, COSMOS legacy  
(Elvis et al. 2009; Civano et al. 2014), that together cover the same 2 deg$^2$ area of the \xmm\ observation, with a homogeneous exposure time of $\sim160$ ks
(flux limit F$_{2-10}$ $\sim7 \times 10^{-16}$ \cgs).
This will allow us to use the deeper \xray\ data to evaluate, {\it a posteriori}, the efficiency of our XMM-based CT AGN  selection method (see Sec.~4.1),
with the final goal to extend, in the future, these studies to the deeper full \chandra\ catalog in COSMOS.
Given the difficulty of unambiguously identifying CT AGN beyond the local Universe, 
this is a valid alternative approach to simulations, to evaluate the efficiency of the selection method.

Furthermore, a great wealth of information has been made available in the latest years,
regarding multi-wavelength, host and SMBH properties of the sources in the XMM-COSMOS catalog:
Trump et al. (2009), Merloni et al. (2010) and Rosario et al. (2013) present SMBH masses for large samples of type-1 AGN  
(182, 89 and 289, respectively, the latter computed from a compilation of all the spectra available);
Mainieri et al. (2011) present multi-wavelength properties of 142 obscured QSOs;
Bongiorno et al. (2012, B12 hereafter) present host properties (star formation rate, stellar mass etc.) of a sample of 1700 AGN;
Lusso et al.  (2012, L12 hereafter) present bolometric luminosities and Eddington ratios of a sample of 929 AGN, both type-1 and type-2,
and SMBH masses estimated trough scaling relations for 488 type-2 AGN.
Delvecchio et al. (2014, D14 hereafter) performed broad band SED decomposition
for all the $160~ \mu m$ Herschel detected COSMOS sources.
The availability of all this information is crucial in order to study the role of CT AGN in the context
of AGN/galaxy co-evolution.


\subsection{Sample selection}

   \begin{figure}[!t]
   \centering
      \includegraphics[width=8cm]{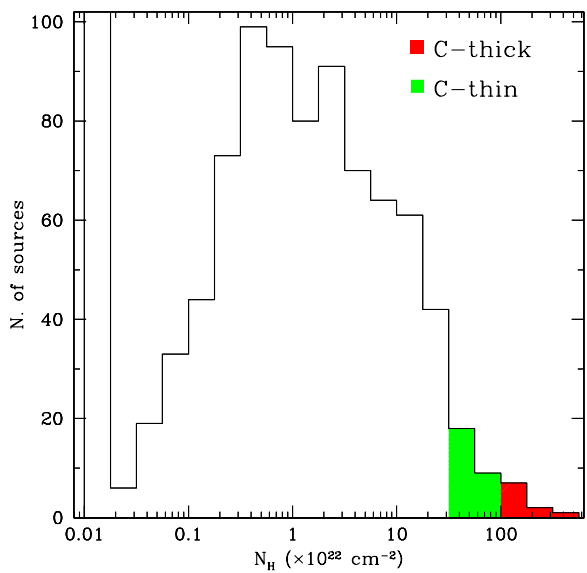}
   \caption{Distribution of \nh\ (in units of $10^{22}$ cm$^{-2}$) in the XMM-COSMOS survey. Only extragalactic sources with $>30$ pn net counts are shown.
   The green (red) shaded histogram show the sample of CTN$_i$ (CTK$_i$) sources. The first bin includes the 390 sources for which the \nh\ is an upper limit
(for clarity the y-axis is rescaled in the range 0-100). }%
   \label{histonh}
   \end{figure}

The \xmm\ spectra have been extracted as described in Mainieri et al. (2007).
The original spectral extraction was performed only for the EPIC pn-CCD (pn) camera (Str\"uder et al. 2001).
All the spectral fits are performed with {\it Xspec} v. 12.7.1 (Arnaud et al. 1996).
We analyzed the 1625 X-ray pn spectra of the identified extragalactic sources, using the same automated fit procedure presented in Lanzuisi et al. (2013a).
This procedure makes use of the C-statistic (Cash 1979), especially developed to model spectra with a small number of counts.
It requires the simultaneous fit of the background (BKG) and very limited counts binning (minimum of 1 counts per bin, in order to avoid empty channels). 

The global \xmm\ pn background between 0.3 and 10 keV
is complex, and comprises an external, cosmic BKG, passing through the telescope mirrors, and therefore convolved with the instrumental Auxiliary Response File (ARF),
and the internal, particle induced BKG, which is not convolved by the ARF.
The external BKG, which dominates at soft energies (below $\sim1$ keV) is modeled with two thermal components (Xspec model {\sc APEC}), with solar abundances, 
one for the local hot bubble (kT$\sim0.04$ keV) and the second for the Galactic component  (kT$\sim0.12$ keV), 
produced by the ISM of the Galactic disk (Kuntz \& Snowden 2000), plus a power-law reproducing the Cosmic X-ray Background ($\Gamma\sim1.4$),
mostly due to unresolved discrete sources (Comastri et al. 1995, Brandt et al. 2002).
The particle induced BKG is well reproduced by a flat power-law ($\Gamma\sim0.5$), plus several strong emission lines at energies of 1.5, 7.4, 8.0 and 8.9 keV,
due to Al, Ni, Cu and Zi+Cu K$\alpha$ lines, respectively (Freyberg et al. 2004).

We adopted as source model a simple double power-law: the primary power-law, 
modified by intrinsic absorption at the source redshift, representing the 
transmitted component, plus a second power-law
to account for the soft emission commonly observed 
in local highly obscured sources (e.g. Done et al. 2003).
This emission can be due to unobscured flux leaking out in the soft band (through scatter or partial covering), thermal emission related to star-formation,
or the blend of emission lines from photo-ionized circumnuclear gas (Guainazzi \& Bianchi 2007) or a combination of these components.
Given the relatively poor photon statistics and the lack of high spectral resolution,
distinguishing between these different origins is not possible here,
and the second power-law is only used as a simple description of the observed spectra in the soft band.
However, the presence of a second, soft component is key to recover a correct estimate of the intrinsic absorption affecting the primary power-law,
especially for highly obscured sources (Brightman et al. 2014; Buchner et al. 2014).

The normalization of the soft component is constrained to be less than 10\% with respect to the hard component.
In well studied nearby AGN, a few \% is the typical flux contribution of the soft component with respect to the {\it unobscured} primary one (e.g. Turner et al. 1997).  
The photon index of the soft component is linked to that of the primary power-law,
in order to minimize the number of free parameters.
Both power-laws are absorbed by a Galactic column density of $1.7\times10^{20}$ cm$^{-2}$
(Kalberla et al. 2005), observed in the direction of the COSMOS field\footnote{In {\sc XSPEC} 
notation the above model is expressed as {\sc wa*(po+zwa*po)}.}. 
The energy range in which the fit is performed is 0.3-10 keV.

   \begin{figure*}[!t]
   \centering
      \includegraphics[width=8cm,height=8cm]{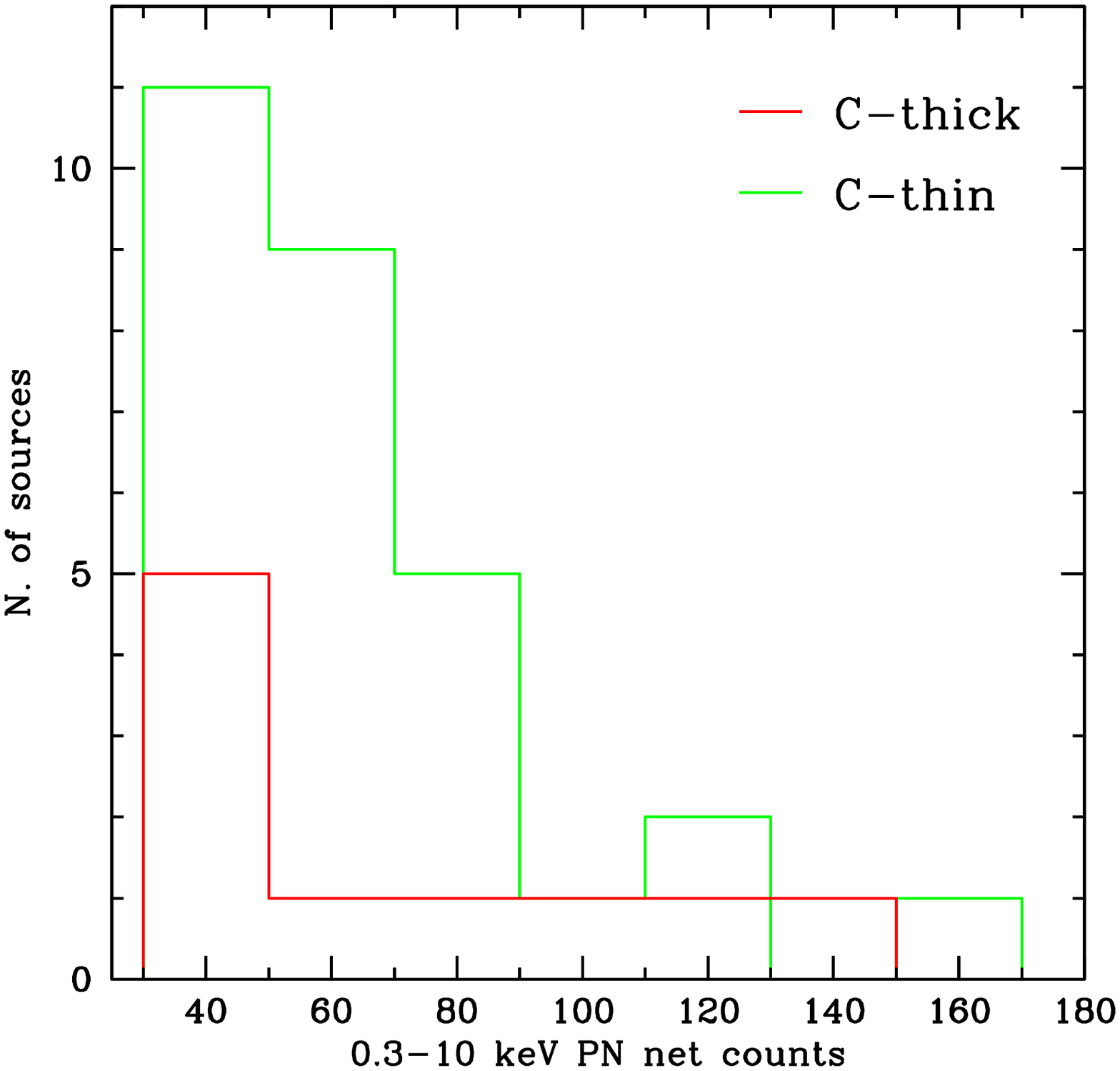}\hspace{0.5cm}
      \includegraphics[width=8cm,height=8cm]{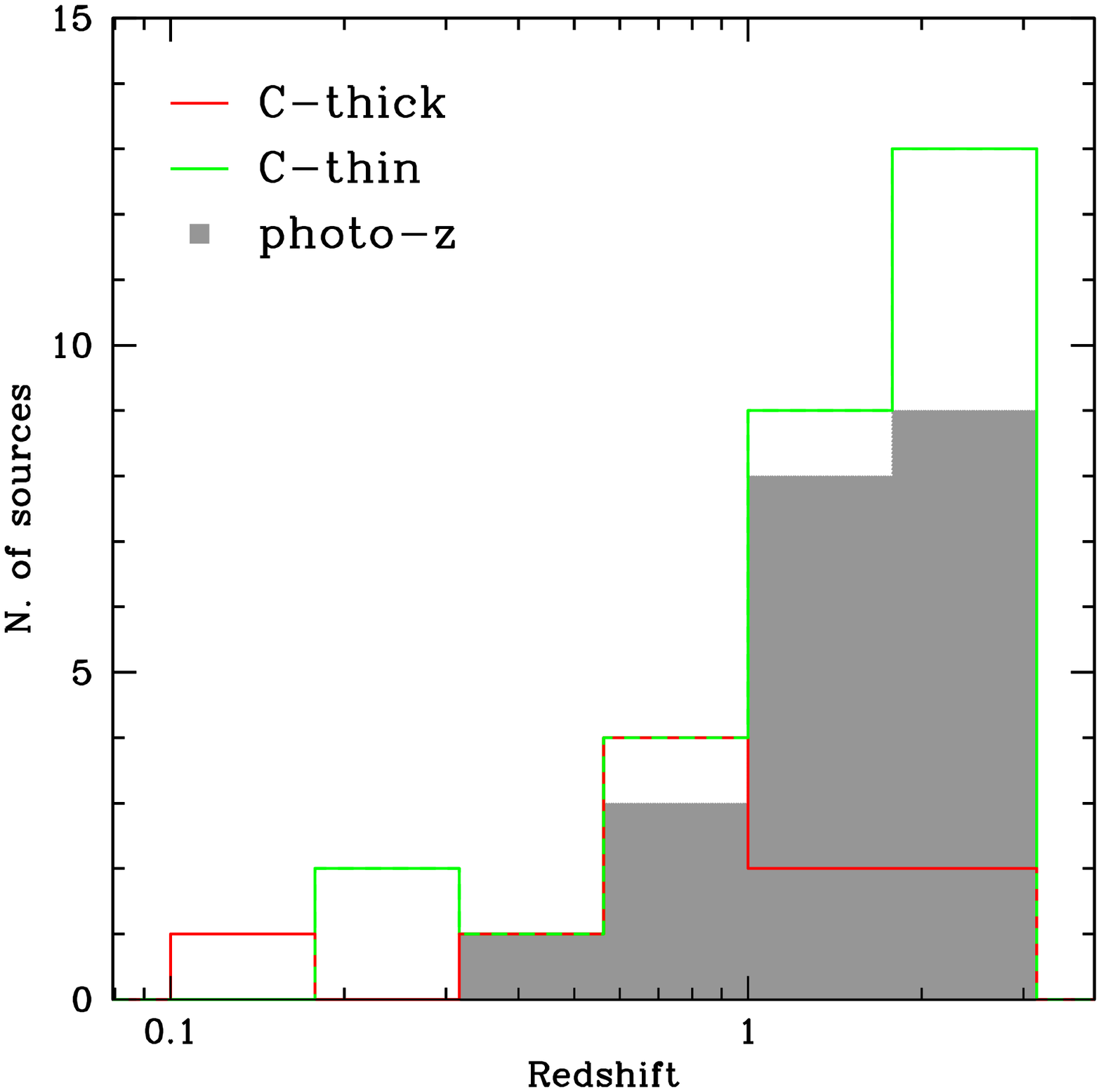}
   \caption{{\it Left panel:} 0.3-10 keV net counts distribution for CTK$_i$ (red) and CTN$_i$ (green) sources.
  {\it Right panel:} Redshift distribution for CTK$_i$ (red) and CTN$_i$ (green) sources. The gray shaded histogram shows the distribution of photometric redshifts.}
   \label{histoz}
   \end{figure*}
We note that the minimum number of counts for which this kind of analysis can be applied
is constrained only by the maximum relative error that one wants to allow for the free parameters in the fit.
Because here we are mainly interested in recovering the intrinsic absorption and luminosity of our sources, we 
fixed the photon index $\Gamma$ to 1.9 for sources with less than 100 net counts (70\% of the sources have less than 100 counts).
This value is the one typically found in AGN at any luminosity level (Nandra \& Pounds 1994; Piconcelli et al. 2005; Tozzi et al. 2006).
In this way the number of free parameters is three (\nh\ and the two power-law normalizations) 
and we are able to constrain, with a typical relative error smaller than 60\%, 
the \nh\ for sources with more than 30 net counts in the full 0.3-10 keV band. 
We discuss in appendix A the detection limits of the XMM-COSMOS survey in the $z$-\lum\ plane, 
for highly obscured sources with at least 30 net counts in full band,
compared with the one for the full XMM-COSMOS catalog.

Fig. 1 shows the distribution of the intrinsic column density \nh\ for the 
1184 extragalactic sources with $>$30 counts. 
The first bin at \nh$=10^{20}$ cm$^{-2}$ includes the 390 sources for which the \nh\ is an upper limit
(for clarity the y-axis of the plot is rescaled in the range 0-100).
With this simple spectral fit,
we are able to select 10 CT AGN candidates (best fit \nh$\geq1\times10^{24}$ cm$^{-2}$, initial Compton thick sample, CTK$_i$ hereafter),
plus a larger sample of 29 highly obscured 
(\nh$>3\times10^{23}$ cm$^{-2}$), 
but nominally not CT (\nh$<1\times10^{24}$ cm$^{-2}$,  initial Compton thin sample, CTN$_i$ hereafter) AGN.
The global properties of all the 39 sources comprised in both sample are described in the next section,
before performing more detailed spectral analysis.


\subsection{The obscured sample}

Fig. 2 (left) shows the distribution of 0.3-10 keV net counts for the  CTK$_i$ (red) and CTN$_i$ (green) sources.
As expected, all these highly obscured sources are faint, and 40-50\% of them have between 30 and 50 counts, just above the threshold chosen for the \xray\ spectral analysis,
with the CTK$_i$ sources having typically fewer counts.
As shown by this plot, the ability to push the analysis to very low counts is crucial,
in order to recover a sizable sample of such highly obscured sources in rather shallow X-ray catalogs.

The total number of available spectroscopic redshifts in the obscured sample is 19.
The majority of them have optical spectra from the 
zCOSMOS survey (11 out of 19). Seven sources have been observed with the IMACS spectrograph on Magellan,
and one has a SDSS spectrum. They are all classified as Narrow Line AGN (see Sec.~6.1 for the optical classification of the final CTK$_i$ sample).
Sources without spectroscopic classification are classified on the basis of their spectral energy distribution (SED) best fit template
from Salvato et al. (2009, 2011): 15 are classified as type-2 and five as type-1 (the photometric classification is however affected by large uncertainties).
The redshift distribution for CTK$_i$ (in red) and CTN$_i$ (in green) sources is shown in Fig. 2 (right)
The gray shaded histogram shows  the distribution of photometric redshifts, where we note that the fraction of sources with photometric redshift 
increases at higher redshifts, being $\sim30\%$ for $z<1$ and $\sim60\%$ for $z>1$.
We also underline that CTN$_i$ sources tend to have higher redshift with respect to CTK$_i$ ones.

The \xmm\ pn spectra analyzed here are obtained as the sum of \xray\ counts collected in different observations, performed during a period of $\sim3.5$ years. 
X-ray spectral variability, due to occultation by broad line clouds,
and producing large variation in observed column densities, has been found to be common in local AGN 
(e.g. Risaliti et al. 2009, see Torricelli-Ciamponi et al. 2014 and Markowitz, Krumpe \& Nikutta 2014 for systematic studies).
We checked that all the sources in the CTK$_i$ and CTN$_i$ samples can be considered not variable in the 0.5-2 and 2-10 keV bands
at 95\% confidence level (variability estimators for the full XMM-COSMOS catalog are available in Lanzuisi et al. 2013b)
through the 3.5 years of observation of the XMM-COSMOS survey.
Therefore the merged X-ray spectra used in the analysis should be considered as representative of the typical spectrum for each source.


\section{CT spectral modeling}
   \begin{figure*}[!t]
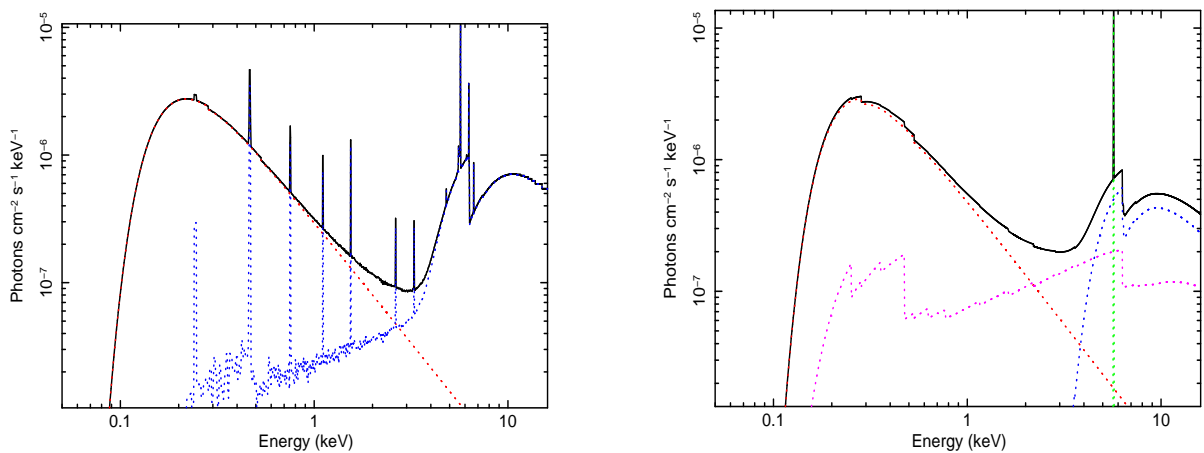

   \centering
      \includegraphics[width=6cm,height=7.5cm,angle=-90]{mo_torus.ps}\hspace{1.0cm}
      \includegraphics[width=6cm,height=7.5cm,angle=-90]{mo_plcabs.ps} 
   \caption{As an example, the two models used in the fit of source 2608 at z=0.125 are shown:  \tor\ (left) and \plcabs\ (right).
   In the \tor\ model the blue line includes the primary power-law, the reflection component and fluorescence lines,
   all self-consistently computed as function of \nh\ and geometry. In the \plcabs\ model 
   the blue line represents the primary power-law, modified by photoelectric absorption and taking into account Compton scattering (PLCABS component in Xspec),
   The magenta line represents the PEXRAV component (i.e. reflection produced by a slab of infinite optical depth material), and the green line represents a Gaussian emission line
   reproducing the fluorescent Fe K$\alpha$ line.
   In both models the red line represents the secondary power-law, reproducing scattered light and/or thermal emission in the soft band.}
   \label{model}
   \end{figure*}

Studies of the X-ray spectra of local Seyferts have identified different
components associated with different physical process, such as
reflection, scattering and absorption of the direct \xray\ emission.
Properly modeling these components is important for meaningful parameter
derivation, especially in the case of heavily obscured and CT sources.
Indeed, building a general model for the \xray\ spectrum of CT AGN is difficult, 
because of the complex interplay between photoelectric absorption, Compton scattering, reflection and fluorescence emission lines
along different line of sights (see e.g. Murphy \& Yaqoob 2009 for a detailed discussion).

In addition to the primary power-law, obscured up to very high energies (typically up to 7-10 keV rest frame),
a rather flat reflection component is usually present, together with a strong (EW $\simgt 1$ keV) Fe K$\alpha$
line, which is considered to be the unambiguous sign of the presence of a CT absorber (Matt et al. 2000),
and even used as one of the most reliable CT selection techniques for sources with medium-good quality \xray\ spectra (Corral et al. 2014). 
Several recent studies showed that a second component in the soft band, and a reflection component in the hard, 
are generally both needed by the data, even in the case of low quality \xray\ spectra (Brightman \& Nandra 2011; Brightman \& Ueda 2012; Buchner et al. 2014).

Because of the complex spectral shape and typical low flux levels,  
the definition of a source as CT, beyond the local Universe,
is in general model dependent and therefore not univocal: see for example the differences in 
CT samples in \chandra\ Deep Field South (CDFS) data from Tozzi et al. (2006), Georgantopoulos et al. (2008), Brightman \& Ueda (2012), 
Georgantopoulos et al. (2013), summarized in Castell{\'o}-Mor et al. (2013, the disagreement is typically of the order of $\sim50$\%), 
or the case of IRAS 09104+4109, a rather bright and low redshift (z=0.442) hyper-luminous IR galaxy 
for which the classification as CT has been debated for a decade (Franceschini et al. 2000; Piconcelli et al. 2007; Vignali et al. 2011; Chiang et al. 2013).

Given the low statistics available for the \xray\ spectra in our sample, and the need to minimize the number of free parameters, 
we will use two rather simple models, in which the geometry is fixed, based on what we know from local CT sources.
This will allow us to constrain the amount of obscuration, the intrinsic luminosity, the strength of the scattered emission, and the EW of the Fe K$\alpha$ line.

The first model (\tor\ hereafter) uses the TORUS template presented in Brightman \& Nandra 2011, which self-consistently reproduces 
photoelectric absorption, Compton scattering and line fluorescence, as function of \nh\ and system geometry (Fig. 3 left).
The second (\plcabs\ hereafter) is built up with an obscured power-law (PLCABS, Yaqoob et al. 1997, that properly takes into account Compton scattering up to 15-20 keV),
a PEXRAV component (Magdziarz \& Zdziarski 1995) 
contributing with only the reflected component, and a redshifted Gaussian line, with line energy fixed at 6.4 keV, to reproduce the Fe K$\alpha$ line (Fig. 3 right).
In both models, a second unobscured power-law is included. 

Since for highly obscured sources the primary power-law is observed only at the highest energies, if observed at all,
in both models we fixed its photon index at the typical value of 1.9 for all the sources (including the four with $>100$ counts).
Also the photon index of the secondary component is fixed at 1.9,
 while the ratio of its normalization
with respect to the primary component is a free parameter, and is
used to estimate the intensity of the scattered component\footnote{
The secondary component must be a redshifted power-law (ZPOWERLW in Xspec)
in order to have the normalization relative to 1 keV {\it rest-frame}, as for the TORUS template and the PLCABS model.}, expressed as \% of the flux emitted  
by the soft component with respect to the flux emitted by the {\it unobscured} primary component.
As mentioned in Sec.~2.2, the scattered component is constrained to be $<10\%$ of the primary component.

For the \tor\ model we have to make an assumption on the geometry of the system, using fixed torus half-opening and inclination angles 
($\theta_{tor}=60^{\circ}$ and $\theta_{i}=85^{\circ}$ respectively).
These correspond to a situation in which the line of sight intercepts the torus itself, and the torus has an intermediate opening angle.
The use of a smaller half-opening angle for the torus (i.e. $\theta_{tor}=30^{\circ}$), increases the amount of reflection, which 
contributes to the flux below the absorption cut-off (see Fig. 3 left),
and requires higher \nh\ values to reproduce the same data points. 
Therefore, the \nh\ values reported in the following analysis, could be underestimated, 
if the absorbing material geometry is closer to a sphere (see Brightman et al. 2014).

For the \plcabs\ model, we fixed the ratio between the reflection and the Gaussian line normalizations, so that the EW of the emission line 
is EW$ = 1$ keV with respect to the reflected continuum. This choice is made in order to minimize the number of free parameters, and is justified by observational evidences,
both at low and high redshift (Matt et al. 1997; Guainazzi et al. 2000; Feruglio et al. 2011), and by model predictions 
(Ghisellini, Haardt, \& Matt 1994; Ikeda et al. 2009, Murphy \& Yaqoob 2009, see discussion in Sec. 4.1).

Therefore the number of free parameters of the \tor\ model is three: the obscuring \nh\ and normalization of the primary component (the TORUS template)
and the normalization of the secondary component (ZPOWERLW).
The number of free parameters for the \plcabs\ model is four: the obscuring \nh\ and normalization of the primary component (PLCABS),
the normalization of the  the secondary component (ZPOWERLW) and the normalization of the reflection plus emission line complex (PEXRAV+ZGAUSS).

We stress that the \tor\ model is expected to give more accurate \nh\ and intrinsic luminosities for these highly obscured sources.
This is because its reflection component is computed for a more physically motivated geometry with respect to the PEXRAV component, 
in which the reflection is computed as produced by an infinite slab of neutral material with a given inclination. 
The PEXRAV component was indeed introduced to reproduce reflected emission from a face-on accretion disk,
and not reflection from an obscuring torus.
The \plcabs\ model is meant to mimic, as much as possible, the global shape of the \tor\ model, and to compute independently 
the EW of the \feka\ line, which is not an output in the \tor\ model.
Then, the \feka\ line EW can be used as a further evidence of the presence of CT absorbing material, and its evolution, e.g. with \nh, 
can be compared with model predictions (see Sec.~4.1). 

The PEXRAV component is also known to reproduce a different global continuum shape and Compton hump, with respect to CT dedicated modeling, i.e. 
Monte Carlo simulations that assume a toroidal geometry and self-consistently include reflection and scattering in their ansatz (e.g. the Torus 
template from Brightman et al. 2014 or MYTORUS model from Murphy \& Yaqoob 2009).
The difference is significant enough to potentially impact fitting results for high signal-to-noise data. 
For our low signal-to-noise data, we checked that the overall continuum properties obtained from the two models used here are in good agreement.
We show in appendix B the results of this comparison.

\section{Results}

\begin{table*}[!t]
\caption{Results of the \xmm\ pn spectral fit for the highly obscured sources, sorted for decreasing \nh.
 The first group includes the ten sources belonging to the CTK$_i$ sample, the second and third groups include the 29 sources belonging to
the CTN$_i$ sample, with the five sources in the second group being border line.}
{\footnotesize
\label{tab:xray}
\begin{center}
\begin{tabular}{ccccccccccccc}
\hline\hline\\
\multicolumn{1}{c} {XMM ID}&
\multicolumn{1}{c} {z}&
\multicolumn{1}{c} {Cl.}&
\multicolumn{1}{c} {C}&
\multicolumn{1}{c} {\nh}&
\multicolumn{1}{c} {F$_{2-10}$}&
\multicolumn{1}{c} {Log(L$_{I}$)}&
\multicolumn{1}{c} {Log(L$_{O}$)}&
\multicolumn{1}{c} {Sc. \%}&
\multicolumn{1}{c} {EW}&
\multicolumn{1}{c} {Cstat/d.o.f.} \\
(1) & (2) & (3) & (4) & (5) & (6) & (7) & (8) & (9) & (10) & (11) \\                
\\
\hline
\\
60361$^*$ &     1.73  & Ph & 49 & $6.35_{-4.80}^{+...}$    &  $0.92_{-0.13}^{+0.05}$  & $45.15_{-0.61}^{+...}$   & $43.46_{-0.03}^{+1.50}$    &0.36$\pm$0.14  &   $2.29_{-1.79}^{+2.29}$   &    175.6/201 \\ 
70082$^*$ &     2.429 & Ph & 34 & $3.19_{-1.48}^{+5.53}$   &  $1.47_{-0.16}^{+0.19}$  & $45.34_{-0.26}^{+0.43}$  & $43.65_{-0.03}^{+0.15}$   &$<$2.10        &   $<$3.07                  &    238.7/255 \\ 
60314     &     1.107 & Ph & 49 & $3.15_{-2.28}^{+4.05}$   &  $2.53_{-0.28}^{+0.29}$  & $45.34_{-0.69}^{+0.35}$  & $43.38_{-0.08}^{+0.08}$   &$<$5.20        &   $<$0.39                  &    156.9/157 \\ 
217       &     0.66  & Sp & 56 & $2.82_{-2.53}^{+...}$    &  $1.92_{-0.19}^{+0.23}$  & $44.47_{-1.25}^{+...}$   & $42.97_{-0.06}^{+0.04}$   &$<$5.60        &   $0.61_{-0.43}^{+0.77}$   &    144.7/157 \\ 
60152$^*$ &     0.579 & Sp & 74 & $1.97_{-1.21}^{+...}$    &  $3.36_{-0.42}^{+0.27}$  & $44.48_{-0.43}^{+...}$   & $43.04_{-0.06}^{+0.03}$   &$<$0.80        &   $0.57_{-0.43}^{+0.63}$   &    164.0/162 \\ 
60342$^*$ &     0.941 & Ph & 33 & $1.48_{-1.40}^{+2.21}$   &  $1.34_{-0.35}^{+0.25}$  & $44.67_{-1.32}^{+0.39}$  & $43.04_{-0.05}^{+0.05}$   &$<$1.22        &   $<$0.32                  &    145.4/143 \\ 
5511$^*$  &     1.023 & Ph &101 & $1.26_{-1.19}^{+2.24}$   &  $2.03_{-0.31}^{+0.21}$  & $44.70_{-1.29}^{+0.44}$  & $43.42_{-0.04}^{+0.03}$   &4.10$\pm$3.50  &   $<$0.38                  &    215.5/246  \\
54514$^*$ &     0.707 & Sp &121 & $1.23_{-0.88}^{+0.55}$   &  $4.45_{-0.76}^{+0.45}$  & $44.51_{-0.56}^{+0.17}$  & $43.57_{-0.06}^{+0.07}$   &1.50$\pm$0.70  &   $0.35_{-0.25}^{+0.43}$   &    214.0/220  \\
60211$^*$ &     0.511 & Sp & 38 & $1.11_{-0.21}^{+2.87}$   &  $0.88_{-0.51}^{+0.21}$  & $43.94_{-0.26}^{+0.55}$  & $42.64_{-0.23}^{+0.06}$   &1.42$\pm$0.45  &   $<$0.23                  &    122.6/144 \\ 
2608$^*\dag$ &  0.125 & Sp &142 & $1.10_{-0.26}^{+0.59}$   &  $4.72_{-0.96}^{+0.79}$  & $43.30_{-0.16}^{+0.19}$  & $42.17_{-0.10}^{+0.05}$   &0.24$\pm$0.08  &   $0.50_{-0.25}^{+0.33}$   &    256.8/299 \\  
\\
\hline\\
202       &     1.32  & Sp &115 & $1.06_{-0.28}^{+0.37}$   &  $1.84_{-0.19}^{+0.21}$  & $44.65_{-0.14}^{+0.13}$  & $43.73_{-0.04}^{+0.05}$   &1.87$\pm$1.32  &   $<$0.92                  &    233.0/270 \\ 
60043     &      1.73 & Ph & 35 & $0.86_{-0.59}^{+2.10}$   &  $0.94_{-0.12}^{+0.10}$  & $44.51_{-0.53}^{+0.54}$  & $43.33_{-0.07}^{+0.11}$   &$<$3.80        &   $<$1.31                  &    187.0/200 \\ 
54490$^*$ &     0.908 & Sp & 63 & $0.82_{-0.40}^{+1.08}$   &  $3.60_{-0.63}^{+0.46}$  & $44.69_{-0.31}^{+0.37}$  & $43.69_{-0.07}^{+0.05}$   &$<$2.60        &   $<$0.93                  &    196.5/206  \\
258       &     1.748 & Ph & 66 & $0.80_{-0.36}^{+0.93}$   &  $2.30_{-0.46}^{+0.38}$  & $45.09_{-0.27}^{+0.34}$  & $44.10_{-0.05}^{+0.07}$   &$<$2.62        &   $<$1.51                  &    198.8/203  \\
70145$^*$ &     2.548 & Ph & 50 & $0.79_{-0.39}^{+1.24}$   &  $0.87_{-0.13}^{+0.11}$  & $44.69_{-0.32}^{+0.41}$  & $43.96_{-0.08}^{+0.10}$   &1.82$\pm$0.87  &   $<$3.67                  &    135.5/111  \\               
\\
\hline\\
60024     &   1.147   & Sp & 55 & $0.70_{-0.38}^{+0.50}$   &   $1.44_{-0.35}^{+0.21}$ & $44.10_{-0.37}^{+0.24}$  &  $43.48_{-0.08}^{+0.06}$  &$<$3.59        &  $<$0.88                   &    175.2/165  \\
5006      &   2.417   & Sp & 80 & $0.67_{-0.29}^{+0.40}$   &   $2.25_{-0.32}^{+0.22}$ & $45.21_{-0.26}^{+0.21}$  &  $44.24_{-0.06}^{+0.08}$  &$<$2.82        &  $<$1.55                   &    293.4/293  \\
5357      &   2.189   & Ph & 56 & $0.63_{-0.19}^{+0.33}$   &   $1.96_{-0.44}^{+0.37}$ & $45.03_{-0.17}^{+0.20}$  &  $44.13_{-0.06}^{+0.09}$  &$<$3.60        &  $<$0.72                   &    129.5/123  \\
5185      &   2.19    & Ph & 32 & $0.57_{-0.38}^{+0.90}$   &   $1.94_{-0.40}^{+0.45}$ & $45.05_{-0.52}^{+0.42}$  &  $44.12_{-0.10}^{+0.12}$  &$<$5.34        &  $<$2.46                   &    105.7/111  \\  
5130      &   1.553   & Sp & 79 & $0.55_{-0.22}^{+0.39}$   &   $1.97_{-0.27}^{+0.23}$ & $44.71_{-0.23}^{+0.23}$  &  $43.89_{-0.05}^{+0.05}$  &3.05$\pm$1.83  &  $<$1.61                   &    224.5/251  \\
348       &   0.779   & Ph & 36 & $0.52_{-0.24}^{+0.84}$   &   $0.47_{-0.33}^{+0.23}$ & $44.39_{-0.30}^{+0.42}$  &  $43.64_{-0.09}^{+0.07}$  &$<$2.38        &  $<$0.21                   &    179.8/190  \\
54541     &   2.841   & Sp & 35 & $0.51_{-0.28}^{+0.35}$   &   $2.06_{-0.64}^{+0.40}$ & $45.23_{-0.39}^{+0.25}$  &  $44.49_{-0.11}^{+0.14}$  &$<$2.63        &  $<$1.53                   &    114.6/82   \\
5534      &   2.12    & Ph & 33 & $0.50_{-0.33}^{+0.83}$   &   $1.51_{-0.27}^{+0.26}$ & $44.89_{-0.48}^{+0.43}$  &  $44.08_{-0.05}^{+0.07}$  &$<$10.00       &  $<$1.71                   &    134.2/129  \\
60492     &   1.914   & Ph & 41 & $0.48_{-0.30}^{+3.39}$   &   $1.16_{-0.22}^{+0.24}$ & $44.67_{-0.45}^{+0.91}$  &  $43.84_{-0.07}^{+0.08}$  &$<$2.91        &  $<$0.75                   &    193.5/208  \\
70084     &   1.27    & Ph & 54 & $0.46_{-0.23}^{+0.58}$   &   $2.43_{-0.38}^{+0.36}$ & $44.51_{-0.31}^{+0.35}$  &  $43.93_{-0.06}^{+0.05}$  &$<$4.66        &  $<$1.36                   &    132.8/119  \\
5033      &   0.67    & Ph & 87 & $0.46_{-0.18}^{+0.27}$   &   $8.98_{-1.62}^{+1.70}$ & $44.66_{-0.24}^{+0.21}$  &  $43.87_{-0.08}^{+0.08}$  &$<$0.93        &  $<$0.40                   &    132.8/150  \\
473       &   2.148   & Ph & 50 & $0.45_{-0.28}^{+0.97}$   &   $1.00_{-0.18}^{+0.15}$ & $44.68_{-0.44}^{+0.50}$  &  $43.99_{-0.07}^{+0.06}$  &5.80$\pm$4.20  &  $<$2.31                   &     203./188  \\
70135     &   2.092   & Ph & 54 & $0.44_{-0.28}^{+0.72}$   &   $1.68_{-0.32}^{+0.29}$ & $44.88_{-0.46}^{+0.42}$  &  $44.18_{-0.07}^{+0.07}$  &$<$9.93        &  $<$1.08                   &    211.0/197  \\
5496      &   0.694   & Sp & 81 & $0.44_{-0.25}^{+0.35}$   &   $6.86_{-1.47}^{+1.35}$ & $44.48_{-0.41}^{+0.26}$  &  $43.84_{-0.12}^{+0.08}$  &$<$1.76        &  $<$0.53                   &    166.5/152  \\
5007      &   2.390   & Ph & 31 & $0.41_{-0.23}^{+0.47}$   &   $1.24_{-0.21}^{+0.21}$ & $44.89_{-0.37}^{+0.33}$  &  $44.07_{-0.06}^{+0.08}$  &$<$3.63        &  $<$1.65                   &    216.2/198  \\
5222      &   0.332   & Ph &120 & $0.39_{-0.13}^{+0.16}$   &  $10.05_{-1.73}^{+0.96}$ & $44.03_{-0.19}^{+0.15}$  &  $43.37_{-0.07}^{+0.05}$  &1.78$\pm$0.53  &  $<$0.37                   &    243.3/235  \\
70007     &   1.848   & Sp & 38 & $0.39_{-0.24}^{+0.50}$   &   $0.77_{-0.10}^{+0.11}$ & $44.41_{-0.43}^{+0.36}$  &  $43.79_{-0.06}^{+0.05}$  &7.30$\pm$5.15  &  $<$2.15                   &    125.0/109  \\
302       &   0.186   & Sp & 72 & $0.39_{-0.20}^{+0.37}$   &   $1.87_{-0.47}^{+0.33}$ & $42.70_{-0.35}^{+0.29}$  &  $42.16_{-0.10}^{+0.07}$  &4.20$\pm$1.70  &  $<$1.43                   &    157.1/166  \\
5042      &   2.612   & Sp & 90 & $0.39_{-0.16}^{+0.21}$   &   $1.31_{-0.23}^{+0.19}$ & $45.02_{-0.25}^{+0.19}$  &  $44.29_{-0.07}^{+0.06}$  &$<$7.66        &  $<$0.62                   &    227.3/231  \\
60436     &   2.313   & Ph & 48 & $0.37_{-0.18}^{+0.37}$   &   $1.07_{-0.15}^{+0.16}$ & $44.76_{-0.30}^{+0.30}$  &  $44.06_{-0.05}^{+0.07}$  &$<$5.92        &  $<$0.49                   &    182.9/174  \\
5562      &   2.626   & Sp & 64 & $0.32_{-0.19}^{+0.28}$   &   $0.90_{-0.16}^{+0.13}$ & $44.78_{-0.40}^{+0.27}$  &  $44.21_{-0.06}^{+0.06}$  &$<$10.00       &  $<$0.90                   &    145.9/160  \\
5427      &   1.177   & Sp & 55 & $0.32_{-0.13}^{+0.20}$   &   $3.42_{-0.64}^{+0.52}$ & $44.63_{-0.25}^{+0.21}$  &  $43.99_{-0.07}^{+0.06}$  &$<$1.32        &  $<$0.95                   &    148.4/183  \\  
70216     &   1.577   & Ph & 36 & $0.31_{-0.20}^{+0.36}$   &   $0.94_{-0.19}^{+0.17}$ & $44.33_{-0.47}^{+0.34}$  &  $43.70_{-0.07}^{+0.07}$  &$<$4.52        &  $<$0.84                   &    147.5/135  \\
5014      &   0.213   & Sp &155 & $0.31_{-0.09}^{+0.11}$   &  $20.70_{-3.50}^{+2.80}$ & $43.92_{-0.17}^{+0.14}$  &  $43.31_{-0.07}^{+0.05}$  &1.71$\pm$0.35  &  $<$0.54                   &    209.4/218  \\        
\\
\hline         
\end{tabular}\end{center}                 
Column: 
(1) \xmm\ ID number; 
(2) redshift;
(3) Classification (photometric or spectroscopic redshift);
(4) 0.3-10 keV net counts;
(5) Column density in units of $10^{24}$ cm$^{-2}$;
(6) 2-10 keV observed flux in units of $10^{-14}$ erg cm$^{-2}$ s$^{-1}$;
(7) Log of the 2-10 keV rest frame intrinsic luminosity, in units of erg s$^{-1}$. The errors are obtained taking into account the errors on the observed luminosity and on the column density;
(8) Log of the 2-10 keV rest frame observed luminosity, in units of erg  s$^{-1}$;
(9) Intensity of the scatter component, relative to the primary power-law; 
(10) Rest-frame equivalent width of the Fe K$\alpha$ line, in units of keV, from the \plcabs\ fit;
(11) Best fit Cstat/d.o.f. 
$^*$ sources whose CT nature is confirmed by the combined analysis of \xmm\ (pn and MOS) and \chandra\ data (see Tab.~2). 
$^\dag$ source 2608 was already identified as a CT candidate in Hasinger et al. (2007) and Mainieri et al. (2007).
}
\end{table*}                         


\begin{table*}[!t]
\caption{Results of the joint fit of \xmm-pn+MOS+\chandra\ spectra for the CTK$_f$ sample. Sources are ordered by increasing redshift.}
\label{tab:xraytot}
\begin{center}
\begin{tabular}{ccccccccccccc}
\hline\hline\\
\multicolumn{1}{c} {XMM ID}&
\multicolumn{1}{c} {Ch. ID}&
\multicolumn{1}{c} {z}&
\multicolumn{1}{c} {Tot C}&
\multicolumn{1}{c} {\nh}&
\multicolumn{1}{c} {F$_{2-10}$}&
\multicolumn{1}{c} {Log(L$_{I}$)}&
\multicolumn{1}{c} {Log(L$_{O}$)}&
\multicolumn{1}{c} {Sc. \%}&
\multicolumn{1}{c} {EW}&
\multicolumn{1}{c} {C/d.o.f.} \\
(1) & (2) & (3) & (4) & (5) & (6) & (7) & (8) & (9) & (10) & (11)  \\\hline\\ 
2608  &  482  &   $0.125^1$     & 271  & $1.92_{-0.46}^{+0.50}$   & $4.72_{-0.24}^{+0.46}$  &  $43.68_{-0.12}^{+0.10}$ &  $42.10_{-0.04}^{+0.03}$ &0.34$\pm$0.15  &   $0.85_{- 0.45}^{+0.19}$     &  542.6/579  \\  
60211 &  368  &   $0.511^2$     &  85  & $1.16_{-0.21}^{+0.43}$   & $1.17_{-0.22}^{+0.15}$  &  $43.45_{-0.12}^{+0.14}$ &  $42.72_{-0.07}^{+0.04}$ &1.62$\pm$0.90  &   $<$0.61                     &  158.2/217  \\  
60152 &  298  &   $0.579^2$     & 121  & $2.10_{-0.46}^{+0.54}$   & $1.87_{-0.25}^{+0.35}$  &  $44.15_{-0.11}^{+0.12}$ &  $43.07_{-0.03}^{+0.07}$ &$<$0.41        &   $0.89_{- 0.59}^{+0.32}$     &  189.4/236  \\  
54514 &  New  &   $0.707^2$     & 265  & $1.29_{-0.26}^{+0.28}$   & $4.18_{-0.44}^{+0.37}$  &  $44.48_{-0.11}^{+0.09}$ &  $43.57_{-0.05}^{+0.04}$ &1.58$\pm$0.48  &   $0.50_{- 0.32}^{+0.34}$     &  403.9/416   \\ 
54490 &  284  &   $0.908^2$     & 147  & $1.12_{-0.27}^{+0.40}$   & $2.25_{-0.35}^{+0.47}$  &  $44.74_{-0.16}^{+0.15}$ &  $43.48_{-0.10}^{+0.08}$ &$<$1.50        &   $0.54_{- 0.39}^{+0.19}$     &  258.7/261   \\ 
60342 &  576  &   $0.941^3$     &  72  & $1.01_{-0.30}^{+0.69}$   & $1.00_{-0.11}^{+0.12}$  &  $44.11_{-0.17}^{+0.23}$ &  $43.10_{-0.06}^{+0.05}$ &$<$1.22        &   $<$0.71                     &  195.4/196  \\  
5511  &  New  &   $1.023^3$     & 202  & $1.08_{-0.34}^{+0.57}$   & $1.52_{-0.12}^{+0.13}$  &  $44.55_{-0.17}^{+0.18}$ &  $43.55_{-0.03}^{+0.03}$ &6.78$\pm$2.50  &   $<$0.59                     &  401.3/422   \\ 
60361 &  New  &   $1.73^3 $     & 120  & $3.92_{-2.84}^{+... }$   & $1.01_{-0.09}^{+0.10}$  &  $44.75_{-0.57}^{+...}$  &  $43.48_{-0.05}^{+0.04}$ &0.59$\pm$0.45  &   $1.47_{- 0.76}^{+0.50}$     &  335.2/382  \\  
70082 &  747  &   $2.429^{3,*}$ & 102  & $1.20_{-0.47}^{+0.87}$   & $1.47_{-0.13}^{+0.09}$  &  $44.95_{-0.25}^{+0.25}$ &  $43.54_{-0.09}^{+0.11}$ &$<$2.01        &   $<$1.301                    &  402.0/418  \\   
70145 &  708  &   $2.548^3$     & 123  & $1.28_{-0.40}^{+0.54}$   & $0.87_{-0.11}^{+0.23}$  &  $44.63_{-0.18}^{+0.16}$ &  $43.79_{-0.06}^{+0.05}$ &2.72$\pm$0.93  &   $0.89_{- 0.57}^{+0.67}$     &  140.1/138   \\ 
\\
\hline  
\hline  
\end{tabular}\end{center}                 
Column: 
(1) \xmm\ ID number; 
(2) \chandra\ ID;
(3) redshift ($^1$ spectroscopic from SDSS, $^2$ spectroscopic from zCOSMOS-20k, $^3$ photometric from Salvato et al. 2011, $^*$ tentative
spectroscopic redshift of $z=2.710$ from zCOSMOS-deep, see sec.~6.1);
(4) Total 0.3-10 keV net counts;
(5) Column density in units of $10^{24}$ cm$^{-2}$;
(6) 2-10 keV observed flux in units of $10^{-14}$ erg cm$^{-2}$ s$^{-1}$;
(7) Log of the 2-10 keV rest frame absorption corrected luminosity, in units of erg s$^{-1}$. 
The errors are obtained taking into account the errors on the observed luminosity and on the column density;
(8) Log of the 2-10 keV rest frame observed luminosity, in units of erg  s$^{-1}$;
(9) Intensity of the scattered component, relative to the primary power-law; 
(10) Equivalent width of the Fe K$\alpha$ line, in units of keV, from the \plcabs\ fit;
(11) best fit Cstat/d.o.f.
\end{table*}

\begin{figure*}[!t]
\centering
\vspace{0.5cm}
\includegraphics[width=5.5cm,height=4.5cm]{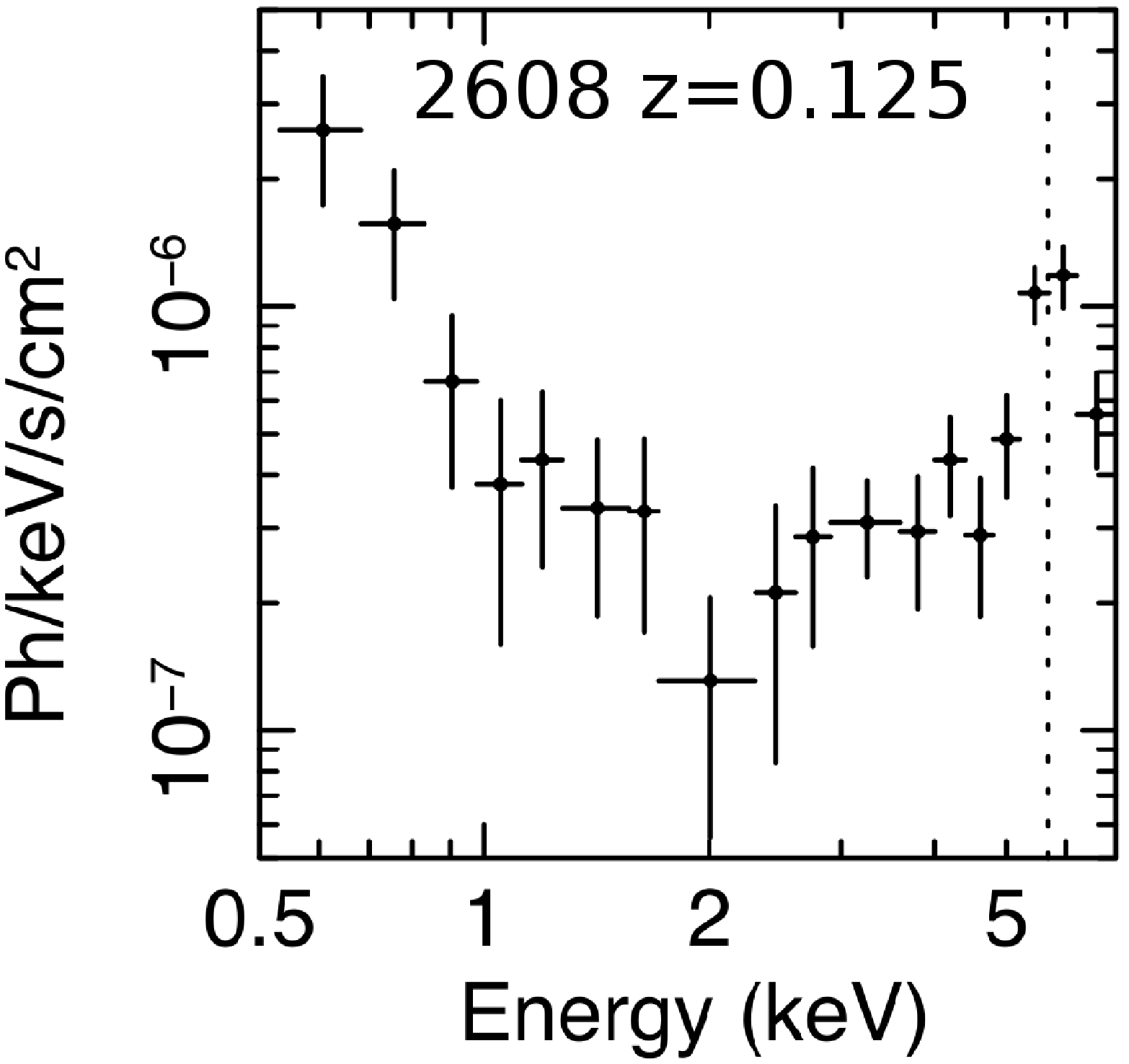}\hspace{0.5cm}\includegraphics[width=5.5cm,height=4.5cm]{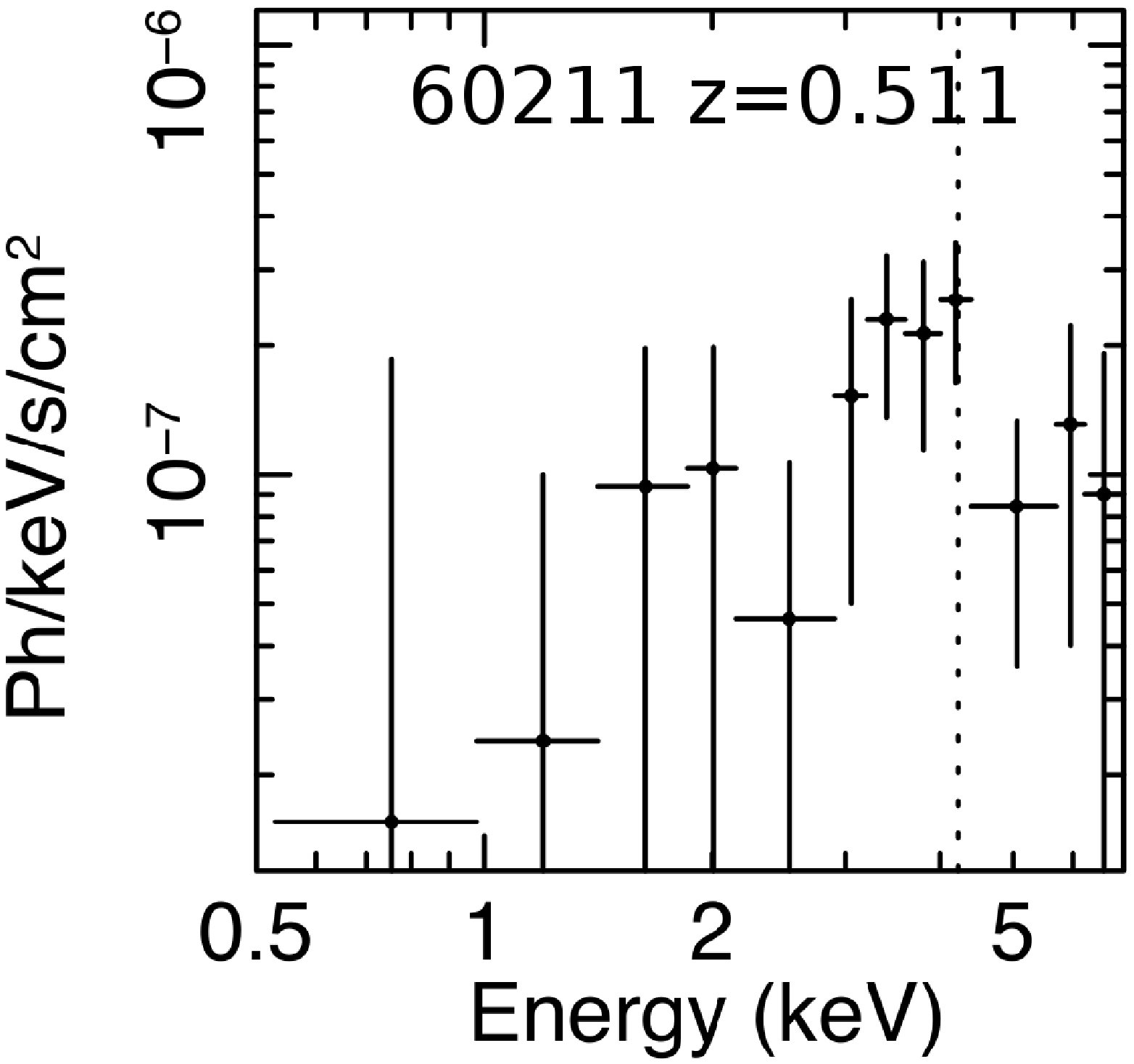}\hspace{0.5cm}\includegraphics[width=5.5cm,height=4.5cm]{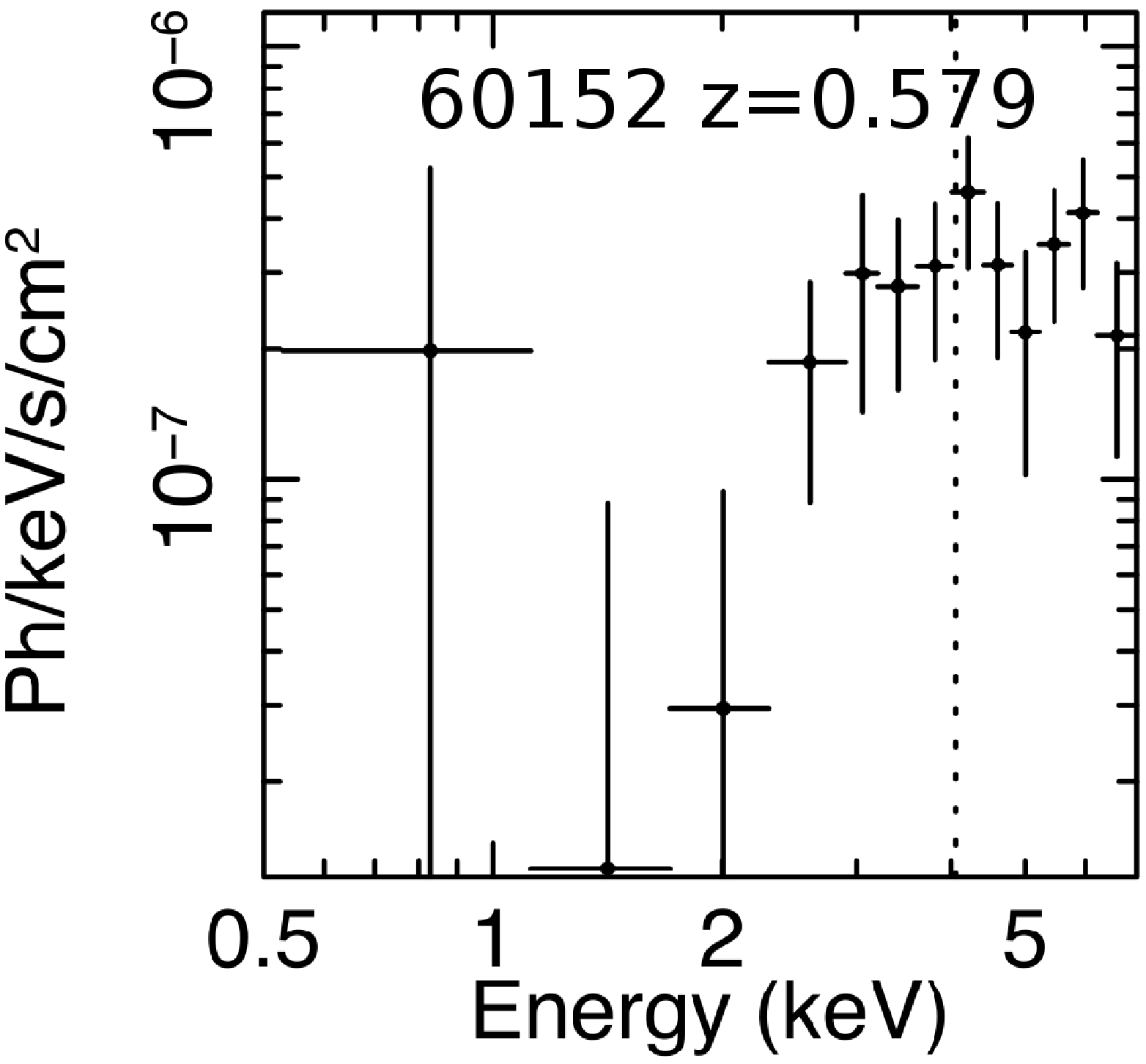}
\vspace{0.5cm}
\includegraphics[width=5.5cm,height=4.5cm]{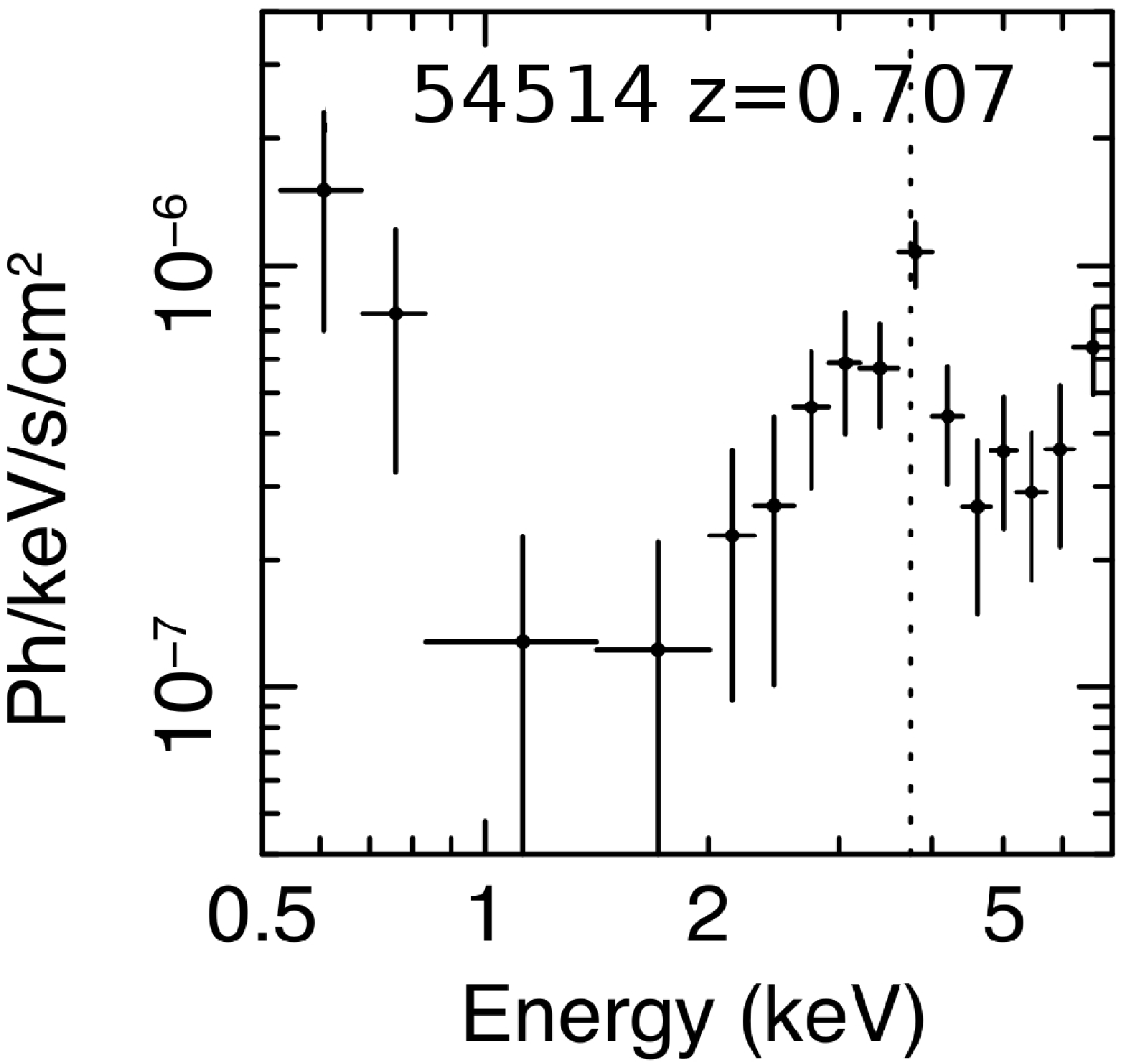}\hspace{0.5cm}\includegraphics[width=5.5cm,height=4.5cm]{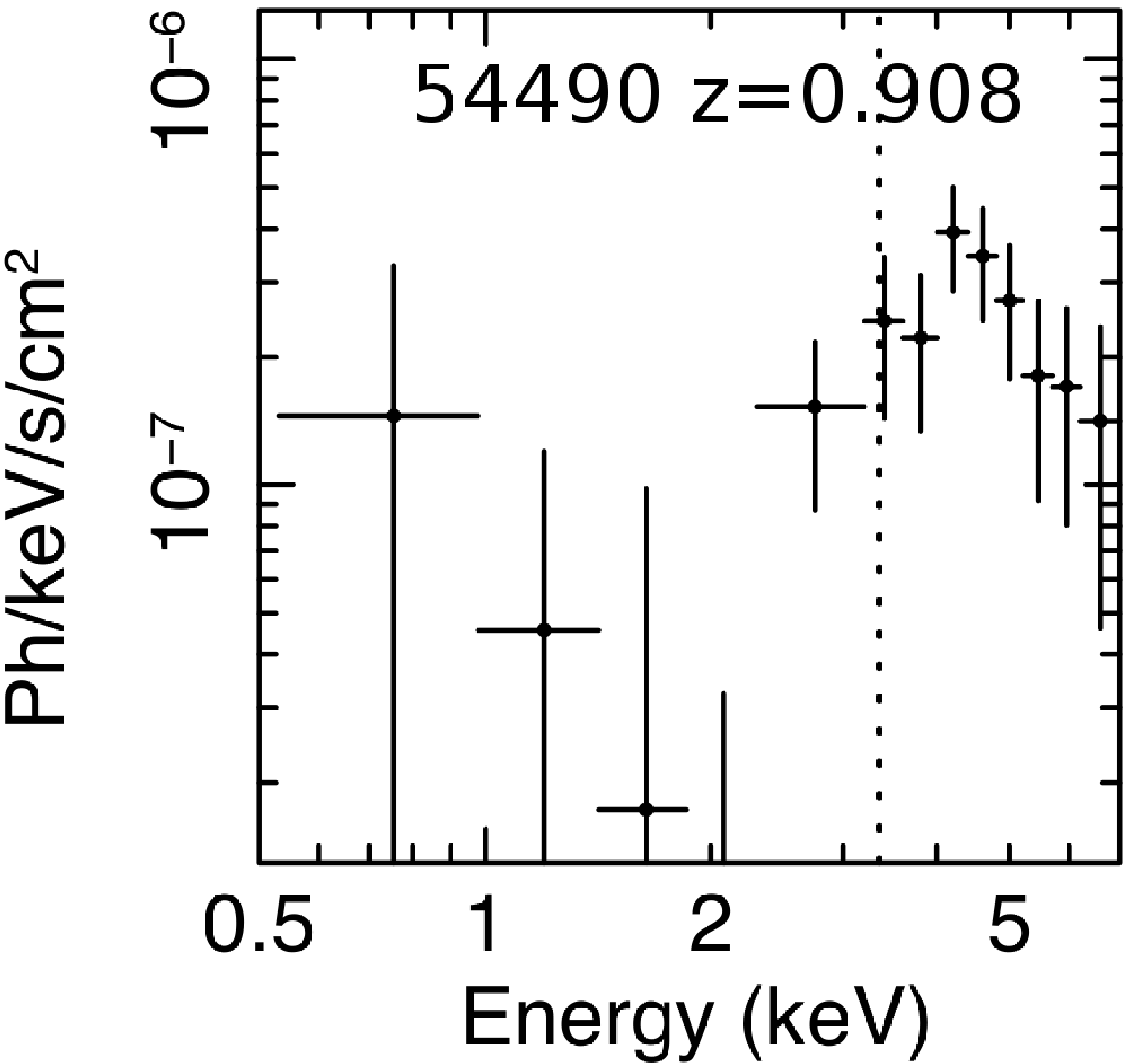}\hspace{0.5cm}\includegraphics[width=5.5cm,height=4.5cm]{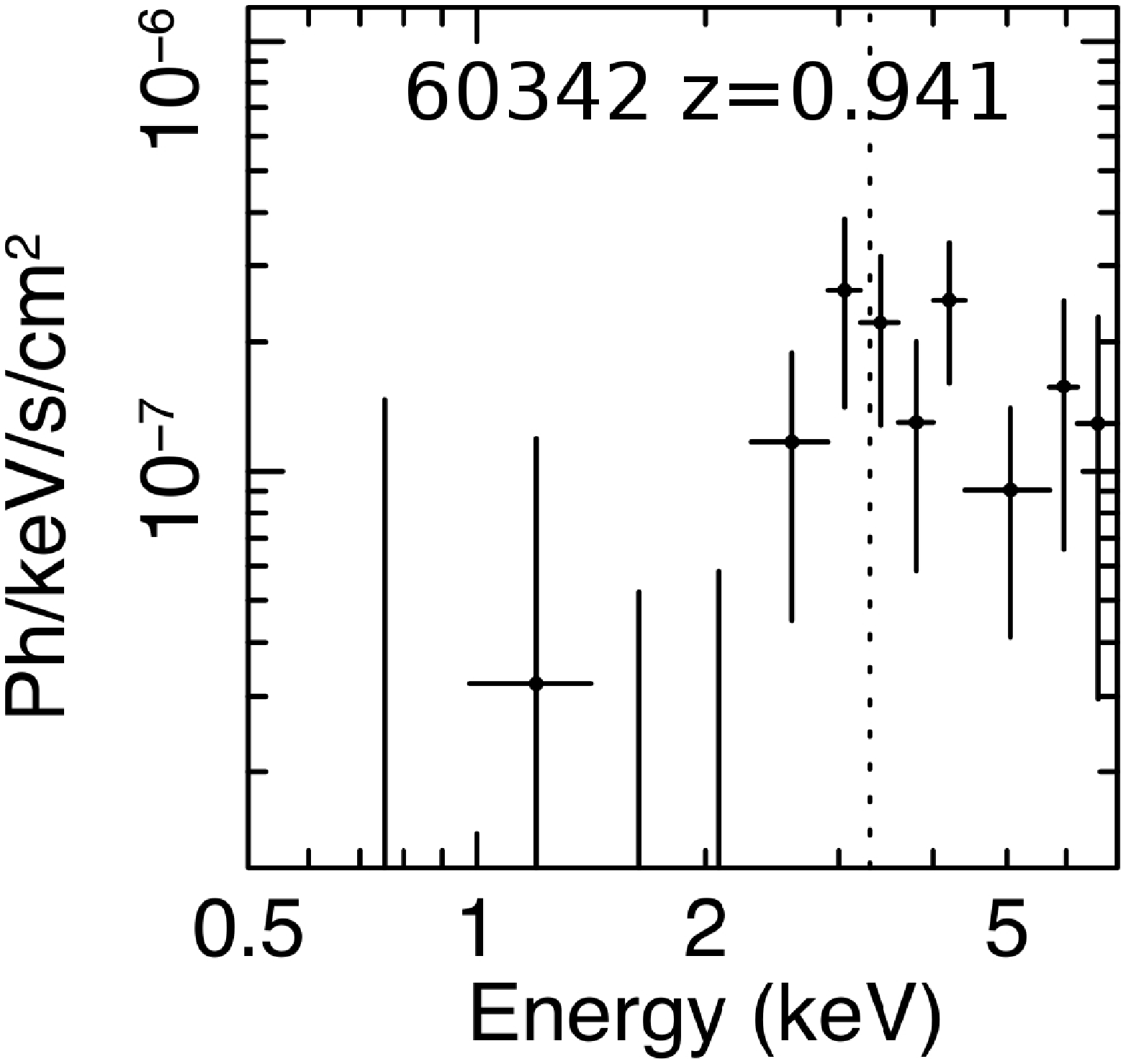}
\vspace{0.5cm}
\includegraphics[width=5.5cm,height=4.5cm]{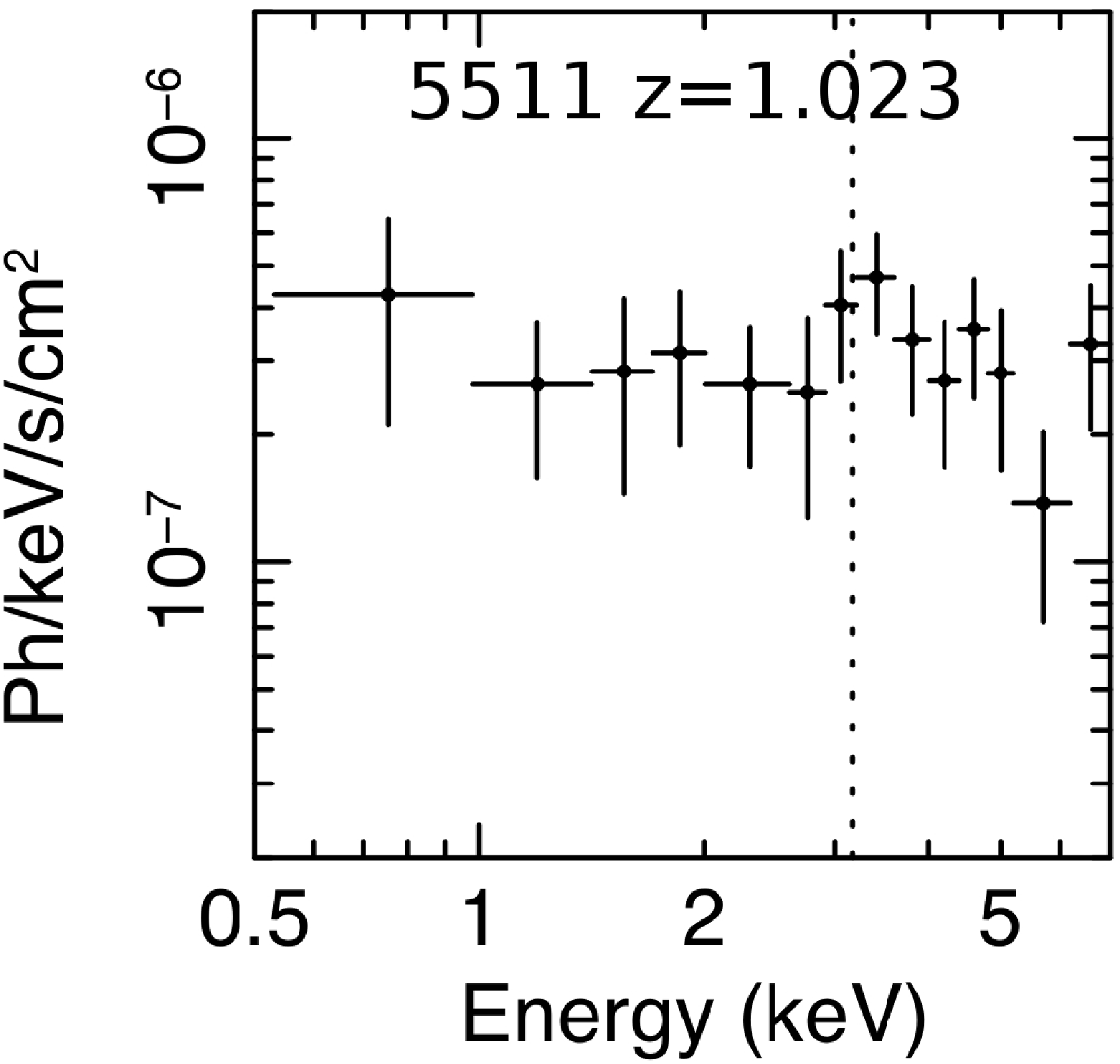}\hspace{0.5cm}\includegraphics[width=5.5cm,height=4.5cm]{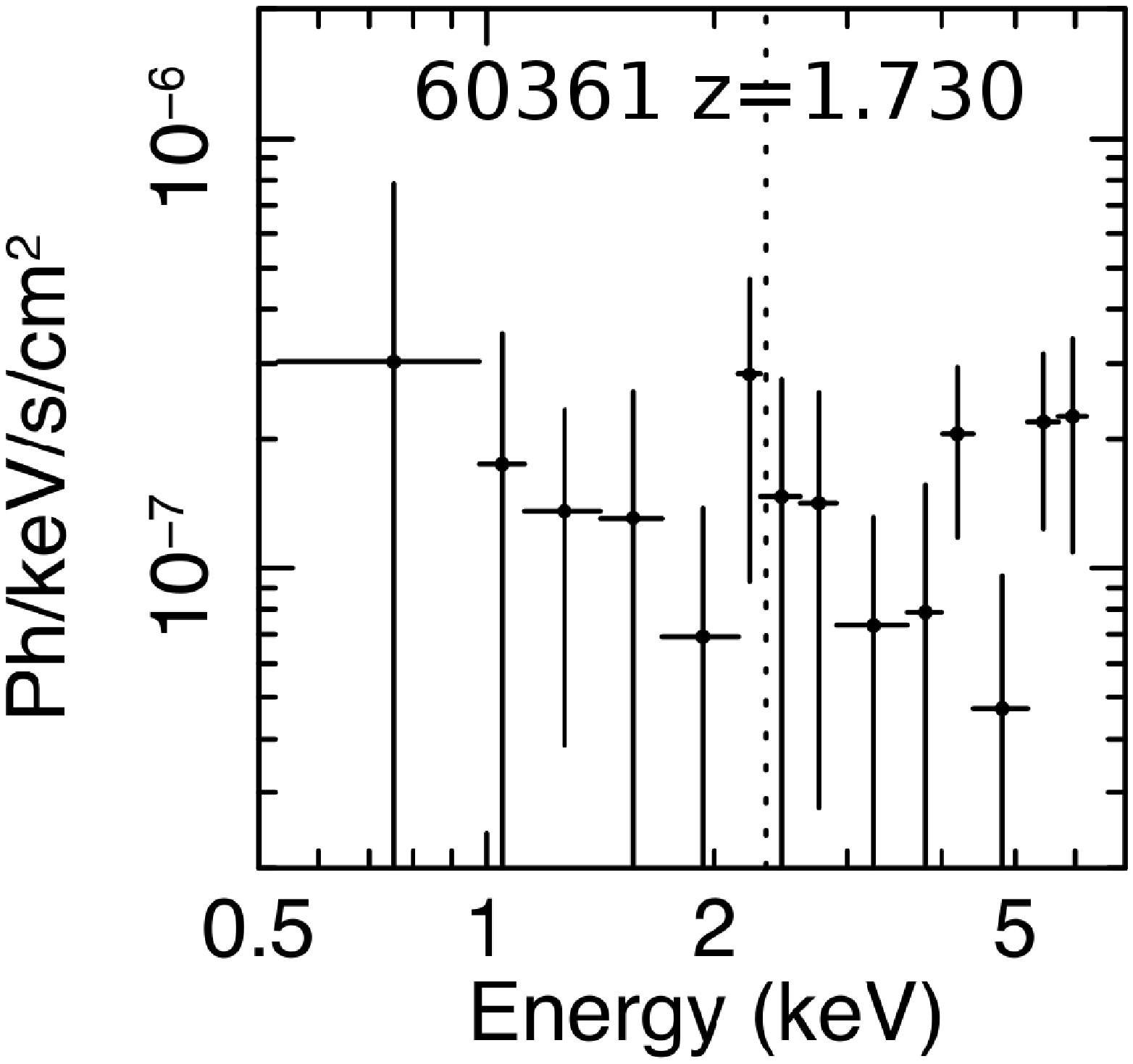}\hspace{0.5cm}\includegraphics[width=5.5cm,height=4.5cm]{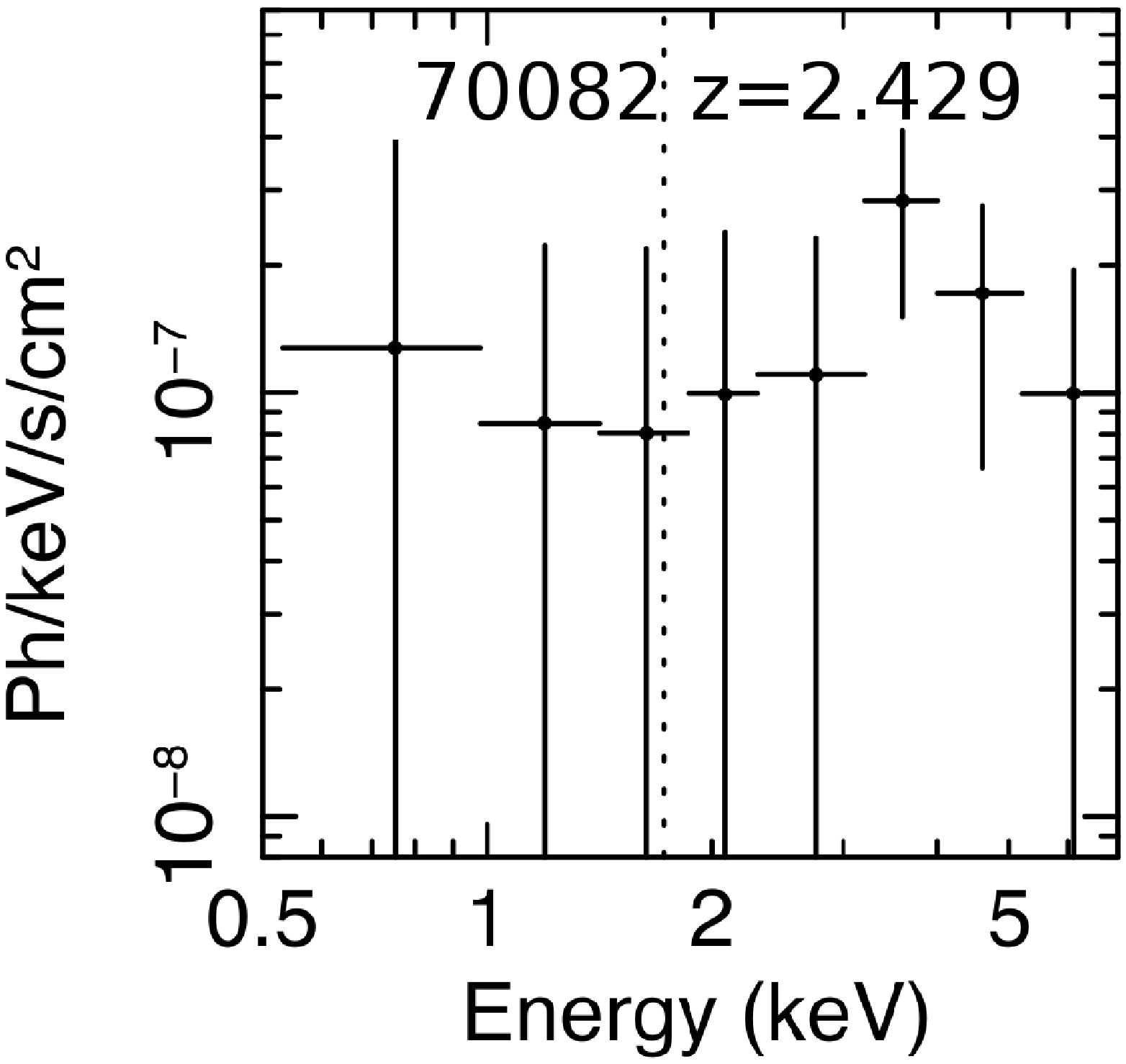}
\vspace{0.5cm}
\includegraphics[width=5.5cm,height=4.5cm]{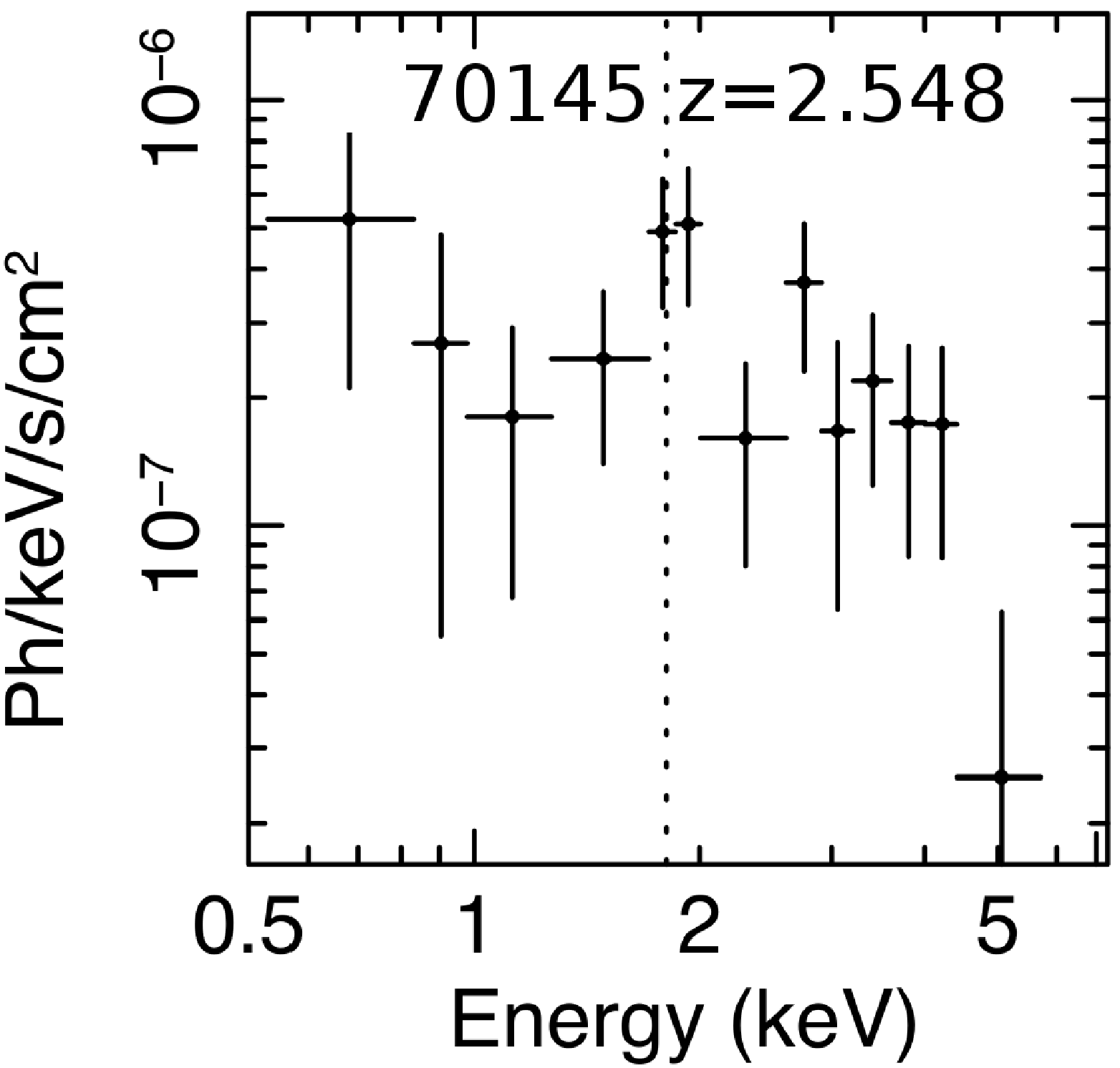}
\caption{'Fluxed', merged pn+MOS+\chandra\ spectra of the CTK$_f$ sources, ordered by increasing redshift. The XMM source ID and redshift are labeled in each panel.
The dashed line marks the expected location of the 6.4 keV Fe K$\alpha$ line.}
\label{xrayspec}
\end{figure*}

The fit results are summarized in Table 1 for the entire obscured sample.
All the parameters, except for the Fe K$\alpha$ line EW, are derived from the \tor\ model fit.
The ten sources belonging to the CTK$_i$ sample (the first group of sources in table 1) show a CT absorber from both fits, 
and five of them have a detection of an emission Fe line with EW consistent with the CT nature 
(i.e. EW $\simgt 1$ keV rest frame).
For three sources, the \nh\ is unconstrained at the upper end (the \nh\ upper limit of the TORUS template is $10^{26}$ cm$^{-2}$).
Source XID 2608, the lowest redshift, brightest CT candidate, was identified as a CT candidate already in Hasinger et al. (2007, the so called 'pink source', 
for its peculiar \xray\ 'colors'), and in Mainieri et al (2007).

Five sources (the second group of sources in table 1) are border line: they are either seen as CT from one model but not by the other (see Appendix B), 
or they are in both models very close to the dividing line (\nh$=10^{24}$ cm$^{-2}$)
and with large upper error bars, that make them fully consistent with being CT. The remaining 24 sources are highly obscured but in the Compton thin regime.
All of the latter have only an upper limit for the EW of the Fe line.

Given the relatively poor photon statistics available for all our sources, 
and the availability of deeper \xray\ data in the COSMOS field, 
we decided to look more carefully into the \xray\ properties of the small sample of CTK$_i$ sources, plus the 5 border-line sources. 
The idea is to collect all the available \xray\ data for this subsample, with the aim of
improving the available statistics, and possibly confirm, with better data, the CT nature of these sources.

\subsection{\xmm\ pn, MOS and \chandra\ spectra}

For the majority of our sources, \xmm\ MOS1 and MOS2 data are available, each with roughly the same net exposure of pn.
The sum of MOS1 and MOS2 spectra typically increases the available number of counts by a factor 1.5.
We extracted the spectra for the candidate CT sources from the MOS1 and MOS2 cameras in each observation, following the procedure developed in Ranalli et al. (2013) 
for the XMM-CDFS. We then merged the spectra together (MOS1+MOS2) producing average matrices using standard HeaSOFT tools\footnote{http://heasarc.nasa.gov/lheasoft/}.

Furthermore, merging the original C-COSMOS data (Elvis et. al 2009) with the new Legacy data (Civano et al. 2014, observations concluded in April 2014), 
we have a \chandra\ counterpart for all our sources.
The \chandra\ observation of the COSMOS field is much deeper ($\sim160$ ks average) and affected by much lower background levels.
Therefore, even if the net number of counts available from the \chandra\ spectra is 
typically of the same order as the \xmm\ pn one (in the range 30-100, thus doubling the number of counts available),
the S/N ratio is always much higher.
The results of the joint fit (pn+MOS+\chandra) are indeed more reliable and the parameters better constrained 
(average error on \nh\ of 35\% instead of 56\%),
and they can be used as an {\it a-posteriori} test of the reliability of the XMM-pn only fits.
The systematic differences, in flux and photon index, between \chandra\ and \xmm, found for faint sources (Lanzuisi et al. 2013a) 
are too small to have any measurable effect, given the data quality of the sample discussed here.

We performed a joint fit of the three spectra for each source, using the same models described in Sec.~3.
For 8 out of 10 sources, classified as CT from the \xmm\ pn spectrum alone, we can confirm, through the simultaneous fit of \chandra\ and \xmm\ pn and MOS spectra, 
that the \nh\ is indeed $>1\times10^{24}$ cm$^{-2}$, and therefore in the CT regime. Therefore they will be included in the final CTK sample (CTK$_f$ hereafter).
Two sources from the original CTK$_i$ sample, namely XID 217 and XID 60314, have instead a best fit value of \nh$\sim3-4\times10^{23}$ cm$^{-2}$:
when observed with deeper \xray\ data they fall into the Compton thin regime and are excluded from the CTK$_f$ sample, and will be included in the final CTN sample (CTN$_f$ hereafter).
   
Furthermore, two of the border-line sources, namely XID 54490 and XID 70145 show, in the joint fit, a column density  which is actually higher
than the one measured by the \xmm\ pn fit alone, and in the CT regime, 
and a strong Fe K$\alpha$ line, again consistent with these sources being CT. We therefore include them in the CTK$_f$ sample, see Tab.~2.
The other 3 border-line sources show high levels of obscuration, but in the Compton thin regime, and are therefore not included in the CTK$_f$ sample.

\begin{figure}[!h]
\centering
\includegraphics[width=8cm,height=8cm]{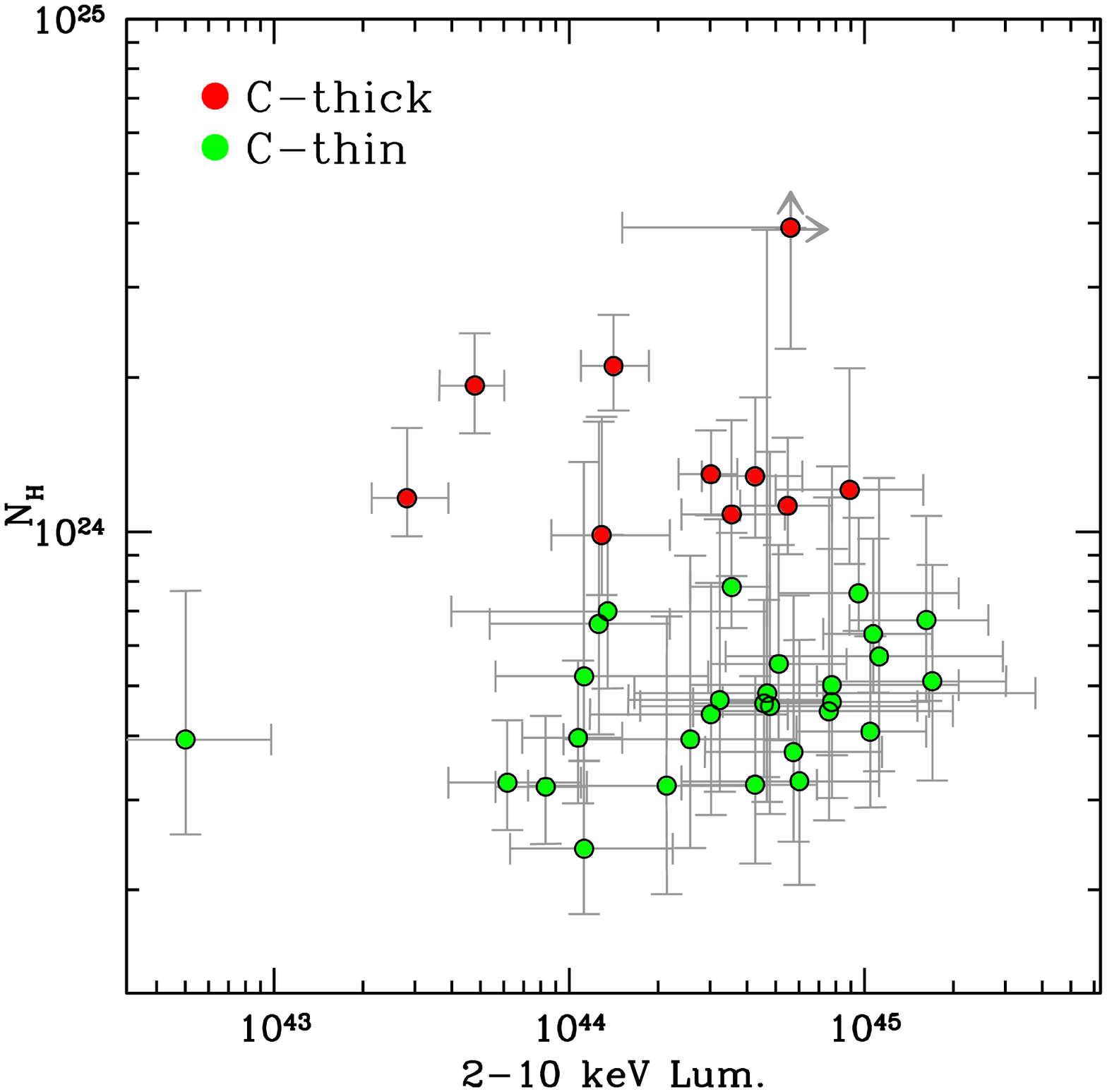}
\caption{2-10 keV rest frame, absorption corrected luminosity vs. column density. In red are shown CTK$_f$ sources, and in green CTN$_f$ sources.}
\label{lumnh}
\end{figure}

We conclude that the selection method, based on the simple two power law model, applied to \xmm\ spectra alone
has a selection efficiency of $\sim80\%$:
given the above results, we indeed measure a 20\% contamination by non CT sources in the sample, 
and 20\% of missed CT out of the sample.
This can be compared with results from Brightman \& Ueda (2012): through simulations, they estimated a similar, rather constant selection efficiency,
for sources with $z>1$ (in this \nh\ regime the shift of the intrinsic spectrum through the observed band helps in identifying CT sources),
while the selection efficiency at lower redshift decreases strongly with the number of counts. 
Indeed our lowest redshift source (XID 2608, z=0.125) is also the one with the largest number of counts, and would be probably classified as a soft source (see the spectrum in Fig. 4),
in case of shallower \xray\ data.

Table 2 summarizes the final results for the CTK$_f$ sample (8 from the CTK$_i$ sample and 2 from the border-line sample), ordered by increasing redshift. 
Column 2 shows the ID of the \chandra\ counterpart, if already available in the catalog of Elvis et al. 2009,
column 3 show the redshift (and its origin) for each source,
while column 4 shows the total number of net counts (pn+MOS+\chandra).
The rest of the columns show the same parameters of Table 1, this time obtained from the joint fit of all the instruments.

Finally, we co-added all the available counts, following Iwasawa et al. (2012).
We show in Fig.~\ref{xrayspec} the resulting total spectra for all the CTK$_f$ sources.
The spectra are de-convolved for each instrumental response, then merged together, and shown in units of $photons/keV/s/cm^{2}$.
In this way we show the intrinsic spectrum, free from the distorting effects of the instrumental response, 
without imposing any preferred model, and therefore providing a more objective visualization of the 'fluxed' spectra\footnote{ See Appendix C
for an example of a spectrum plus model plot.}.
All sources appear to be strongly absorbed, up to rest frame energies of 7-10 keV.
Furthermore, more than half of the sources show some evidence of an emission line in correspondence of the expected rest frame Fe K$\alpha$ line energy (dashed line in each panel).
Indeed, 6 out of 10 CTK$_f$ sources have a detection of the  Fe K$\alpha$ line in the joint fit (column 10 of Tab.~2).
Our result is strengthened by the fact that all our sources have an \nh\ constrained to be $>10^{23.7}$ cm$^{-2}$ within the 90\% error-bar.

Fig.~\ref{lumnh} shows the final distribution of 2-10 keV intrinsic, 
absorption corrected luminosity vs. column density, for all the sources in the sample (the CTK$_f$ in red and CTN$_f$ in green).
As expected by the fact that these sources are highly obscured, and that the XMM-COSMOS survey has a shallow X-ray limiting flux,
almost all sources are in the QSO regime (\lum$>10^{44}$ erg s$^{-1}$),
with a few exceptions due to the very low redshift sources.
For the most obscured source of the CTK$_f$ sample (XID 60361) the upper boundary of the \nh\ distribution 
is not constrained, due to the very low 
quality of the spectrum (see Fig. 4). As a result, also the 2-10 keV absorption corrected luminosity has no upper boundary.
However, assuming as upper limit of \nh=$10^{25} cm^{-2}$, we can estimate the resulting Log(\lum) upper limit to be $\sim45.4$ \ergs.

   \begin{figure*}[!t]
   \centering
      \includegraphics[width=8cm,height=8cm]{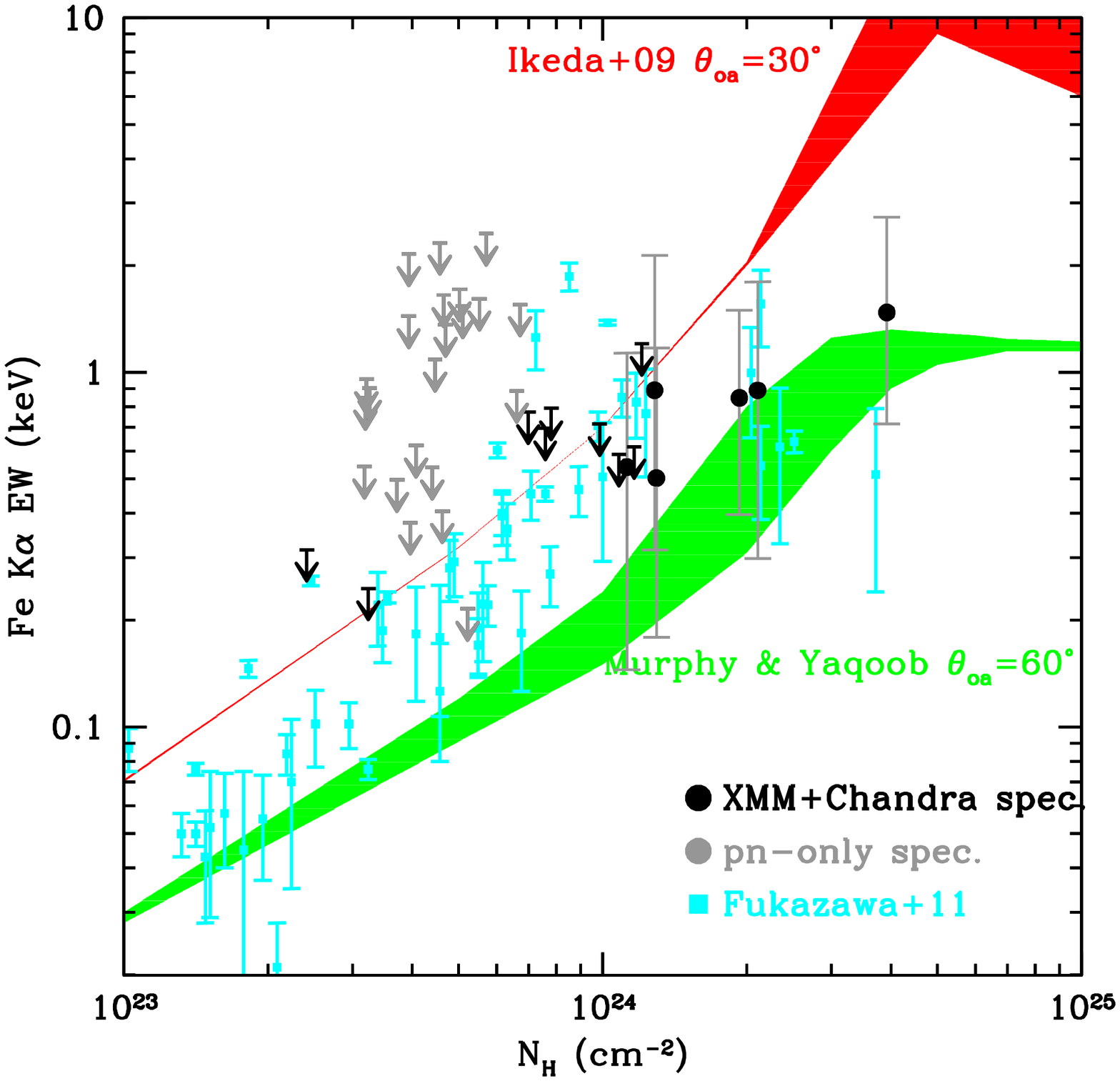}\hspace{0.5cm}
      \includegraphics[width=8cm,height=8cm]{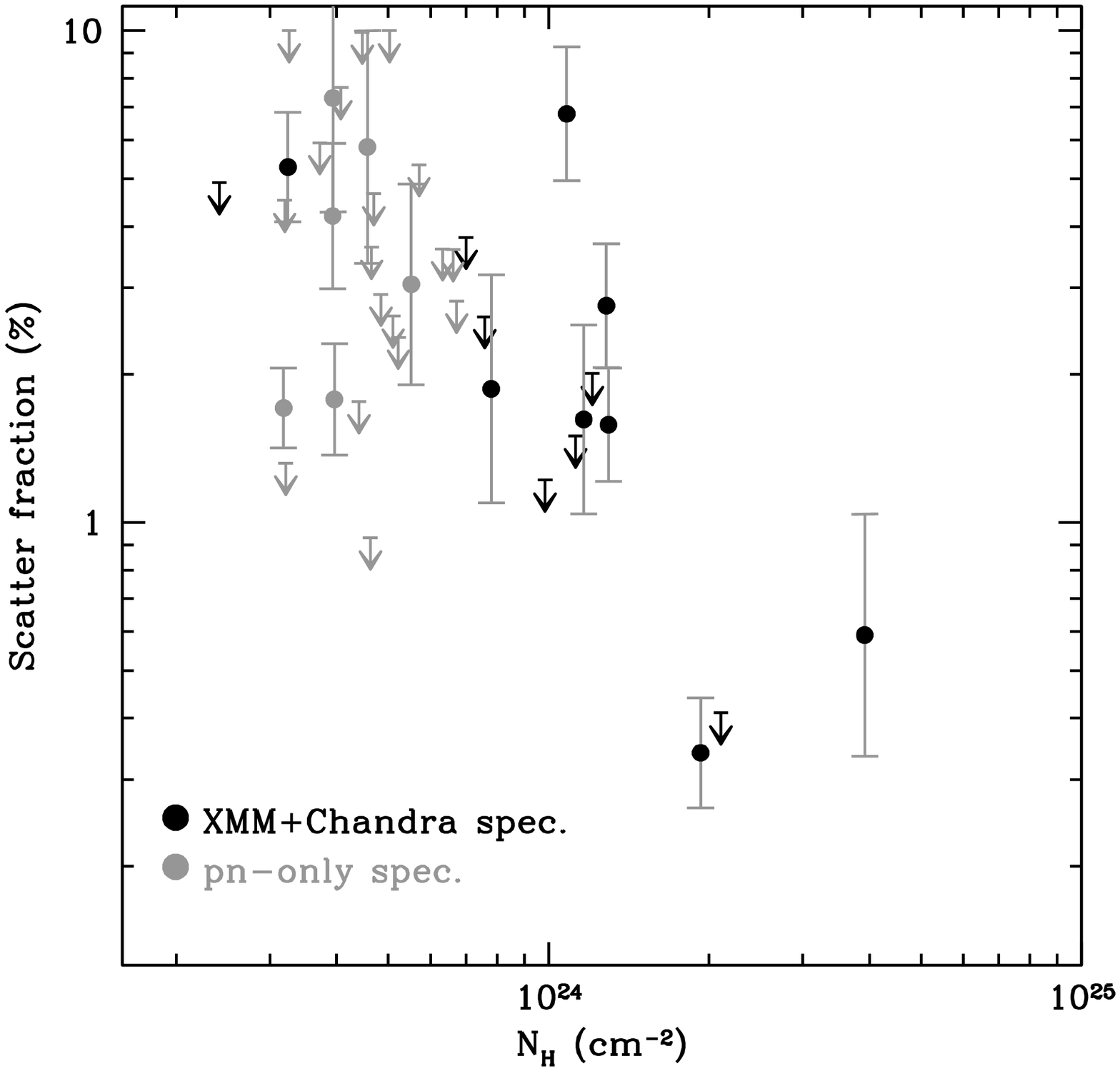}
   \caption{{\it Left panel:} Column density vs. Fe K$\alpha$ EW for the obscured sources with pn+MOS+\chandra\ joint fit (black) and pn only fit (gray). The red (green) shaded areas show the expected
    Fe K$\alpha$ EW as a function of \nh\ as computed in the torus model of Ikeda et al. 2009 (Murphy \& Yaqoob 2009). Cyan points represent the observed \nh\ and EW for a sample of 
    local Seyfert galaxies observed with Suzaku (Fukazawa et al. 2011). 
    {\it Right panel:} Scattered fraction vs. column density for the obscured sources with pn+MOS+\chandra\ joint fit (black) and pn only fit (gray).}
   \label{nhew}
   \end{figure*}

Fig.~\ref{nhew} (left) shows the rest frame equivalent width of the Fe K$\alpha$ line as a function of the column density \nh.
Sources for which the pn+MOS+\chandra\ joint fit has been performed are shown in black, while the remaining are in gray.
It is well known that, as the EW is measured against a suppressed continuum, its value increases with increasing column density,
from few tens eV in type-1 AGN (Bianchi et al. 2007), 
to values of several hundred eV to over 1 keV in Compton-thick sources (Matt et al. 1997; Guainazzi et al. 2000).
The red shaded area in fig.~\ref{nhew} (left) shows the expected Fe K$\alpha$ EW as a function of \nh\ 
as computed in the torus model of Ikeda et al. (2009),
for a torus half-opening angle of $30^{\circ}$ and an inclination angle in the range $60-84^{\circ}$. 
The green shaded area shows the predictions from the torus model of Murphy \& Yaqoob (2009) 
for a torus half-opening angle of $60^{\circ}$ and similar inclination angles ($60-84.26^{\circ}$).
The number of detected lines in our sample is small (and all above \nh=10$^{24}$ cm$^{-2}$).
Therefore we decided to include 55 highly obscured sources (in cyan) taken from a sample of 88 local Seyfert galaxies analyzed in Fukazawa et al. (2011) observed with Suzaku,
to put in a context our result.
The two samples populate differently the EW-\nh\ plane, but are in very good agreement where they overlap (e.g. around \nh$\sim10^{24}$ cm$^{-2}$).
All together, these observational results show that, while for Compton-thin absorbers the observed EW lies between
the predictions of the two models, and closer to the prediction of the Ikeda et al. model at \nh$\sim10^{24}$ cm$^{-2}$ (i.e. EW$\sim0.5-1$ keV), the increase in EW saturates above $10^{24}$ cm$^{-2}$,
and the EW remains in the range $\sim0.5-1.5$ keV even for \nh$=4\times10^{24}$ cm$^{-2}$, closer to the predictions of the Murphy \& Yaqoob model.

Fig.~\ref{nhew} (right) shows the fraction of the soft component, with respect to the primary power-law, as a function of \nh.
If the soft component is indeed produced by scattered light,
and assuming that the covering factor of the torus increases with obscuration,
the scattered fraction is expected to decrease with increasing \nh\ (Brightman \& Ueda 2012). 
There is indeed some evidence that, above \nh$=1.5\times10^{24}$ cm$^{-2}$, the typical scattered fraction is significantly lower (below 1\%) with respect to less severely obscured sources.
However, given the limited sample available, and the large number of upper limits on the scattered fraction, plus the limited dynamical range probed, (especially in \nh) 
we cannot draw any firm conclusion in this case.
A similar but more significant trend of decreasing scattered fraction with increasing \nh\ has been found in Brightman et al. (2014), 
in a much larger sample of CT sources, in two redshift bins.

\subsection{Bolometric luminosity}

   \begin{figure}[!t]
   \centering
   \includegraphics[width=8cm,height=8cm]{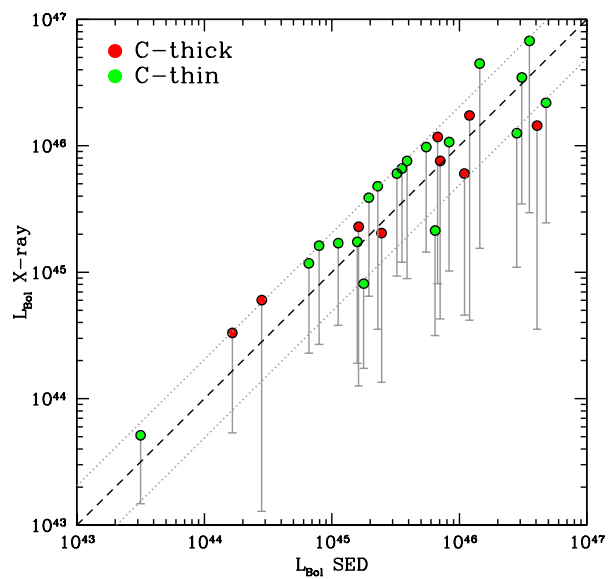}
   \caption{Comparison between the SED-based L$_{bol}$ from L12 or D14 and the \xray based L$_{bol}$ for CTK$_f$ (CTN$_f$) sources in red (green).
   The error-bars show the difference between the L$_{bol}$ computed from the absorption-corrected and the observed \lum.
   The dashed line shows the 1:1 relation, while the dotted lines show the $1\sigma$ dispersion of 0.31 dex.}
   \label{lumbol}
   \end{figure}

Given the wealth of multi-wavelength data available in the COSMOS field,
and the great effort already put in the data analysis (see Sec.~2.1), 
we can rely on the independent measurement of the bolometric luminosity (L$_{bol}$),
available for our sources from SED fitting.
This information can then be used in order to test our selection method for highly obscured sources.

The absorption-corrected \xray\ luminosity, after the application of a bolometric correction k$_{bol}$,
is generally considered a good measure of the AGN bolometric luminosity because it is less affected by  obscuration/reprocessing than other wavelengths
The \xray\ bolometric correction, ranging from factor 	$\sim10$ to $\sim100$, is known to be luminosity dependent 
(i.e. brighter objects have larger \xray\ k$_{bol}$, Marconi et al. 2004; Hopkins et al. 2007; L12).
Even if this general trend is robust, different k$_{bol}$-L$_{bol}$ relations have been derived in the literature for different sample selections of AGN,
and the scatter is usually large.
We decided to use for our sources the 2-10 keV k$_{bol}$-L$_{bol}$ relation computed in L12,
which is derived from a large sample of \xray\ selected type-2 AGN from the same XMM-COSMOS catalog.
The uncertainties in k$_{bol}$ is estimated to be of the order of $\sim0.20$ dex.
The \xray\ based L$_{bol}$ is shown in the y-axis of Fig.~\ref{lumbol}, together with the 
L$_{bol}$ that would be obtained using the observed \lum\ (error-bars).

On the other hand, it is possible to estimate the AGN bolometric luminosity,
assuming that the AGN mid-IR luminosity 
is an indirect probe (through absorption and isotropic re-emission) 
of the accretion disc optical/UV luminosity.
The accretion luminosity is then computed through the SED fitting,
by first decomposing the host and AGN emission and then
integrating the AGN IR luminosity L$_{IR}$, e.g. between 1 and 1000 $\mu m$, correcting the IR emission to account for the obscuring torus geometry and optical thickness (see Pozzi et al. 2007, 2010).
The bolometric luminosity is then obtained adding the absorption corrected total \xray\ luminosity, computed in the 0.5-500 keV energy range\footnote{$L_{0.5-500}$ is 
estimated from the observed $L_{2-10}$ assuming a power-law spectrum with $\Gamma= 1.9$ and an exponential cut-off at 200 keV. 
The value found for the ratio $L_{0.5-500}/L_{2-10}$ is $=4.1$.}.
We collected SED based bolometric luminosities for 9 out of 10 CTK$_f$ sources and 19 out of 29 CTN$_f$ either from L12 or D14.
There is a good agreement between the two quantities,
with a small systematic shift in the direction of higher \xray\ based L$_{bol}$.
The agreement is even more striking considering the large differences between the L$_{bol}$ estimated 
from the observed and absorption corrected \lum, shown with the vertical error-bars. 
The dotted lines in the figure marks the $1\sigma$ dispersion of 0.31 dex.

We stress that the contribution of the \xray, absorption corrected luminosity to the L$_{bol}$ from SED fitting
is very small (of the order of $\sim15\%$),
and therefore the SED based  L$_{bol}$ is largely independent of the \nh\ as estimated from the \xray.
On the other hand, the \xray\ based L$_{bol}$ is extremely sensitive to the \nh, given that an overestimate of the obscuration would lead to an overestimate 
of the \xray\ luminosity, that is finally multiplied by the 10-100 factor of the k$_{bol}$.
Therefore, we can interpret the fact of having a good agreement between the two quantities (within a factor $\sim2$), 
as a further confirmation of the reliability of our \nh\ estimates.

%

\section{SMBH and host properties of highly obscured sources}

\begin{figure*}[!t]
   \centering
      \includegraphics[width=6.0cm,height=6.0cm]{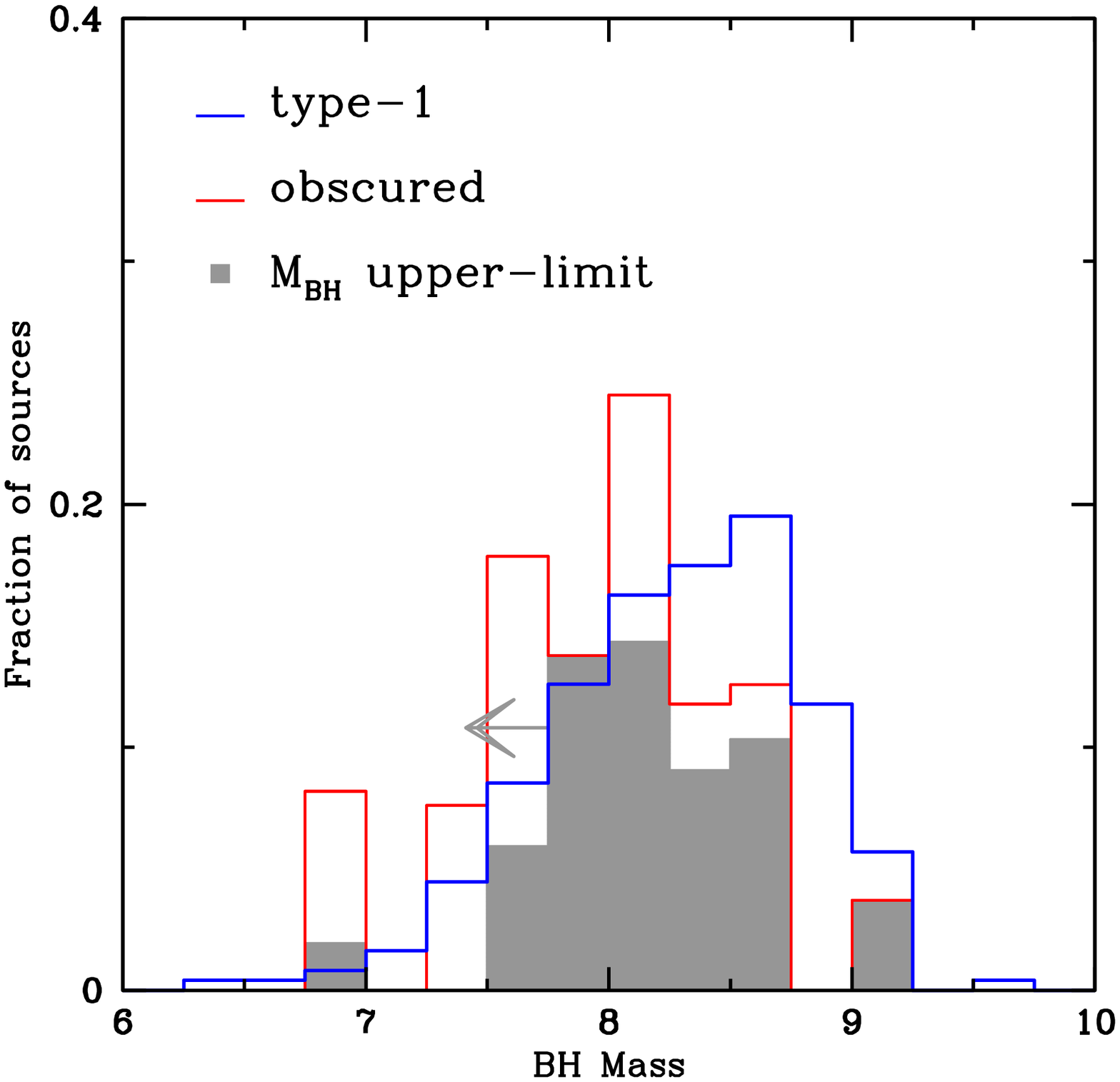}\hspace{0.05cm}
      \includegraphics[width=6.0cm,height=6.0cm]{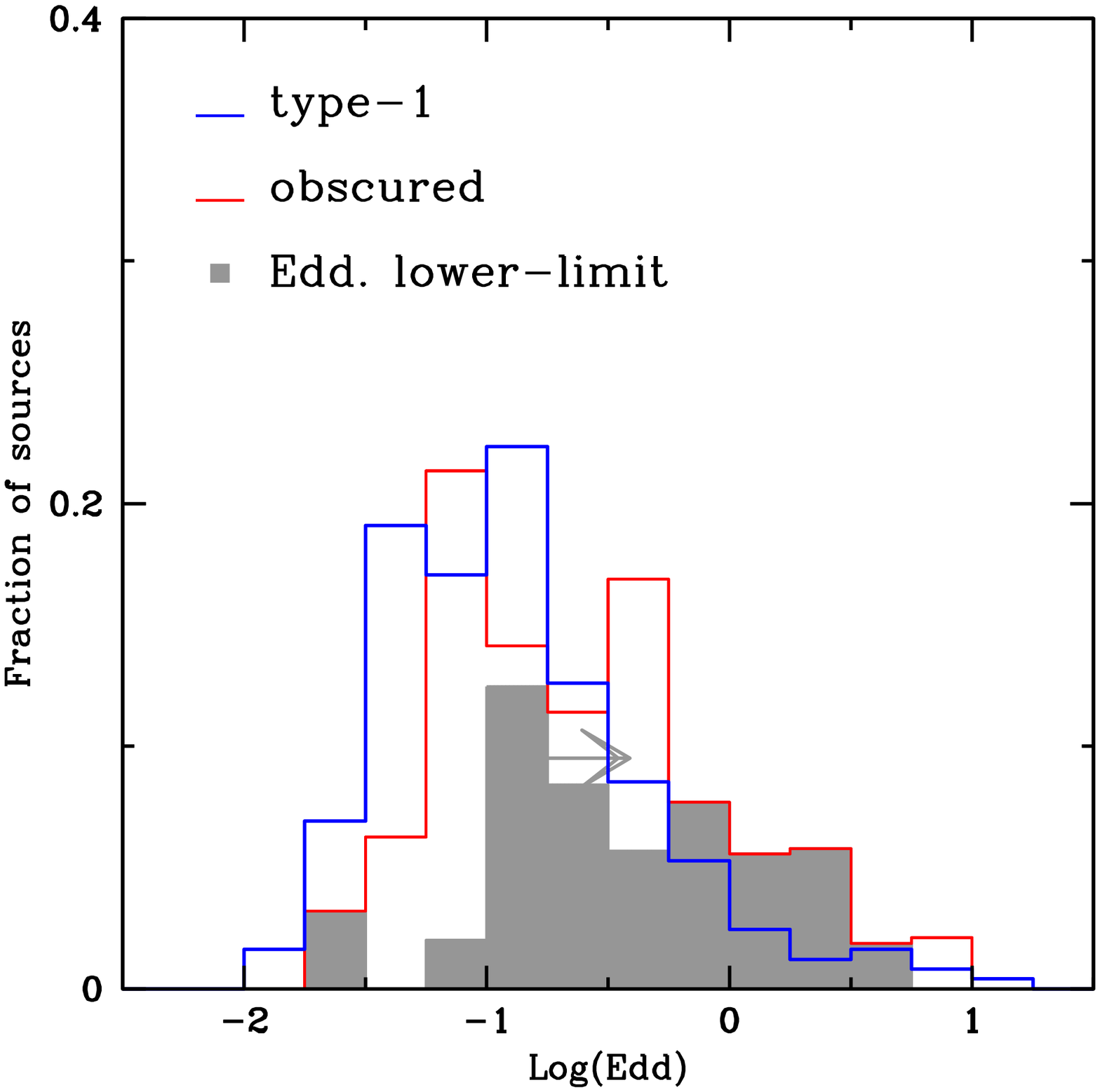}\hspace{0.05cm}
      \includegraphics[width=6.0cm,height=6.0cm]{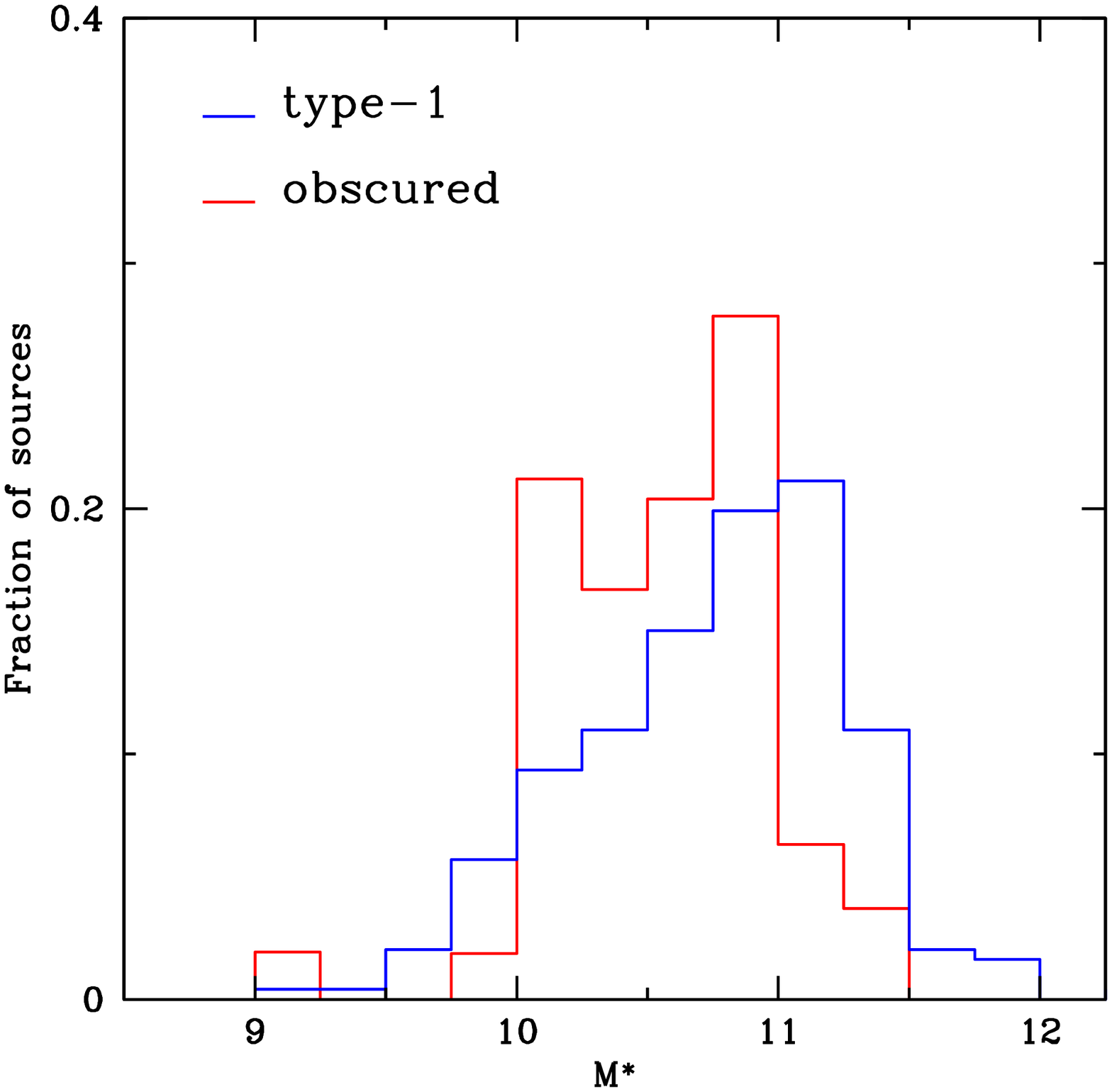}
   \caption{{\it Left panel:} Fractional distribution of Log($M_{BH}$) for type-1 sources (in blue) and highly obscured sources (red).
   The gray shaded histogram and arrow show obscured sources for which the $M_{BH}$ must be considered an upper limit, due to lack of bulge-to-total mass ratio information.
   {\it Central panel:}  Fractional distribution of Log($\lambda_{Edd}$) (same colors of left panel). 
  The gray shaded histogram and arrow show obscured sources for which the $\lambda_{Edd}$ must be considered a lower limit, due to the upper limit in $M_{BH}$.
  {\it Right panel:} Fractional distribution of the total stellar mass, as computed through SED fitting either in L12, B12 or D14 (same colors of left panel).}
   \label{istobh}
   \end{figure*}
An important goal of searches for highly obscured AGN, in the context of AGN/galaxy co-evolution models,
is to understand if they occupy a special place in the proposed evolutionary track that goes from gas inflow 
(driven by mergers or internal processes, e.g. Hernquist 1989; Ciotti \& Ostriker 1997; Hopkins et al. 2006)
to highly accreting/highly star-forming systems, to unobscured quasar in red ellipticals
or if such evolutionary track exists at all.
The key to test these models resides in looking for the distribution of physical SMBH and host properties, 
such as BH mass and Eddington ratio ($\lambda_{Edd}$\footnote{Defined as L$_{bol}$/L$_{Edd}$, where L$_{Edd}$ is the 
Eddington luminosity associated to a given BH mass, i.e L$_{Edd}$= $1.3\times10^{38}$ erg s$^{-1}$ per $M_{\odot}$.}),
stellar mass ($M_*$) and specific star formation rate (SFR/$M_*$, sSFR). 
For example, if this evolutionary scenario is true, we expect to see these highly obscured systems
to have lower BH masses, accrete close to or possibly above the Eddington limit, and reside in hosts with greater star formation compared to unobscured AGN.
On the other hand, if these sources are seen as obscured only for geometrical effects, no such trends are expected.

\subsection{$M_{BH}$, $\lambda_{Edd}$ and $M_*$}

Because the sample of highly obscured sources is rather small, we analyzed CTK$_f$ and CTN$_f$ sources together.
We stress however that the distribution of the parameters discussed in this section is very similar
between CTK$_f$ and CTN$_f$.
We have available $M_{BH}$ (and hence $\lambda_{Edd}$) for 25 out of 39 of the obscured sources, from L12.
These were obtained rescaling the $M_*$, obtained from the SED fitting and corrected for the bulge-to-total mass fraction, $f_{bulge}$
(when morphological information was available, from the ZEST catalog, Scarlata et al. 2007, Sargent et al. 2007).
The scaling relation of H\"aring \& Rix (2004) was adopted, taking also into account its redshift evolution as estimated in Merloni et al. (2010).
Typical uncertainties are estimated to be of the order of $\sim0.5$ dex.

For the remaining sources, we rely on the $M_*$ computed in B12 or D14\footnote{We 
verified that, when in common, the $M_*$ estimates between the 3 catalogs are always in good agreement, within 0.3 dex, with no systematic shift.}, 
and estimated the $M_{BH}$ following the same steps described in L12.
For all the sources without morphological information, from which an estimate of the $f_{bulge}$ need to be inferred,
the $M_{BH}$ is considered to be an upper limit.
We also derived the $\lambda_{Edd}$ for all our sources, using the $M_{BH}$ described above, and the \xray\ based bolometric luminosities illustrated in Sec.~4.2.

We then compared the distribution of these quantities, with the one of a control sample of unobscured AGN.
 $M_{BH}$ for a large sample of type-1 AGN in XMM-COSMOS have been computed using virial estimators, in several papers.
We use the compilation published in Rosario et al. (2013) for $\sim289$ type-1 AGN. 
To compare these distributions, the two samples have to be matched in intrinsic \xray\ luminosity 
($42.7 \simlt $ L$_{2-10} \simlt 45.6$) and redshift ($0.1\simlt z \simlt 2.8$). 
Within these limits, the distributions of L$_{2-10}$ and $z$, between type-1 and our highly obscured sources are very similar:
the average L$_{2-10}$ of type-1 AGN is $\langle$ Log(\lum) $\rangle  =44.39$ erg s$^{-1}$ with dispersion $\sigma=0.48$ and average redshift is
$\langle z \rangle = 1.58$ with dispersion $\sigma=0.70$, while the average L$_{2-10}$ of obscured sources is $\langle$ Log(\lum) $\rangle=44.47$ erg s$^{-1}$ 
with dispersion $\sigma=0.51$ and the average redshift is $\langle z \rangle = 1.47$ with dispersion $\sigma=0.78$.
The final sample of type-1 AGN comprise 247 sources.
We note that highly obscured AGN have typically higher L$_{2-10}$ and z with respect to the total population of type-2 AGN in COSMOS, 
and therefore more similar to the ones of type-1 AGN (Brusa et al. 2010), 
because of the positive K-correction introduced by the shape of their X-ray spectrum (see appendix A).

We also take into account the bias against obscured sources related to the flux-limited nature of the XMM-COSMOS catalog.
In appendix A we describe how we computed the detection limit in the $z$-\lum\ plane for CTK and CTN sources,
compared with the one derived for the full XMM-COSMOS catalog.
We can then compute the weight $w$, defined as the ratio between the maximum volume sampled for obscured and unobscured sources of a given intrinsic 
luminosity, given the flux-area curve of the survey.
The weight range is $w=1.1-3.5$ for CTN$_f$ sources and $w=1.3-5.1$ for CTK$_f$.
This correction assumes that sources with the same luminosity share the same properties (BH mass, Eddington ratio, etc.) and that there is no redshift evolution
of these properties for sources at a given luminosity.
However, we will show that the quantitative results described below are robust, and do not depend on the details of this correction.

The final distributions of $M_{BH}$, $\lambda_{Edd}$ and $M_*$, are shown in Fig.~\ref{istobh}.
The left panel shows the fractional distribution of Log($M_{BH}$), for type-1 AGN, in blue, and for CTN$_f$ and CTK$_f$ together,
in red, after the volume correction. Obscured sources for which the $M_{BH}$ is considered an upper limit, 
due to lack of $f_{bulge}$ information, are shown with the dashed gray histogram.
This plot shows that obscured sources indeed tend to have smaller $M_{BH}$ with respect to non obscured AGN.
The central panel of Fig.~\ref{istobh} shows the distribution of Log($\lambda_{Edd}$)
for the same samples. Obscured sources for which the $\lambda_{Edd}$ is considered a lower limit, due to the upper limit in the $M_{BH}$,
are represented by the gray shaded histogram.
Because the distributions of X-ray luminosities (and therefore of bolometric luminosities) are very similar in the two samples, 
the difference in the distribution of $\lambda_{Edd}$
is clearly driven by the different distributions of $M_{BH}$.
Finally the right panel of Fig.~\ref{istobh} shows $M_*$ computed from SED fitting for the sample of 247 type-1 AGN,
and the 39 obscured sources, from L12, B12 or D14.
Comparing $M_*$ can in principle mitigate the uncertainties related to the fact that masses for type-1 and obscured AGN are computed through completely different methods.
On the other hand, the $M_*$ for type-1 are not always well constrained in the SED 
fitting of bright QSOs at high redshift, given that the central source outshines the host galaxy in the optical (Merloni et al. 2010).

The standard way to quantitatively assess the difference between these quantities,
in presence of lower or upper limits, would be to use
the Kaplan-Meier estimator (KM; Kaplan \& Meier 1958; also known as the
product limit estimator, and also independently introduced to
astronomy by Avni et al.\ 1980) to derive the expected cumulative
distribution functions, and the logrank test to quantify the difference
between two samples.  A weighted version of the logrank test has been
introduced by Xie \& Liu (2005), in which the weights are inversely
proportional to the probability of a source to be selected in any of
the samples. This definition of weights corresponds to the one
we employed above (i.e.\ ratio between the type-1 and obscured AGN
sampled volumes).
Applying the weighted logrank test to our sample, we found that 
obscured AGN have smaller  $M_{BH}$ and higher $\lambda_{Edd}$,
with respect to type-1 AGN, at a confidence level $\sim5\sigma$. 
We stress that this result does not depend on the weights used, since an unweighted logrank test would still indicate a
difference at the same confidence level.
Furthermore, if no redshift evolution is applied in the $M_{BH}$-$M_*$ relation adopted for the obscured sources 
(see Schulze \& Wisotzki 2014), the difference in $M_{BH}$ between type-1 and obscured sources would be even larger, i.e. obscured sources would have even smaller $M_{BH}$.

For $M_*$, the logrank test cannot be applied because the hazard
functions (H(f)) for both samples cross each other\footnote{For a given flux f, 
H(f) gives the probability of detecting an object at
flux $<f$, conditioned on its non-detection at flux $>f$.}.
Since the $M_*$ values
in our sample do not include upper or lower limits, and ignoring the
weights, that do not significantly change the distribution, 
one could employ the Wilcoxon-Mann-Whitney test (Lehmann 2006). This test
would support a significant difference between the $M_*$
distributions for type-1 and obscured AGN only at $\sim2\sigma$ level.
The same result is obtained if a two-sample K-S test is applied.

The fact that the results for $M_{BH}$ (and $\lambda_{Edd}$) and  $M_*$ 
disagree in such strong way is worrying, given that type-1 AGN in the control sample, {\it on average}, follow the same $M_*$-$M_{BH}$
relation used for the obscured sources to derive $M_{BH}$\footnote{
After including a small systematic shift by $\sim0.3$ dex toward smaller $M_{BH}$ at all redshifts, to account for the missing $f_{bulge}$ information for type-1 AGN.}. 
We stress, however, that the results on $M_{BH}$ take into account the upper limits in the 
classical sense of survival analysis, i.e.
that an upper limit represents a value that can be everywhere in the distribution, below the limit value itself
(the same applies to $\lambda_{Edd}$ and its lower limits).
We know, instead, that the $M_{BH}$ of a source without morphological information can only be some fraction $f_{bulge}$ of the $M_{BH}(Tot)$, obtained from the total 
$M_*$, with reasonable $f_{bulge}$ comprised between $\sim0.25$ and 1, with typical value of 0.5. 
Given that the vast majority (28 out of 39) of the $M_{BH}$ of 
obscured sources are indeed upper limits, an incorrect treatment of the limits can lead to significantly incorrect results.
Therefore we further investigated the effect of different assumptions on these limits. 
First, we computed the significance of the difference between the two distributions, if the upper limits are considered as detections.
The same Wilcoxon-Mann-Whitney test used for $M_*$ gives a significance of only $2.6\sigma$ in this case, therefore consistent with what observed for $M_*$. 
Second, we considered all the upper limits as detections, but decreased by 0.3 dex (corresponding to $f_{bulge}=0.5$).
In this case the significance of the difference is $4.5\sigma$, somewhat in between what observed treating the upper limits in the classical way,
and what found in the most conservative case of all detections.
We therefore consider this ($2.6 \simlt \sigma \simlt 5$) as a reasonable interval for the significance of the difference between $M_{BH}$
of highly obscured and type-1 AGN.

\subsection{sSFR}

\begin{figure}[t]
   \centering
   \includegraphics[width=8cm,height=8cm]{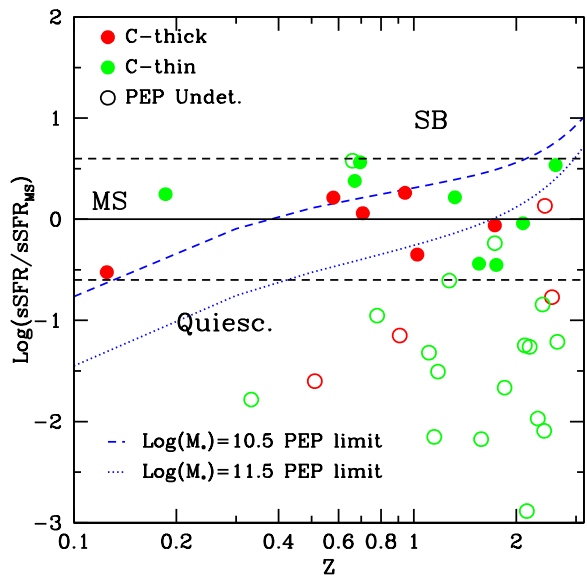}
   \caption{Ratio between the sSFR of highly obscured sources and the sSFR of MS star forming galaxies, as a function of redshift.
   Red (green) circles represent CTK$_f$ (CTN$_f$) sources. Filled (empty) circles represent Herschel PACS detected (undetected) sources. 
   The MS is taken from Whitaker et al. (2012). 
   The dashed lines shows the ratio sSFR/sSFR$_{MS}=4$ times above and below the MS, used as starbursts and quiescent galaxy definition.
   The blue dotted (dashed) curves mark the sSFR/sSFR$_{MS}$ observability limit of the Herschel PEP survey in COSMOS,
   obtained for a Log($M_*$)=11.5 (10.5), using the Seyfert-2 template of Polletta et al. (2007).
}
   \label{ssfr}
   \end{figure}

We collected sSFR derived in D14 for 
all the COSMOS sources with a detection in the $160~\mu m$ Herschel-PEP catalog 
(PACS Evolutionary Probe,  Lutz et al. 2011).
Interestingly, 6 out of 10 CTK$_f$ (8 out of 29 CTN$_f$) sources have a $>3\sigma$ detection in the Herschel-PEP catalog, 
giving a detection rate of $60\pm38\%$ for the CTK and $35\pm9\%$ for the full obscured sample.
The detection fraction for the full XMM-COSMOS catalog is instead $18\pm2\%$ (L12).
A similar detection rate is found ($20\pm2\%$) if we look only at sources having more than 30 counts, in the same redshift and luminosity range
of the highly obscured sample ($42.7 \simlt $ L$_{2-10} \simlt 45.6$ and $0.1\simlt z \simlt 2.8$),
regardless of the classification (type-1 or type-2).
Therefore the Herschel-PEP detection rate of the CTK sample can be considered significantly higher (at ~90\% confidence) 
with respect to the one observed for the full  XMM-COSMOS catalog.

Fig.~\ref{ssfr} shows the distribution of the ratio between the sSFR for the highly obscured sources, 
and the sSFR of main sequence (MS) star forming galaxies as a function of redshift.
We adopted the redshift and $M_*$ dependent MS of star-forming galaxies as derived
in Whitaker et al. (2012). For each source we computed the sSFR$_{MS}$ expected for that specific redshift and $M_*$, 
and hence the ratio with the observed sSFR.
The dashed lines marks the definition of starbursts (sSFR/sSFR$_{MS}>4$) and quiescent galaxy (sSFR/sSFR$_{MS}<1/4$) as proposed in Rodighiero et al. 
(2011, see also Bongiorno et al. 2014).
The sSFR for CTK$_f$ (red) and CTN$_f$ (green) is shown with filled circles in Fig.~\ref{ssfr}.

\begin{figure*}[t]
\centering
\includegraphics[width=14cm,height=16cm]{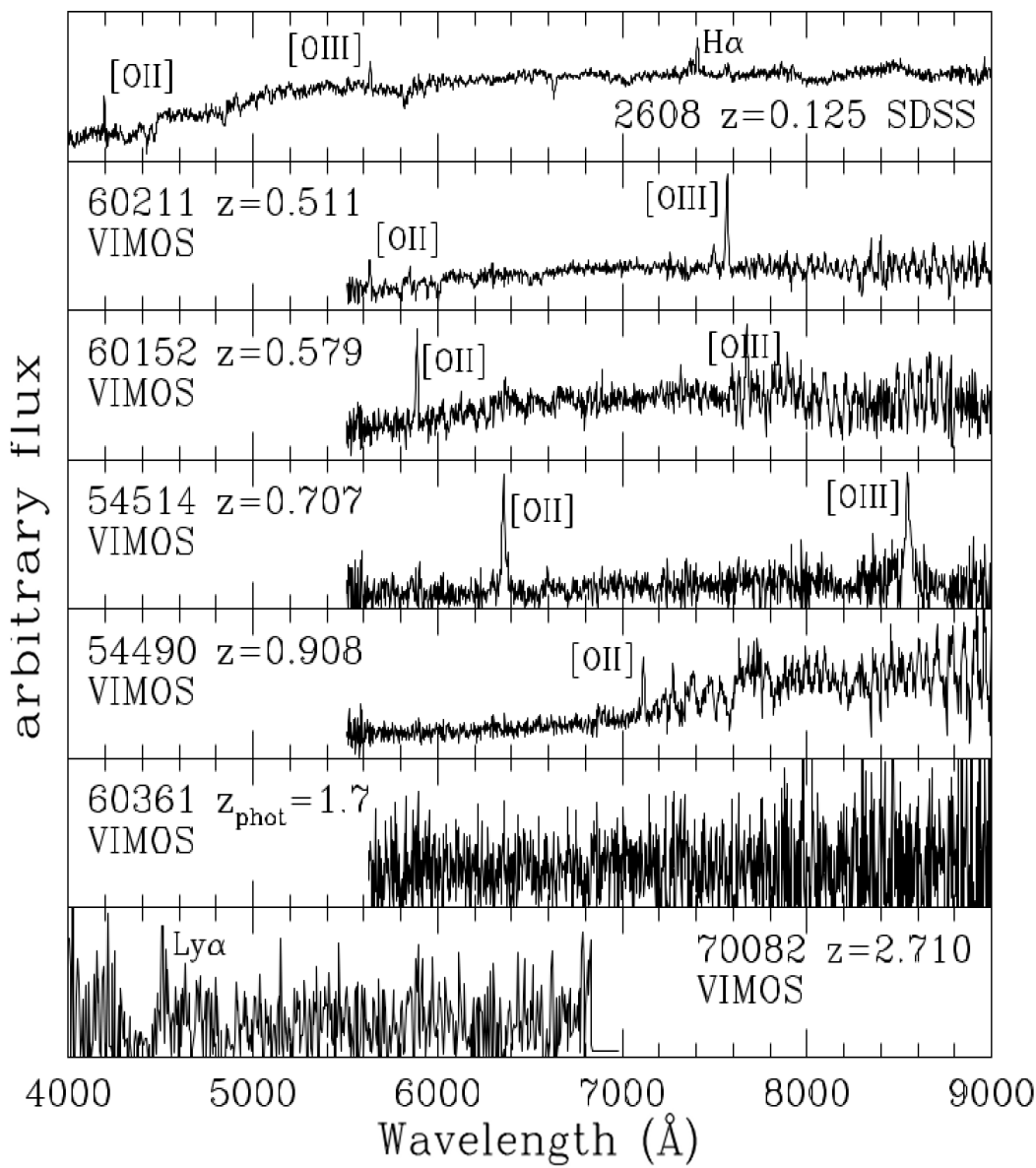}
\caption{Optical spectra of CTK$_f$ sources, ordered by increasing redshift, one from SDSS (top) and six from zCOSMOS. The source ID and redshift is labeled in each panel.}
\label{opt}
\end{figure*}

For 22 sources without Herschel detection we can rely on SFR estimates from SED fitting published in B12. 
However, the lack of far infrared (FIR) detection 
prevent the authors from disentangling the AGN and SF contribution. The SFR is therefore estimated on the 
basis of the UV emission, which traces only the unobscured SF, scaled up by the dust correction factor derived from the full SED\footnote{B12 
compare, in their Appendix B, the SFR estimated from the SED fitting with the one derived using FIR data for the small subsample (10\%) of Herschel detected sources in their sample.
They find that for high star formation rates (above tens of $M_{\odot}$ $yr^{-1}$) the agreement is quite good,
while at lower SFR the disagreement becomes evident, with SFR(FIR) being systematically higher than SFR(SED).}.
The sSFR for these sources is shown with empty circles in Fig.~\ref{ssfr} (red for CTK$_f$ and green for CTN$_f$) and can be considered as a lower limit for their sSFR. 
Indeed almost all of these measurements fall well below the MS region at all redshift.
The upper limit in the ratio sSFR/sSFR$_{MS}$ that can be derived for Herschel undetected sources 
from the flux limit of the PEP coverage in COSMOS (2.0 mJy at $160~ \mu m$), is shown with the blue dotted (dashed) lines, 
assuming Log($M_*$)=11.5 (10.5) and the Seyfert 2 template of Polletta et al. (2007).

Taken at face value, this result would imply that 70\% of the CTK$_f$ sources (7 out of 10) fall within the star forming MS region,
while 68\% of the CTN$_f$ sources (17 out of 25) fall in the quiescent galaxy region.
The fact that CTK$_f$ sources have an exceptionally high Herschel detection fraction with respect to both CTN$_f$ and the full XMM-COSMOS catalog, points in this direction.
However, as the blue lines show, most of the CTN$_f$ sources at $z>1$ would remain undetected, even being in the MS region, due to the shallow flux limit of the PEP catalog.
Indeed, the higher average redshift of the CTN$_f$ sources (see Sec.~2.3) probably plays a major role in the smaller fraction of Herschel detection in this sample.
Interestingly, none of the highly obscured sources presented in this work falls in the starburst region as defined in Rodighiero et al. (2011).
We can therefore conclude that these obscured sources are all consistent with being star forming or even strongly star forming  (four of them have $SFR>100$ $M_{\odot} yr^{-1}$),
but that there is not necessarily a direct connection between the presence of a highly obscured AGN, and the strongest, possibly major merger driven, star-burst activity,
that powers for example ULIRGs and Sub-mm Galaxies ($SFR>1000$ $M_{\odot} yr^{-1}$).
Similar results were obtained for other \xray\ selected samples of obscured QSOs, such as the sample of type-2 QSO selected in the COSMOS field in Mainieri et al. (2011), and the sample of CT AGN 
selected in the CDFS from Georgantopoulos et al. (2013).

\section{Atlas of multi-wavelength properties of CT AGN}

\subsection{Optical spectra}

In Fig.~\ref{opt} the optical spectra of the CT sources are shown, one from SDSS (top)
and six from the zCOSMOS survey (Lilly et al. 2009).
The first five spectra provide secure redshifts ($0.125 < z < 0.908$) and all of them show typical
spectral characteristics of obscured sources, with narrow emission lines (of H$\alpha$, [OII], [OIII])
and continuum dominated by the stellar component of the host galaxy.
The only exception is the source XID 54514, which has strong emission lines with
FWHM$\sim900$ km/s, resolved in the VIMOS spectrum but well in the range expected
for type-2 AGN.

   \begin{figure*}[!t]
   \centering
      \includegraphics[width=5cm,height=4cm]{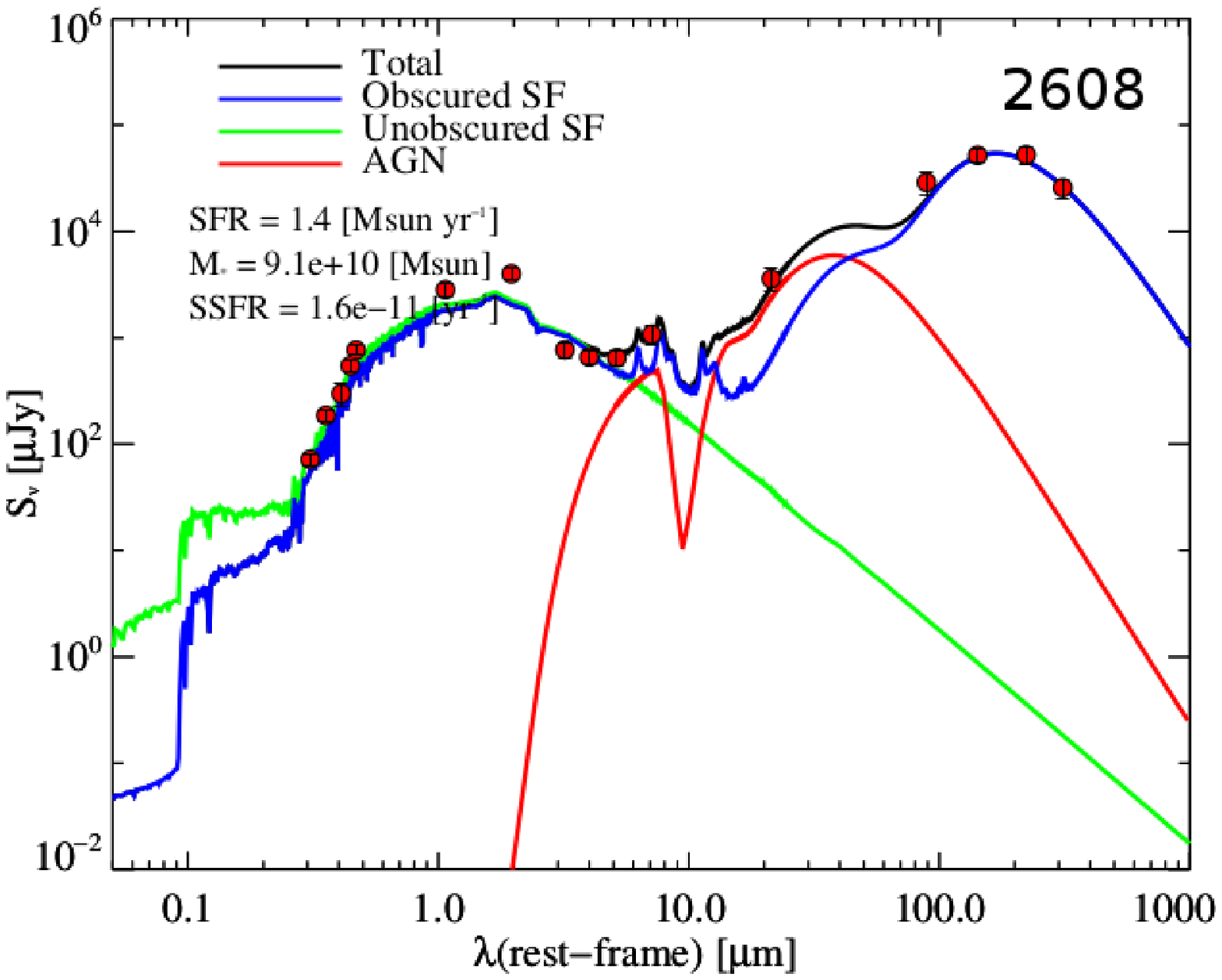}\hspace{0.5cm}\includegraphics[width=5cm,height=4cm]{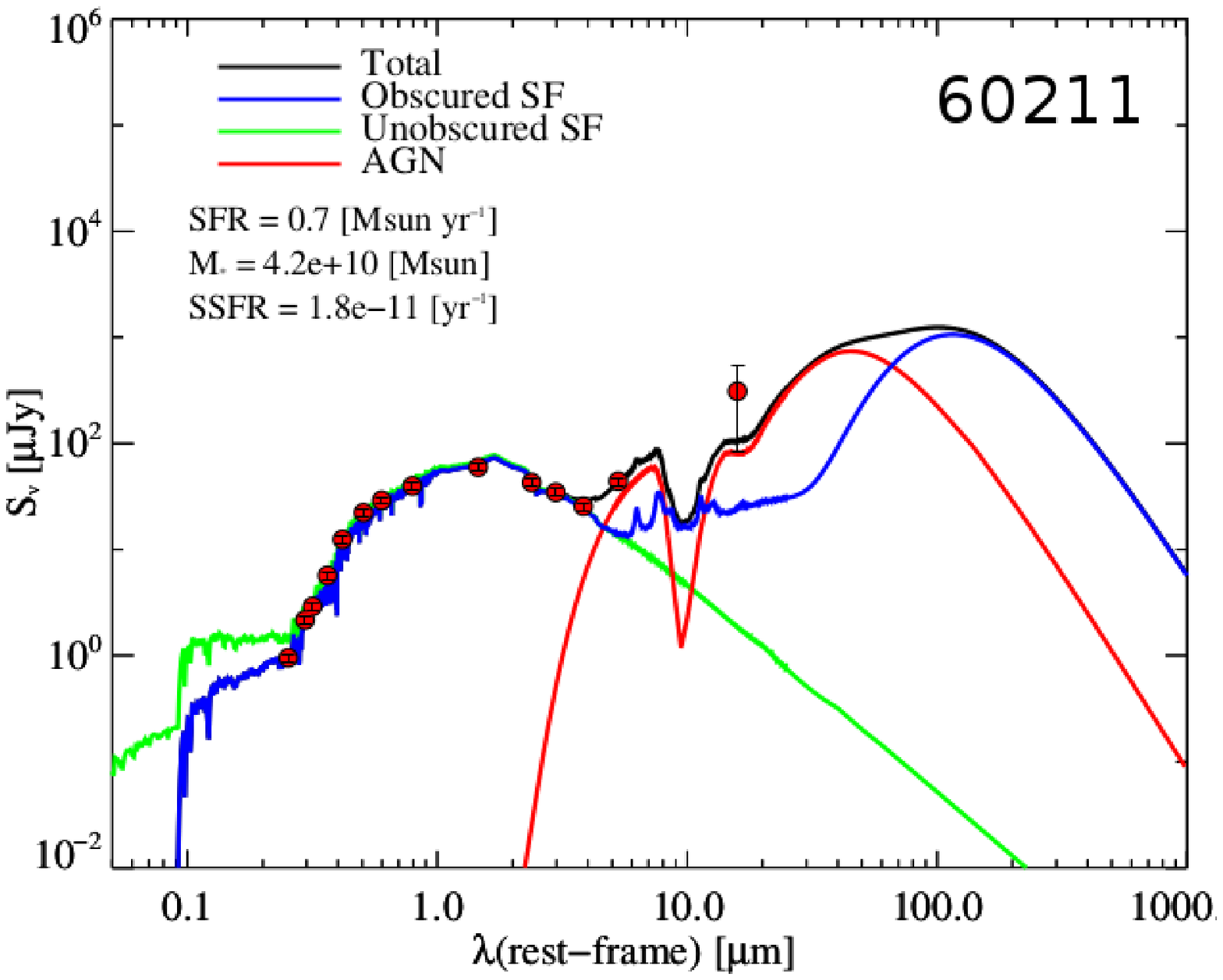}\hspace{0.5cm}\includegraphics[width=5cm,height=4cm]{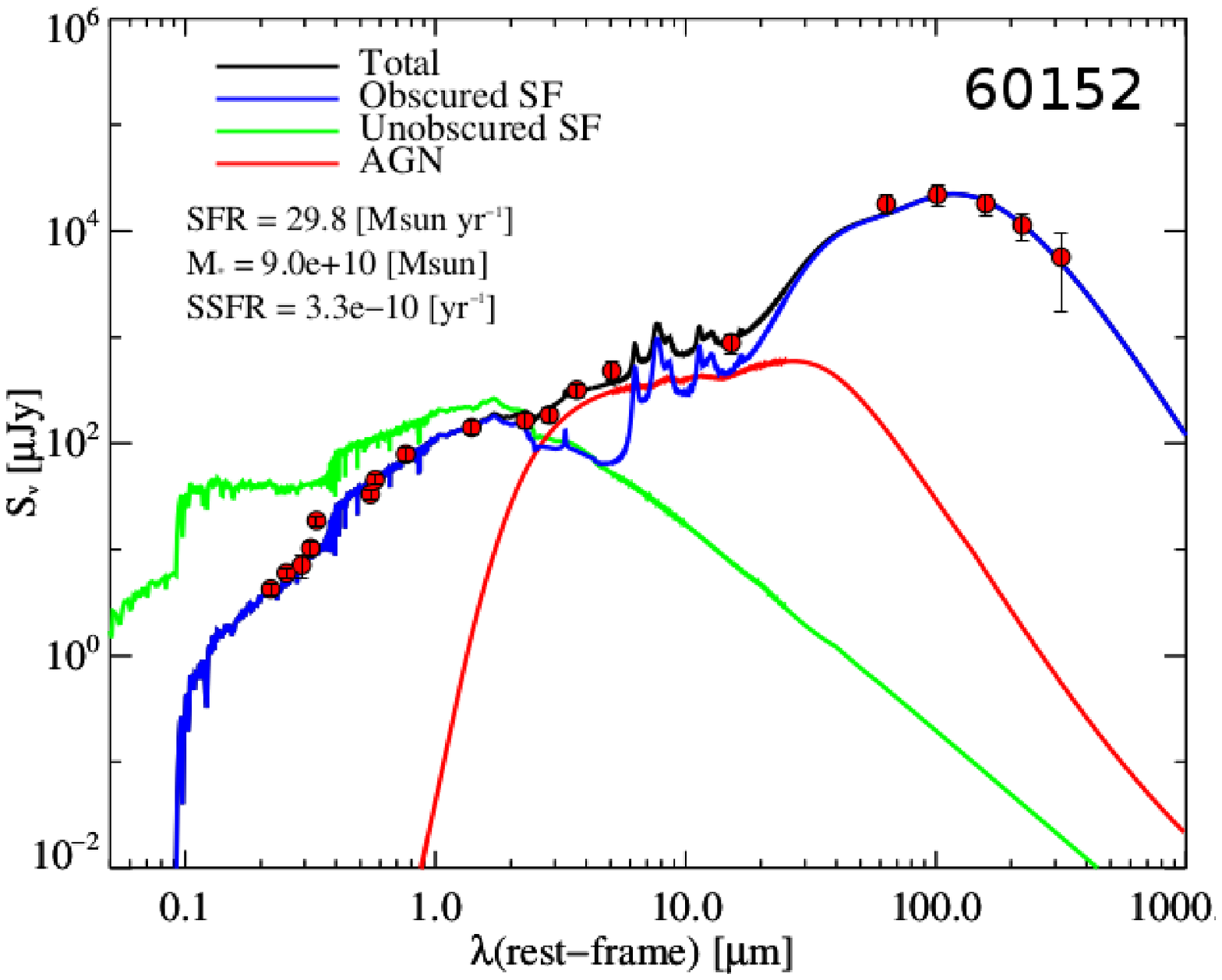}
      
      \vspace{0.5cm}
      
      \includegraphics[width=5cm,height=4cm]{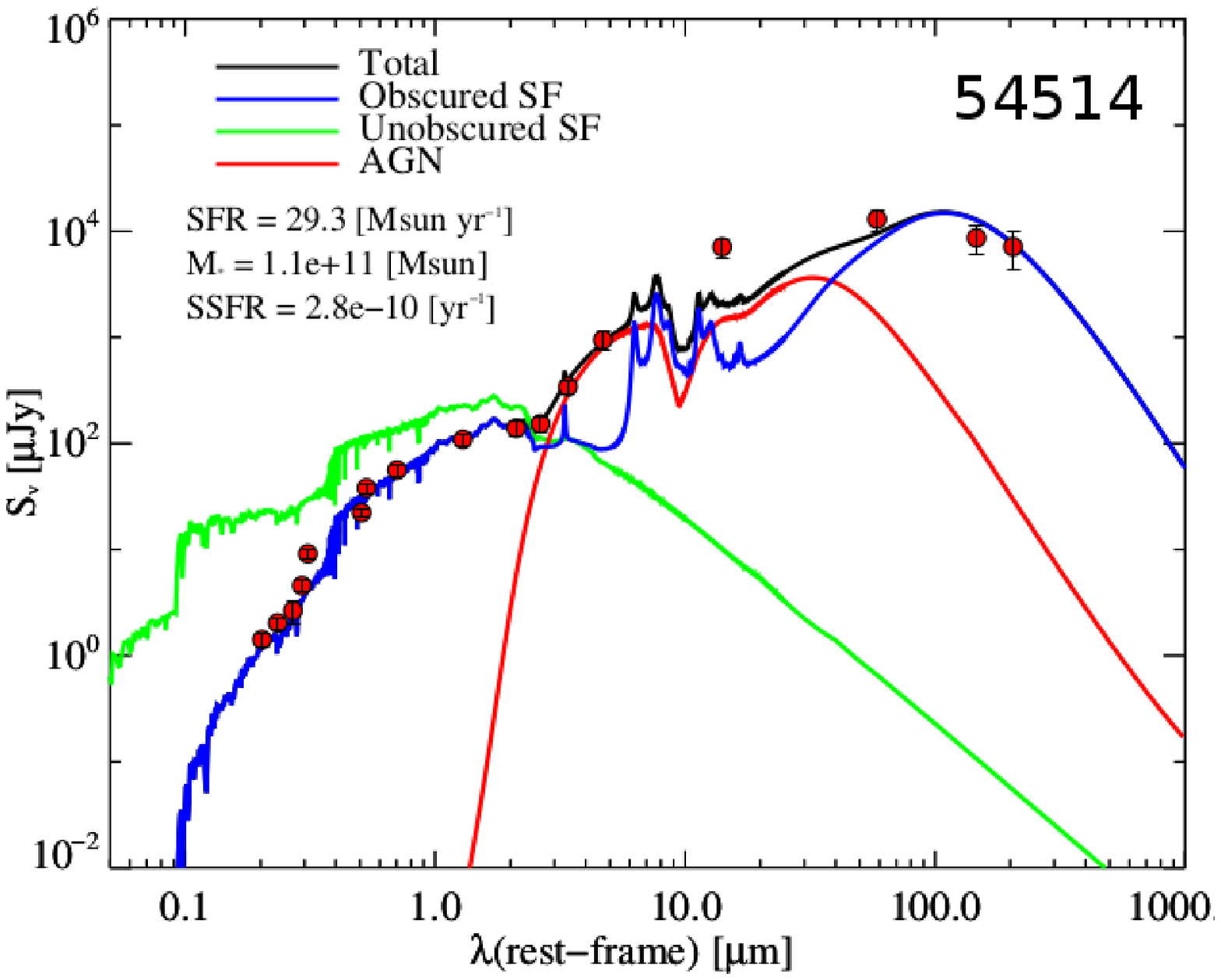}\hspace{0.5cm}\includegraphics[width=5cm,height=4cm]{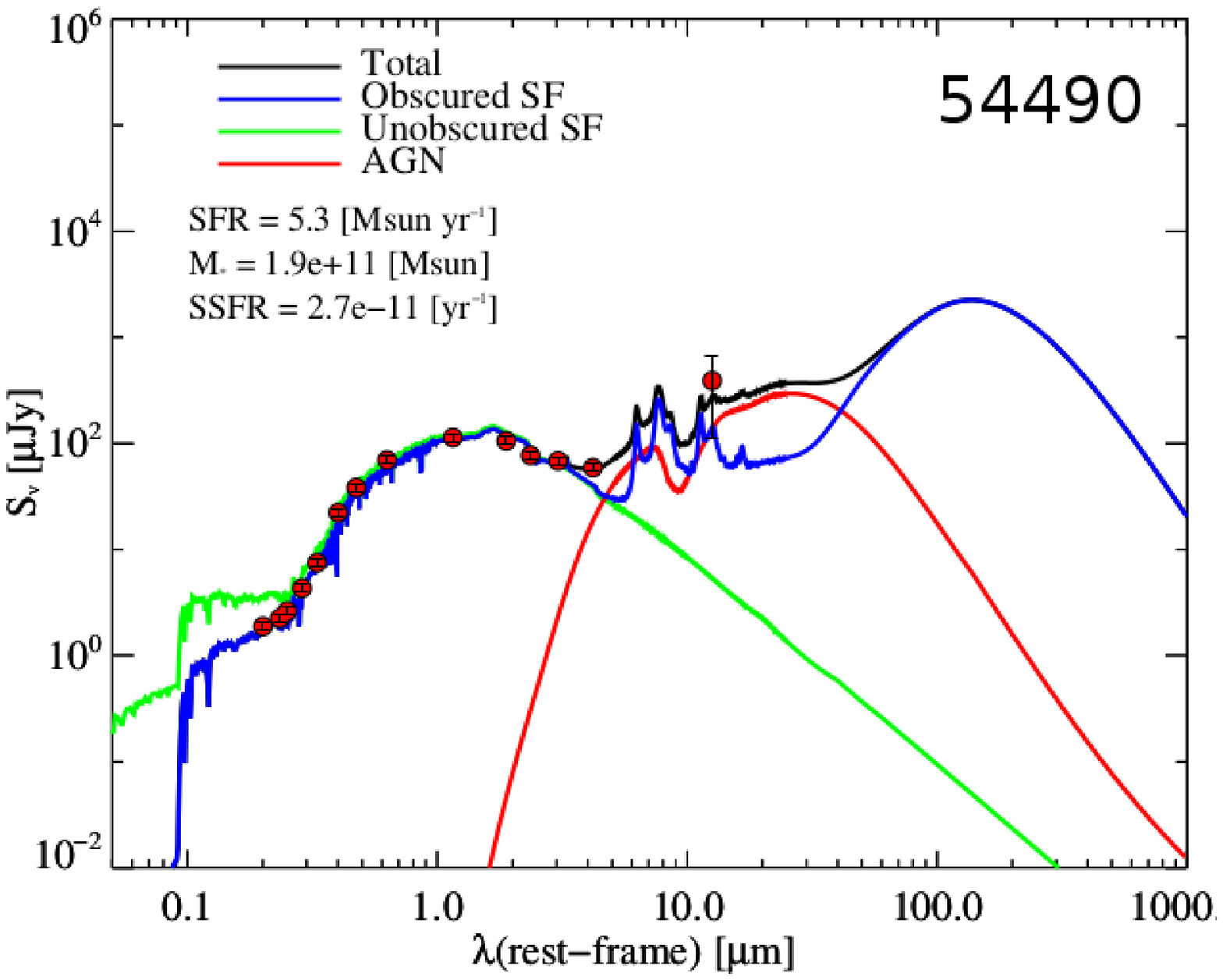}\hspace{0.5cm}\includegraphics[width=5cm,height=4cm]{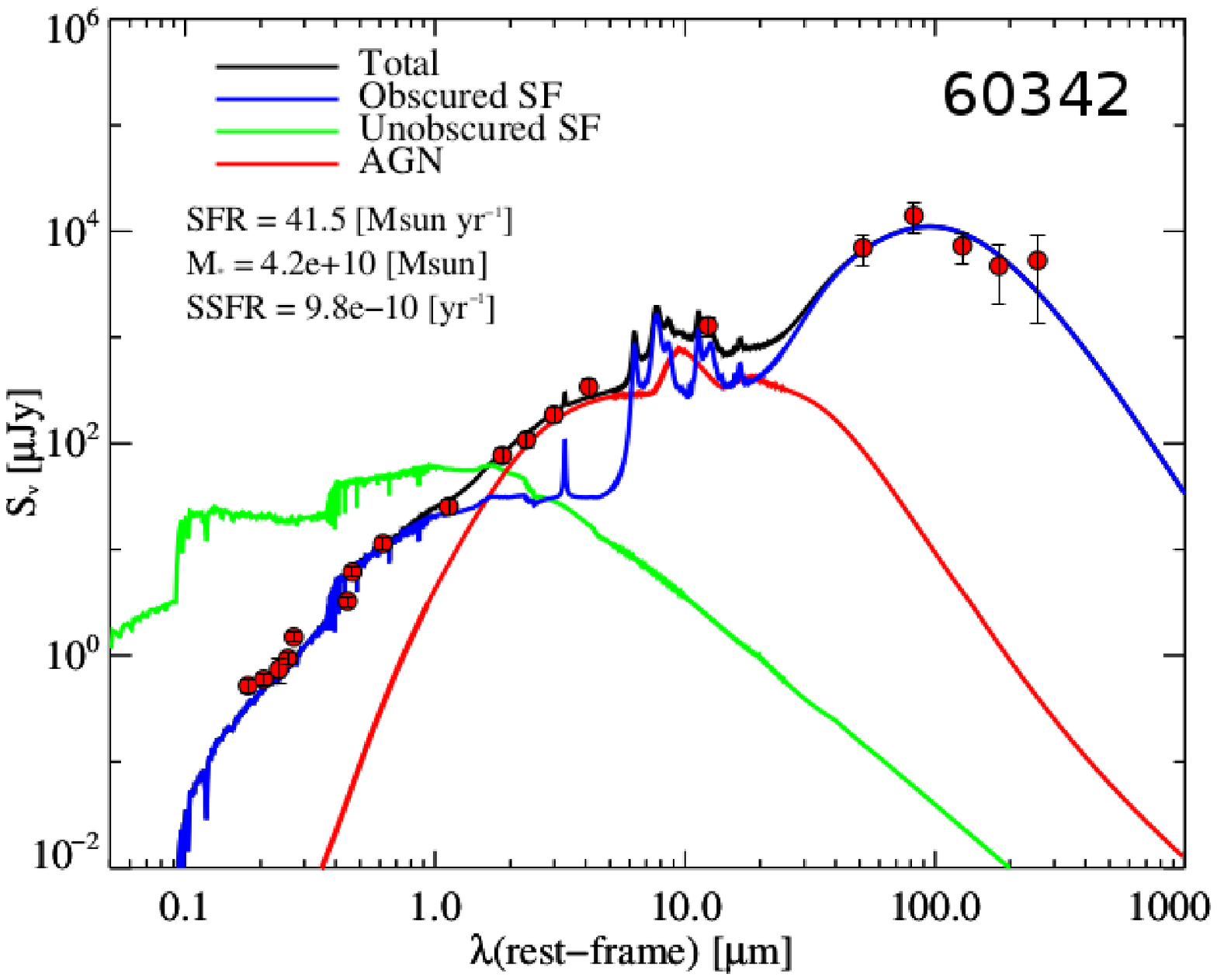}
      
      \vspace{0.5cm}
    
      \includegraphics[width=5cm,height=4cm]{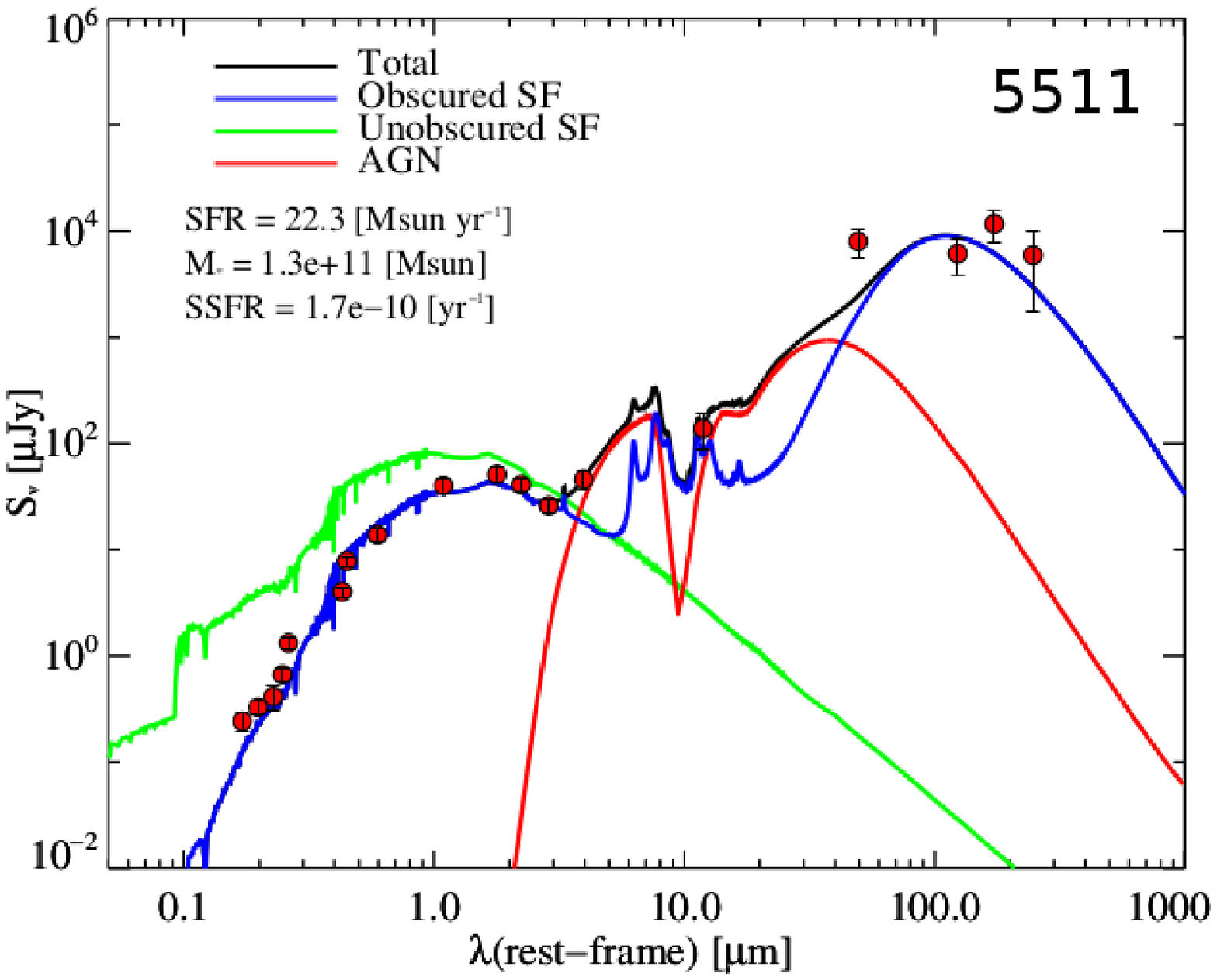}\hspace{0.5cm}\includegraphics[width=5cm,height=4cm]{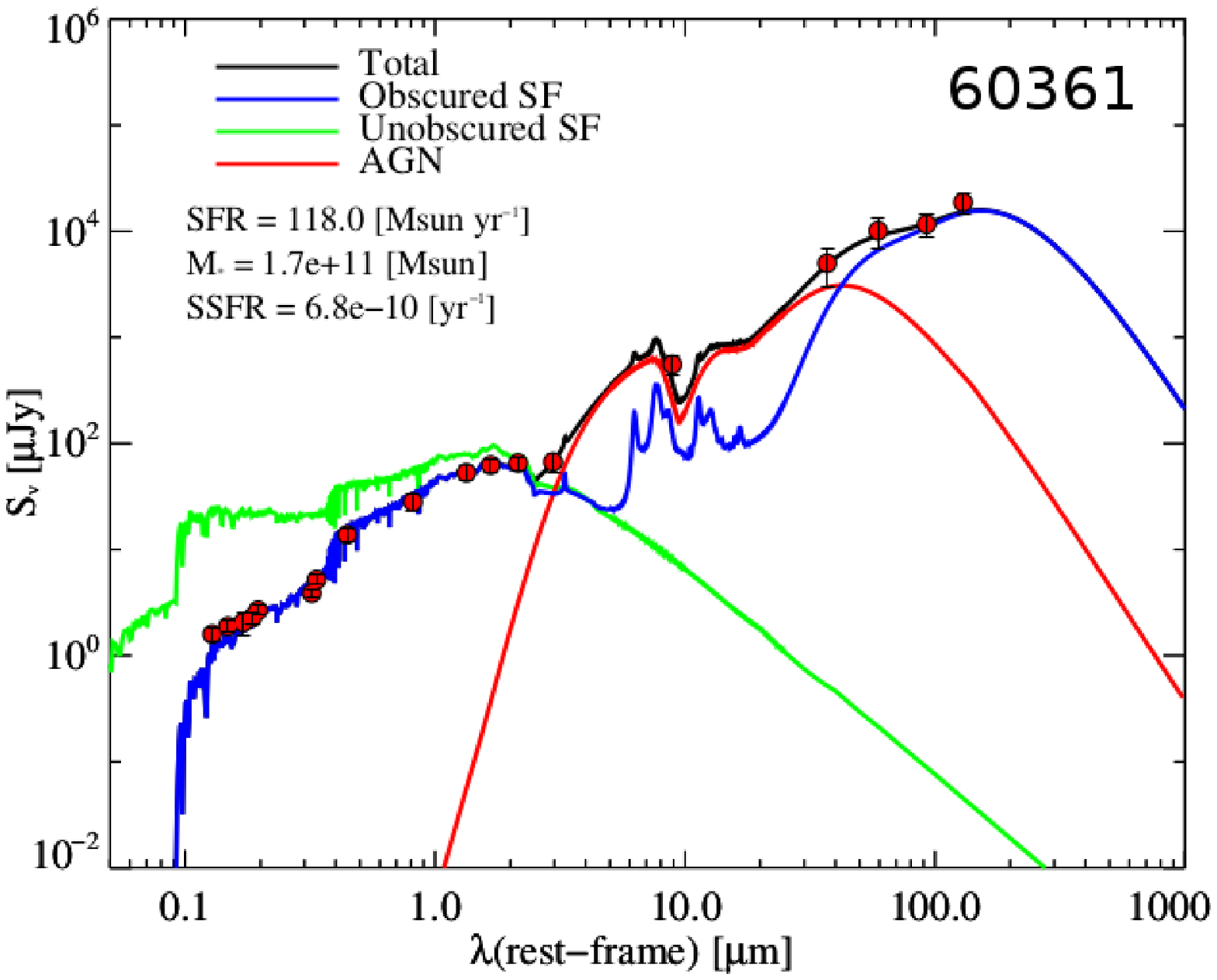}\hspace{0.5cm}\includegraphics[width=5cm,height=4cm]{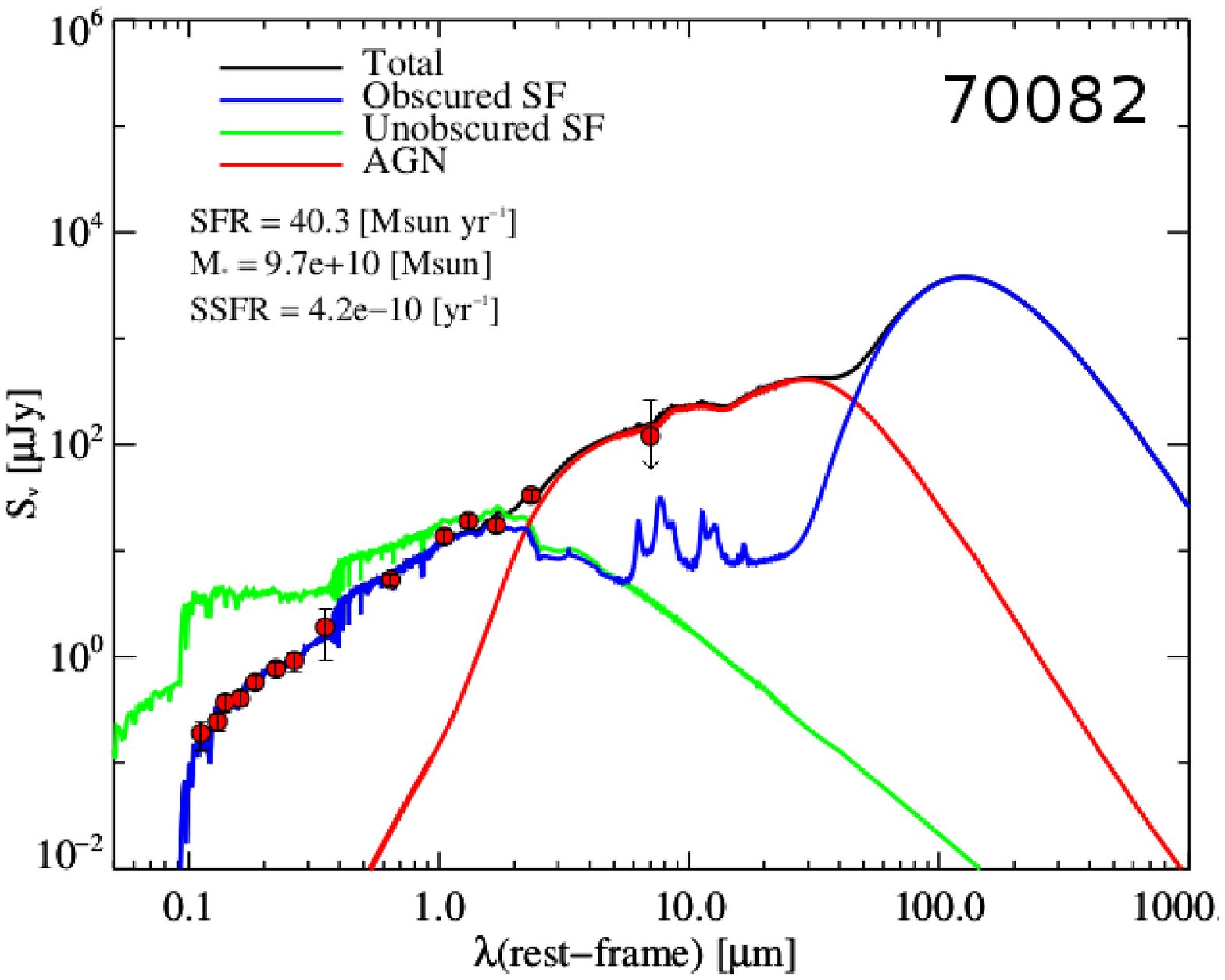}
 
      \vspace{0.5cm}
      
      \includegraphics[width=5cm,height=4cm]{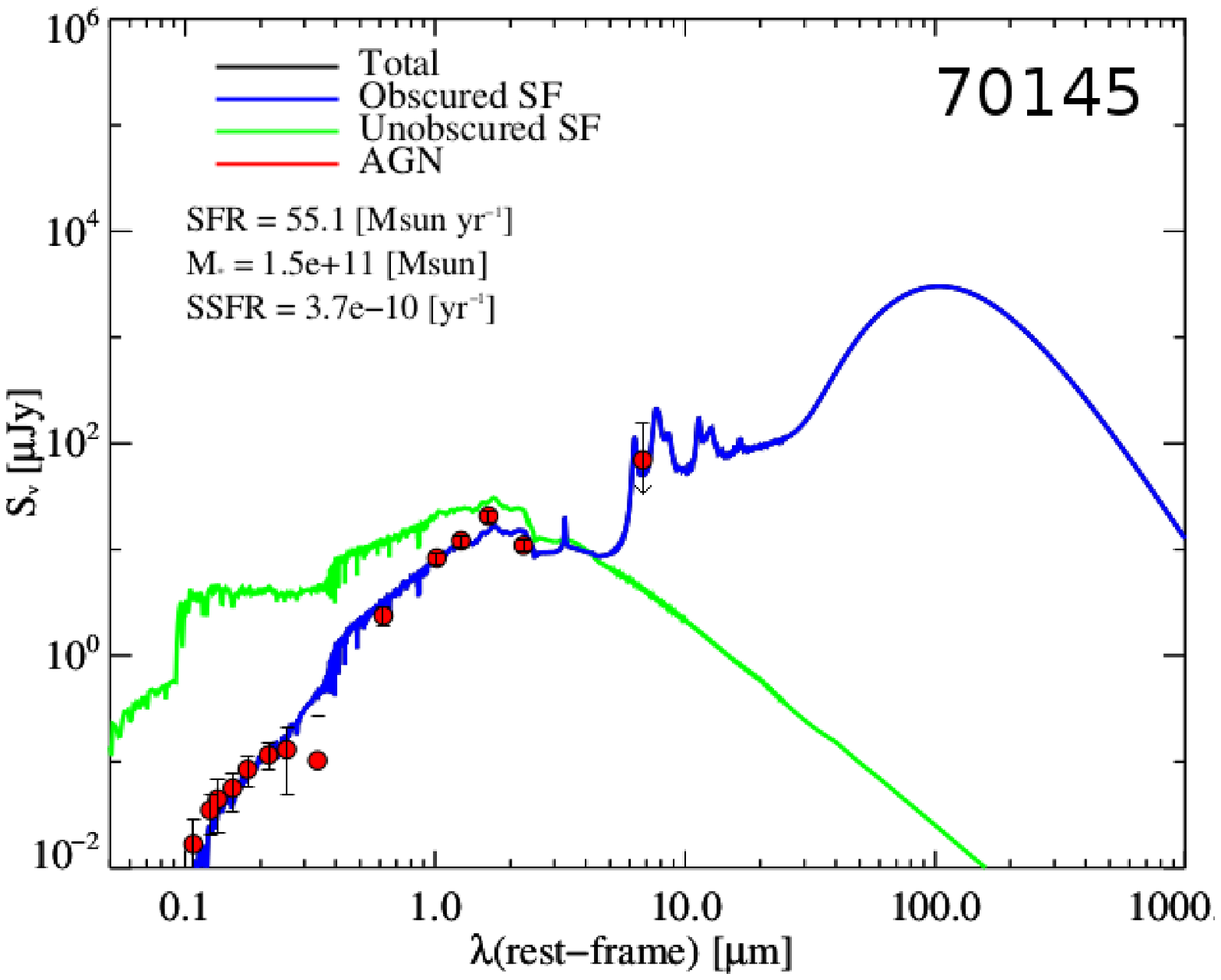}
   \caption{Rest-frame SED fit for the CTK$_f$ sources, ordered by increasing redshift. The fit has been performed as in D14. In green is shown the unextincted stellar emission, in blue 
   the star formation contribution, and in red the AGN contribution.}
   \label{sed}
   \end{figure*}

We computed the [OIII] $\lambda5007$ luminosity of the four sources for which this line falls in the observed band.
The observed $L_{[OIII]}$ are in the range Log($L_{[OIII]}$)=40.4-41.8 erg s$^{-1}$ (after slit loss correction for the zCOSMOS spectra). 
These values are $\sim2.5$ order of magnitudes lower than the intrinsic, absorption corrected \lum. 
To reconcile these values in a self-consistent AGN intrinsic luminosity, 
we need to apply a rather extreme $L_{[OIII]}$-\lum\ correlation, similar to the one found by Heckman et al. (2005) 
for X-ray selected AGN, that gives $Log($\lum$/L_{[OIII]})\sim2$, higher than the typical correction of  $Log($\lum$/L_{[OIII]})\sim1-1.5$ 
(e.g. Panessa et al. 2006; Georgantopoulos \& Akylas 2009).
However, on top of that, we need to invoke also a correction factor of $\sim0.7$ for $Log(L_{[OIII]})$ for dust reddening for 
sources 60152 and 54514, or even more (a factor $\sim1.3$) for 2608. These values are higher than those reported
for samples of e.g. optically selected type-2 AGN ($\sim0.4$, Zakamska et al. 2003, $\sim0.3$, Mignoli et al. 2013). 
Our spectra do not allow to actually compute the Balmer decrement through the $H\alpha/H\beta$ ratio. We stress however that similar or 
even higher extinction values (up to $A_V=6.9$ mag, more typical values of $A_V\sim2-3$) have been found for a sample of CT candidates selected 
in the SDSS (Goulding et al. 2010). All this would imply that, apart from the CT obscuration produced by the torus at pc 
scales around the SMBH, these systems may have strong dust reddening occurring at larger distances, $\sim100-1000 pc$ scales, 
obscuring the narrow line region.

The optical spectrum of the source 60361 does not provide a
redshift, but it is shown in the figure to stress the absence of strong emission
lines in the wide spectral range covered. The photometric redshift adopted through the text for this source is $z=1.73$.
If the photometric redshift is correct, the absence of strong emission lines is a further indication of obscuration,
even if the observing band would allow only the Mg II $\lambda2800$ to be observed.
The narrow component of this line is known to show no tight correlation with the intrinsic AGN luminosity, 
and can be easily reddened (Zakamska et al. 2013, Mignoli et al. 2013).

Finally, source XID 70082 has a tentative spectroscopic redshift of $z=2.71$, based on a faint Ly$\alpha$ in
emission with a continuum break across the line. Because of the uncertain identification,
we used through the analysis the photometric redshift of $z=2.429$.
If the identification of the Ly$\alpha$ is indeed correct, 
the measured \nh\ and \lum\ would be underestimated by a factor $\sim15\%$ and $30\%$, respectively,
i.e. the source would be even more extreme in terms of obscuration and luminosity.

\subsection{Broad band SED}

D14 performed broad band SED fitting
for all the COSMOS sources with a detection in the $100$ or $160~ \mu m$ Herschel-PEP catalog,
with the aim of disentangling the stellar and star-forming emission from the AGN contribution.
The SED fitting is based on a modified version of the
MAGPHYS code (Berta et al. 2013; Da Cunha et al. 2008),
that combines stellar light, emission from dust, heated by stars and a
possible warm dust contribution, heated by the AGN, 
in a simultaneous three-components fit. The adopted libraries are
the stellar models by Bruzual \& Charlot (2003), the star-forming dust
models by Da Cunha and the Torus library by Feltre et al. (2012) and Fritz et
al. (2006).

Fig.~\ref{sed} shows the rest-frame SED decomposition performed by D14 for the 6 CTK$_f$ sources detected with Herschel in the PEP catalog, 
plus the 4 CTK$_f$ not detected with Herschel. 
The green curve represents the unextincted stellar emission (not fitted to the observed data), the blue curve represents the star formation contribution 
re-distributed across the MIR/FIR range in a self-consistent way. 
The red line reproduces the AGN emission and incorporates both the accretion disc and the torus emission.
Data points include photometry from the optical to the Herschel SPIRE bands (250, 350 and 500 $\mu m$) for most of the sources.
In each panel are reported the SFR in $M_{\odot}$ $yr^{-1}$, obtained converting the IR luminosity using a standard Kennicutt (1998) law, 
rescaled to a Chabrier initial mass function (Chabrier et al. 2003), the $M_*$ and the specific SFR.
We note that all the sources need an AGN torus component, in order to reproduce the MIR photometric points, with the only exception of XID 70145, probably due to the 
non-detection at 24 $\mu m$.
Furthermore all the tori used in these fits are highly obscured: except for source 60342 (Log(\nh$(SED)=22.5$), 
the \nh\ derived from the torus template used in the  SED fitting are in the range Log(\nh$(SED)=23-23.8$). 
We stress that Log(\nh$(SED)=23.8$) is the maximum value allowed in the torus library, corresponding to a maximum optical depth of $\tau(9.6 \mu m) = 6$. 

We recall that the flux limits (at 3 $\sigma$) of the PACS and SPIRE imaging in the COSMOS field are 5.0 mJy at 100 $\mu m$, 10.2 mJy at 160 $\mu m$, 
8.1 mJy at 250  $\mu m$, 10.7 mJy at 350 $\mu m$ and 15.4 mJy at 500 $\mu m$.
Even if the upper limits are not taken into account in the SED fitting code, we checked {\it a posteriori}, 
that the best fit model for the four sources undetected in both SPIRE and PACS, 
is always lower than the upper limits that can be placed in these bands, 
and therefore fully consistent with the available observational constraints.

\begin{figure*}[!th]
   \centering
      \includegraphics[width=5cm,height=5cm]{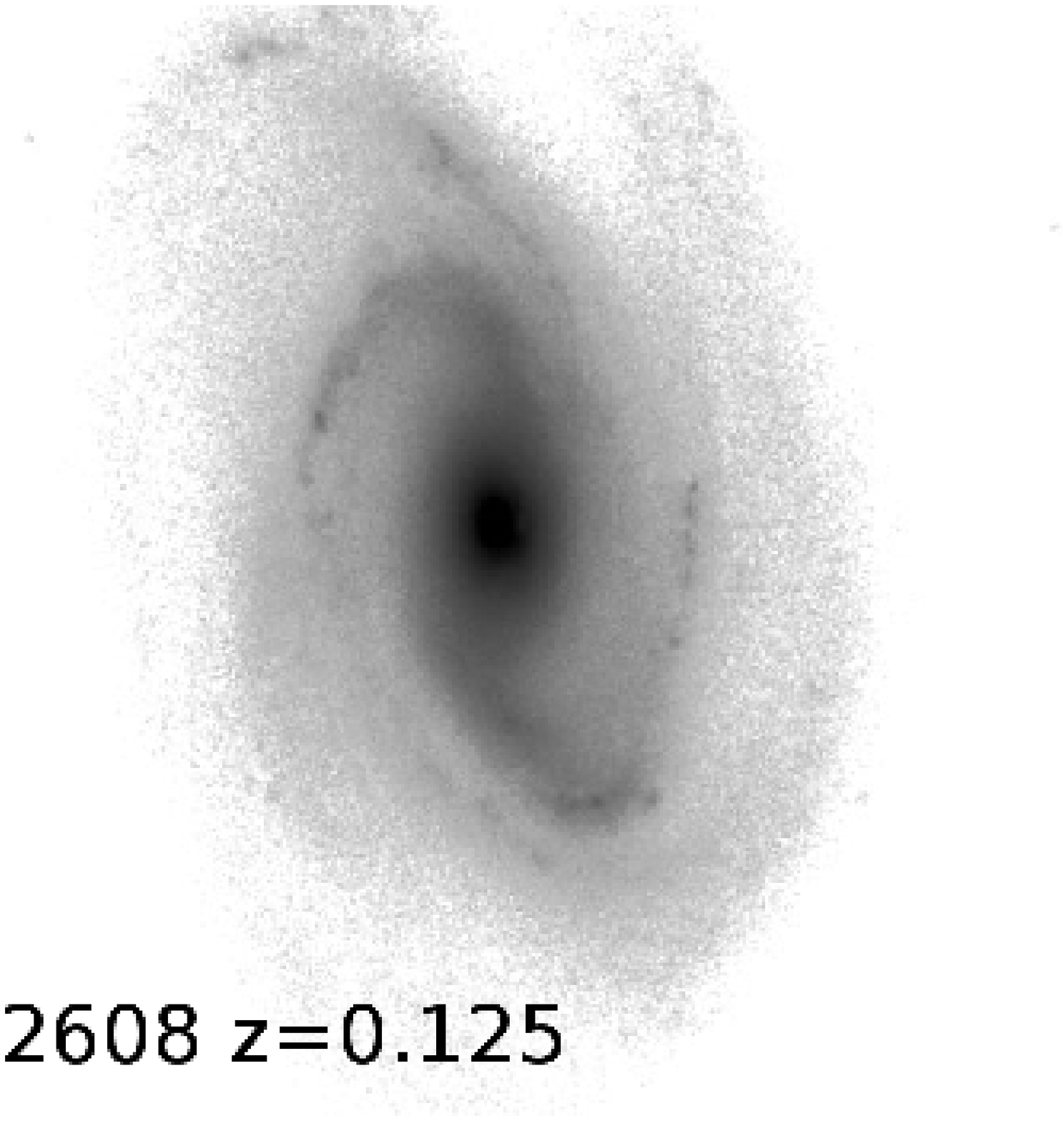}\hspace{0.3cm}\includegraphics[width=5cm,height=5cm]{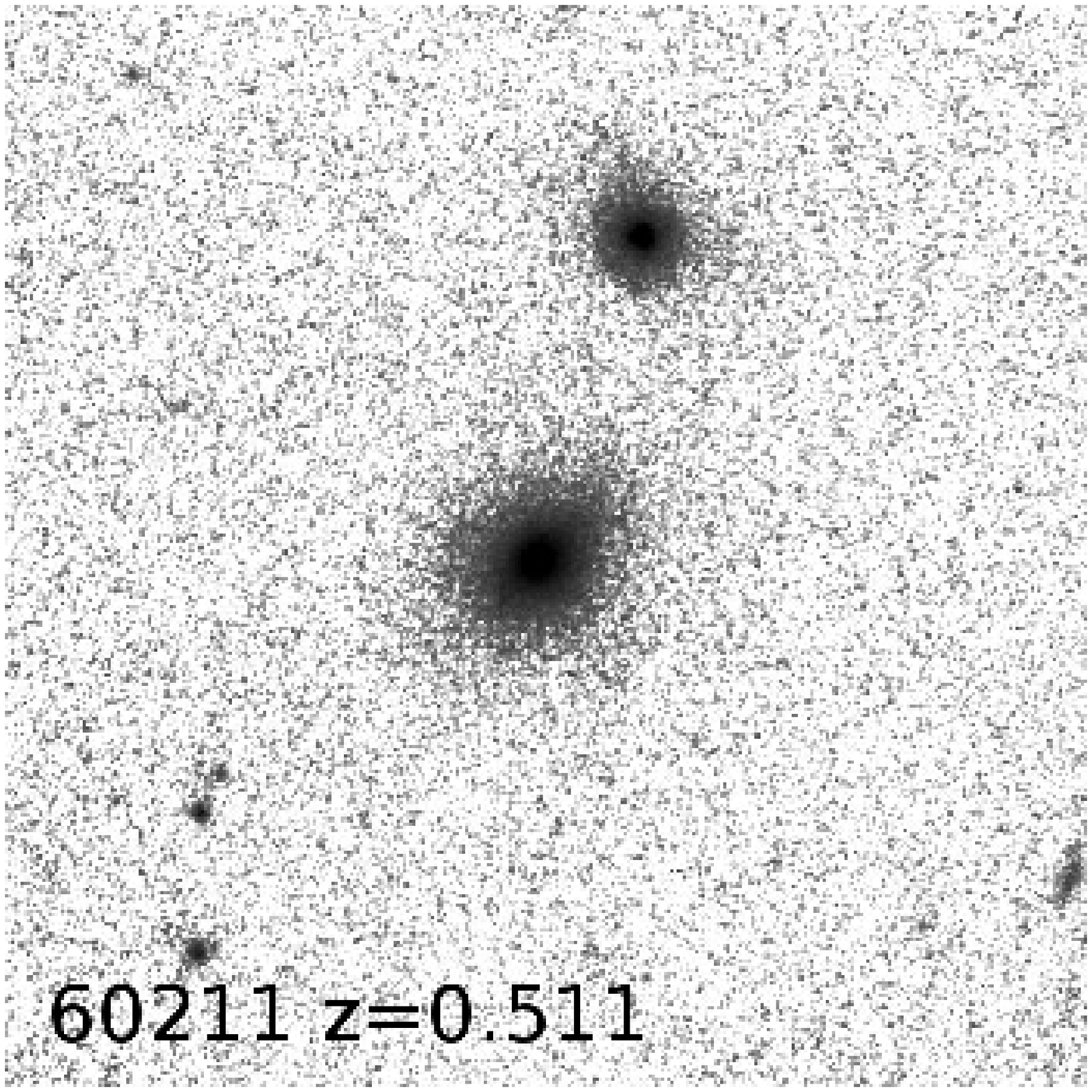}\hspace{0.3cm}\includegraphics[width=5cm,height=5cm]{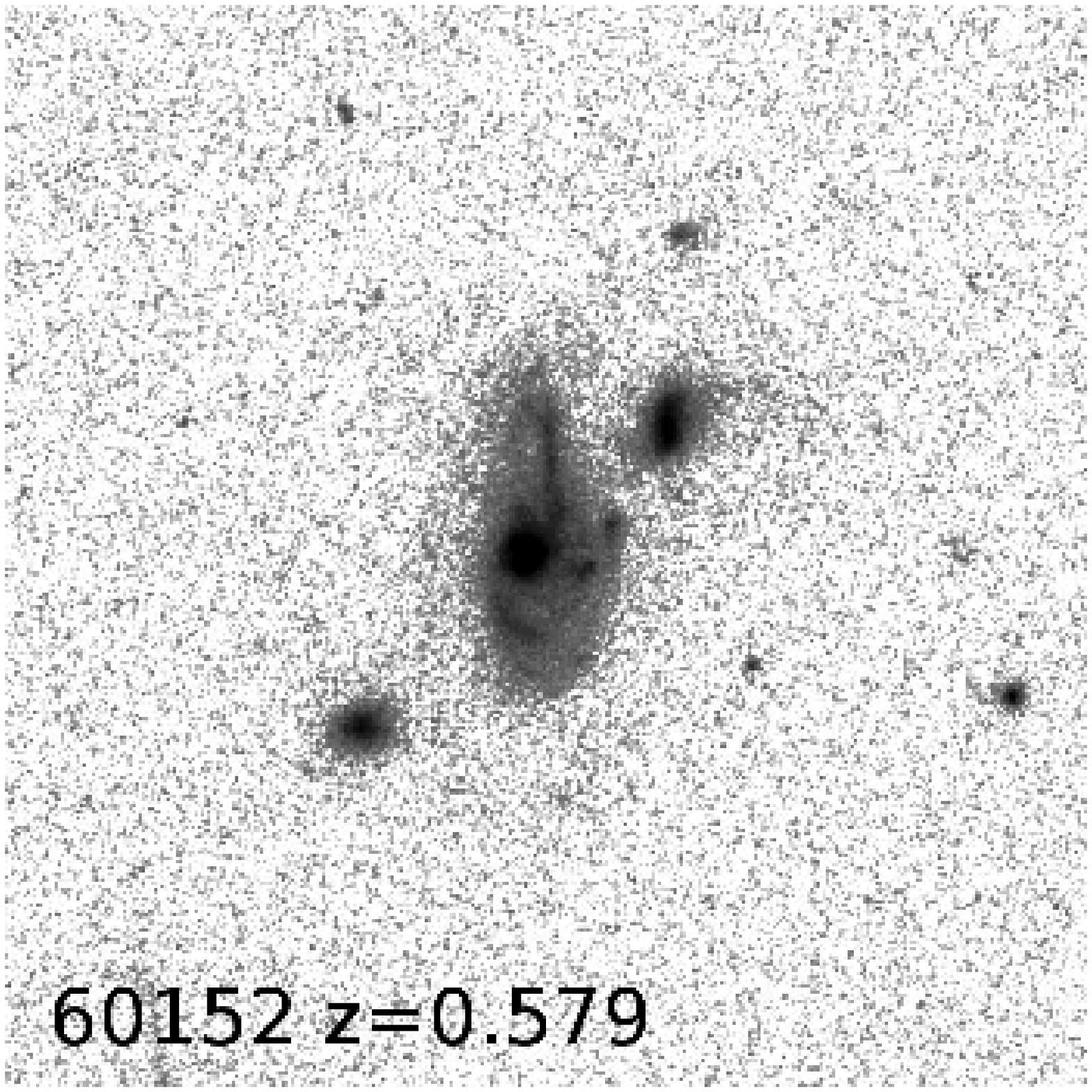}
      
      \vspace{0.3cm}
      
      \includegraphics[width=5cm,height=5cm]{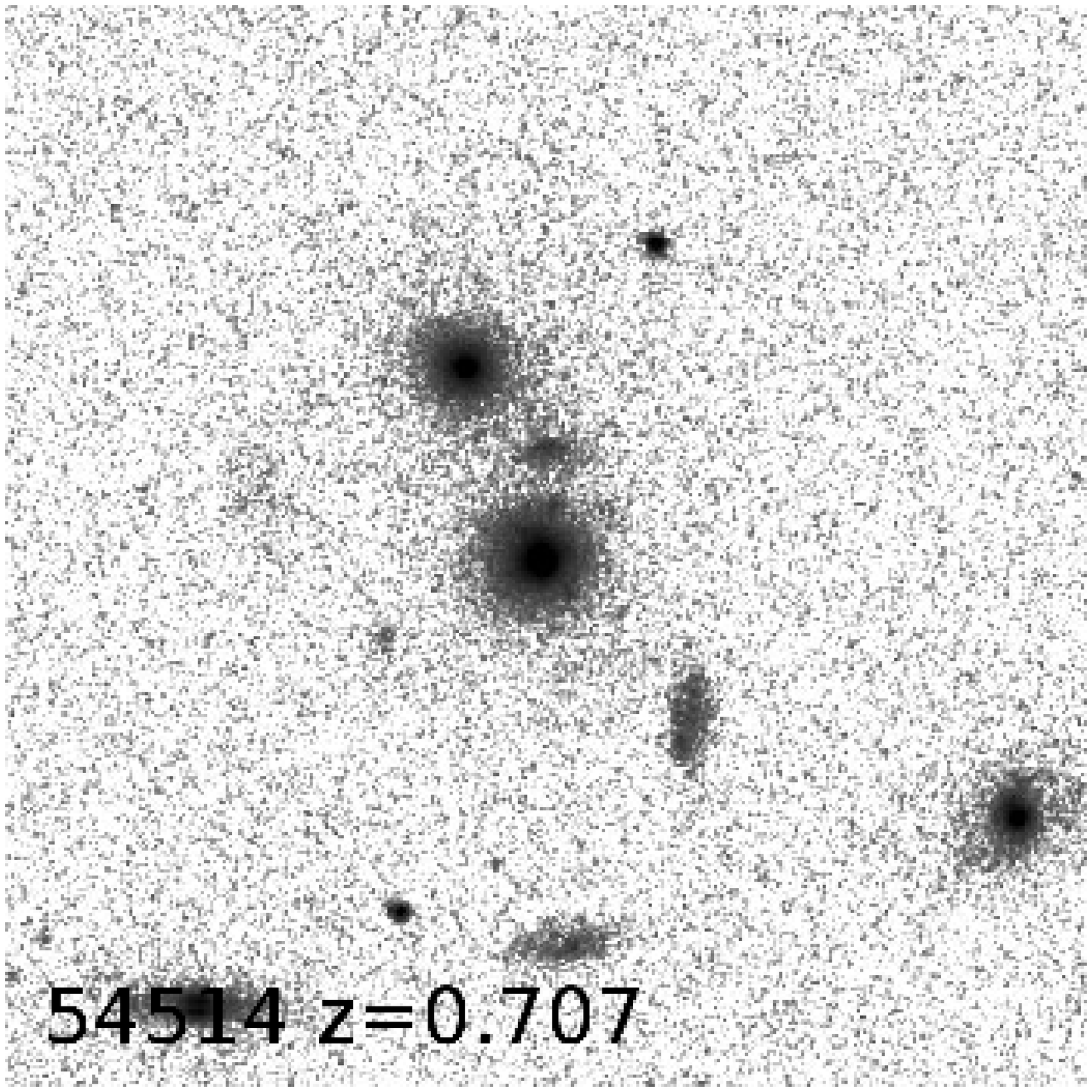}\hspace{0.3cm}\includegraphics[width=5cm,height=5cm]{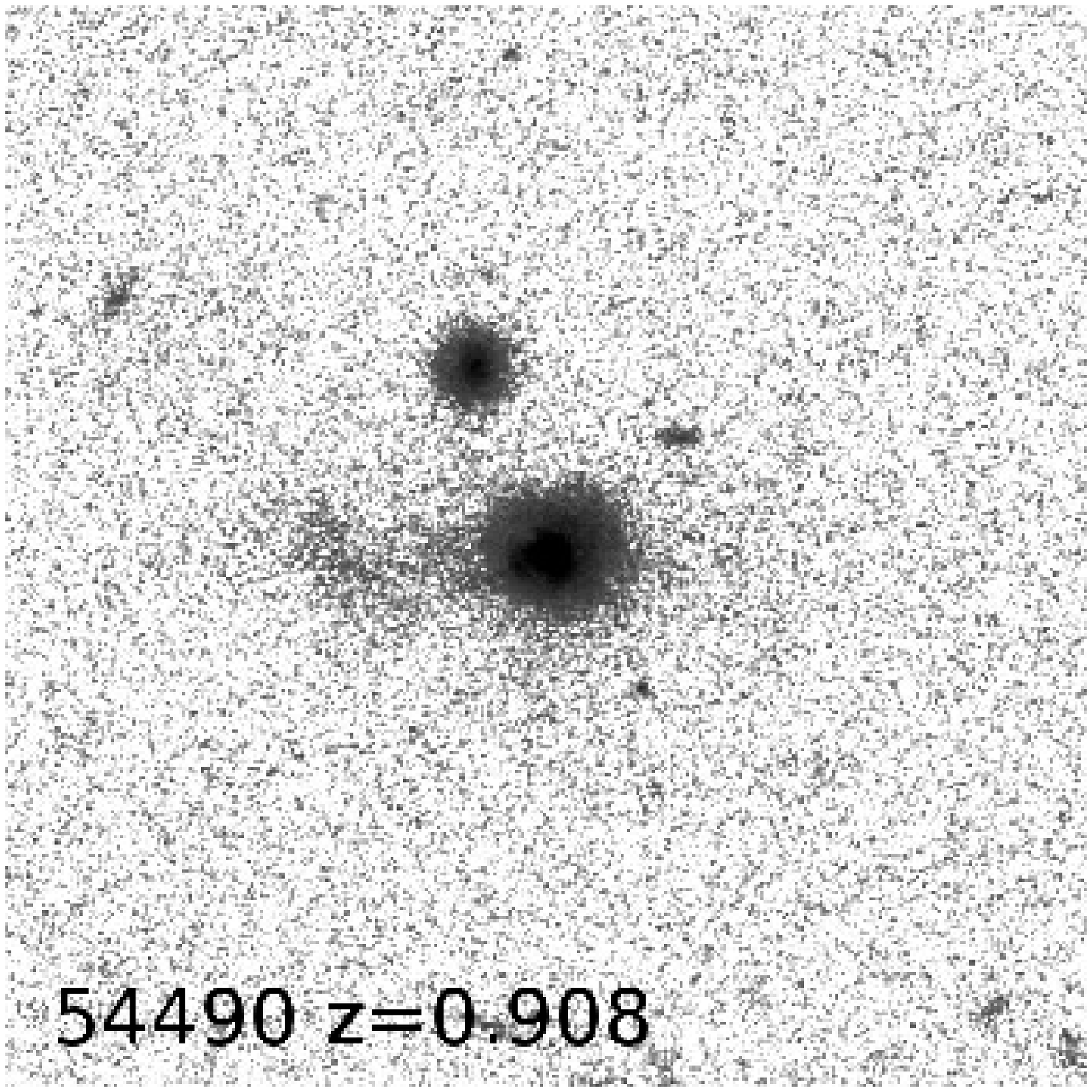}\hspace{0.3cm}\includegraphics[width=5cm,height=5cm]{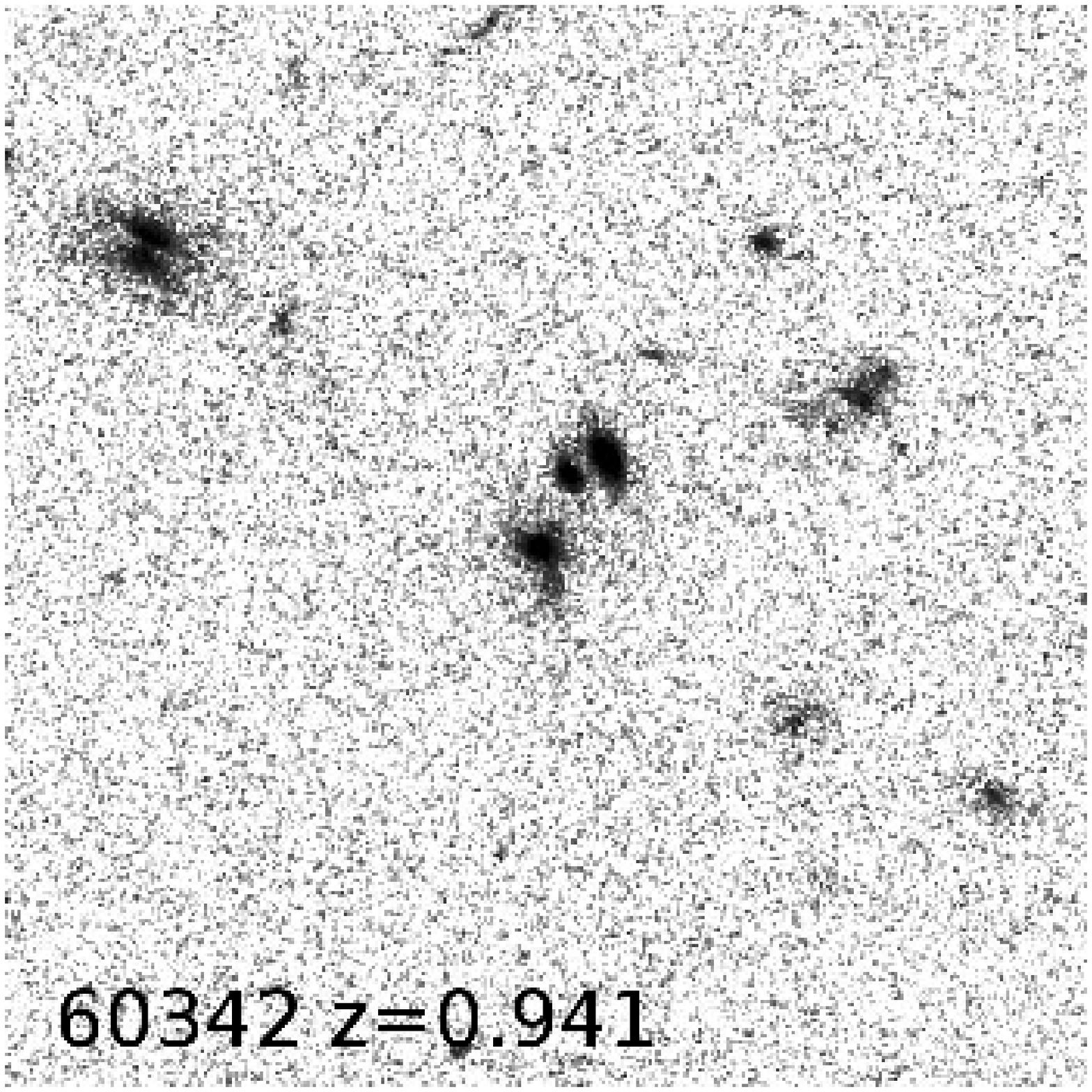}
      
      \vspace{0.3cm}
      
      \includegraphics[width=5cm,height=5cm]{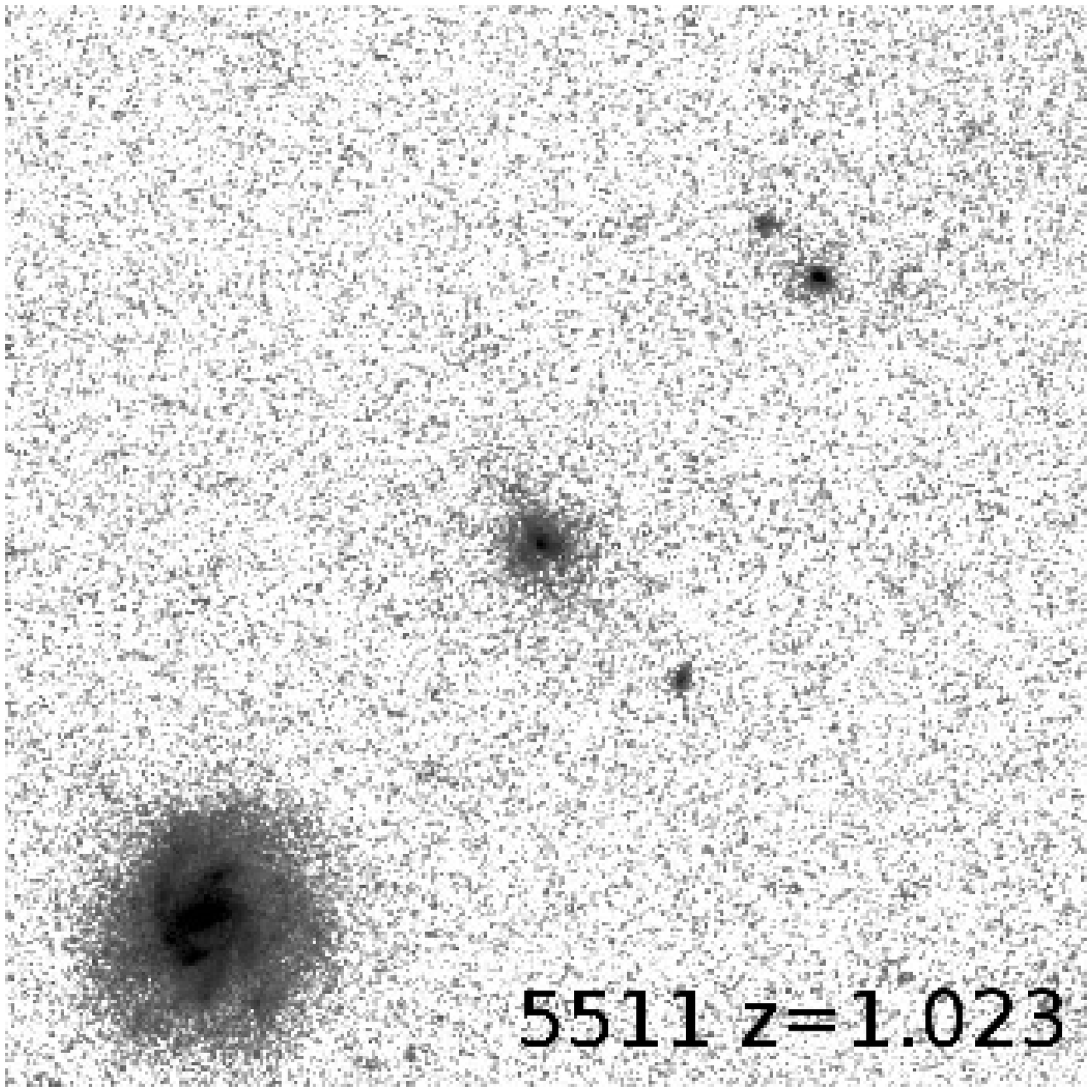}\hspace{0.3cm}\includegraphics[width=5cm,height=5cm]{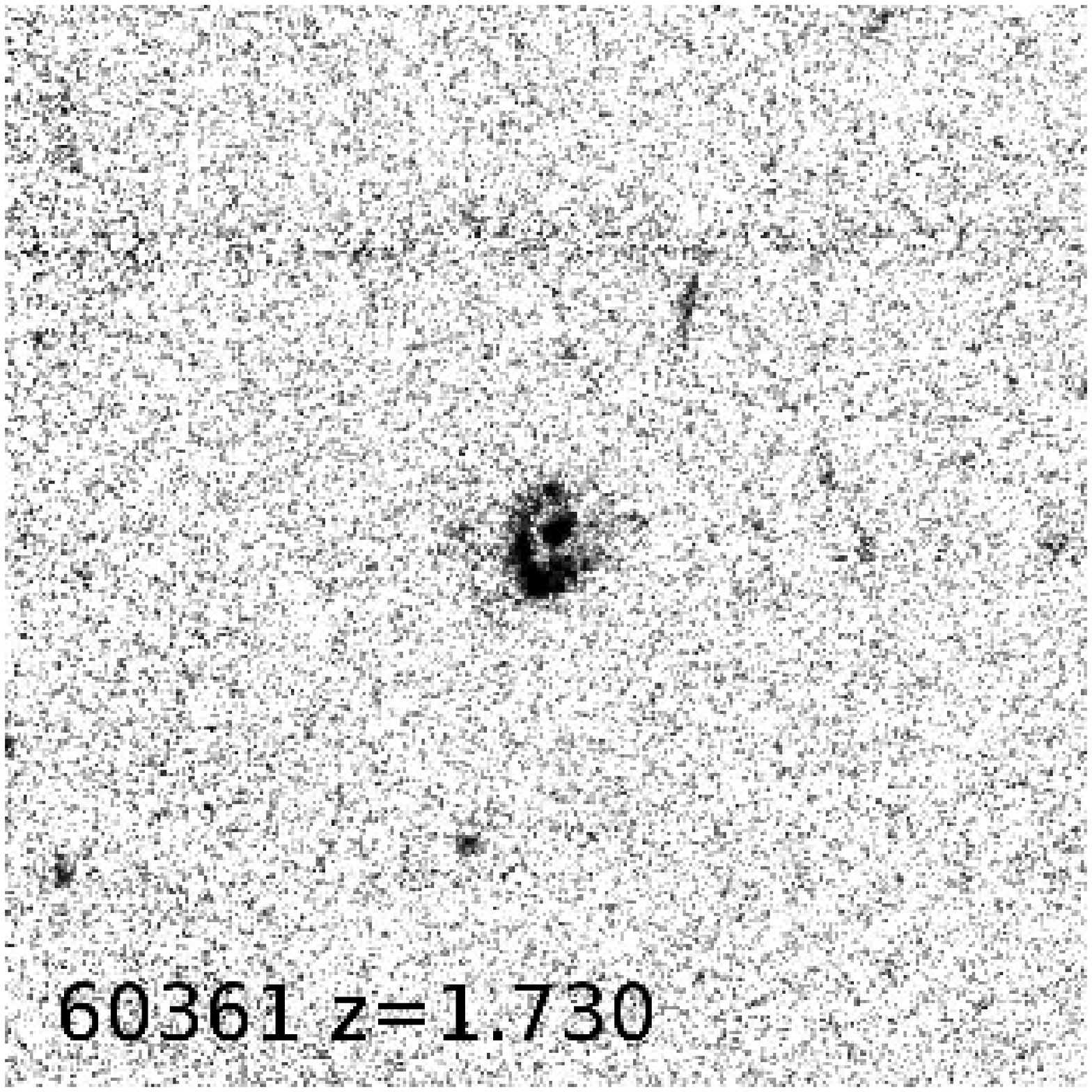}\hspace{0.3cm}\includegraphics[width=5cm,height=5cm]{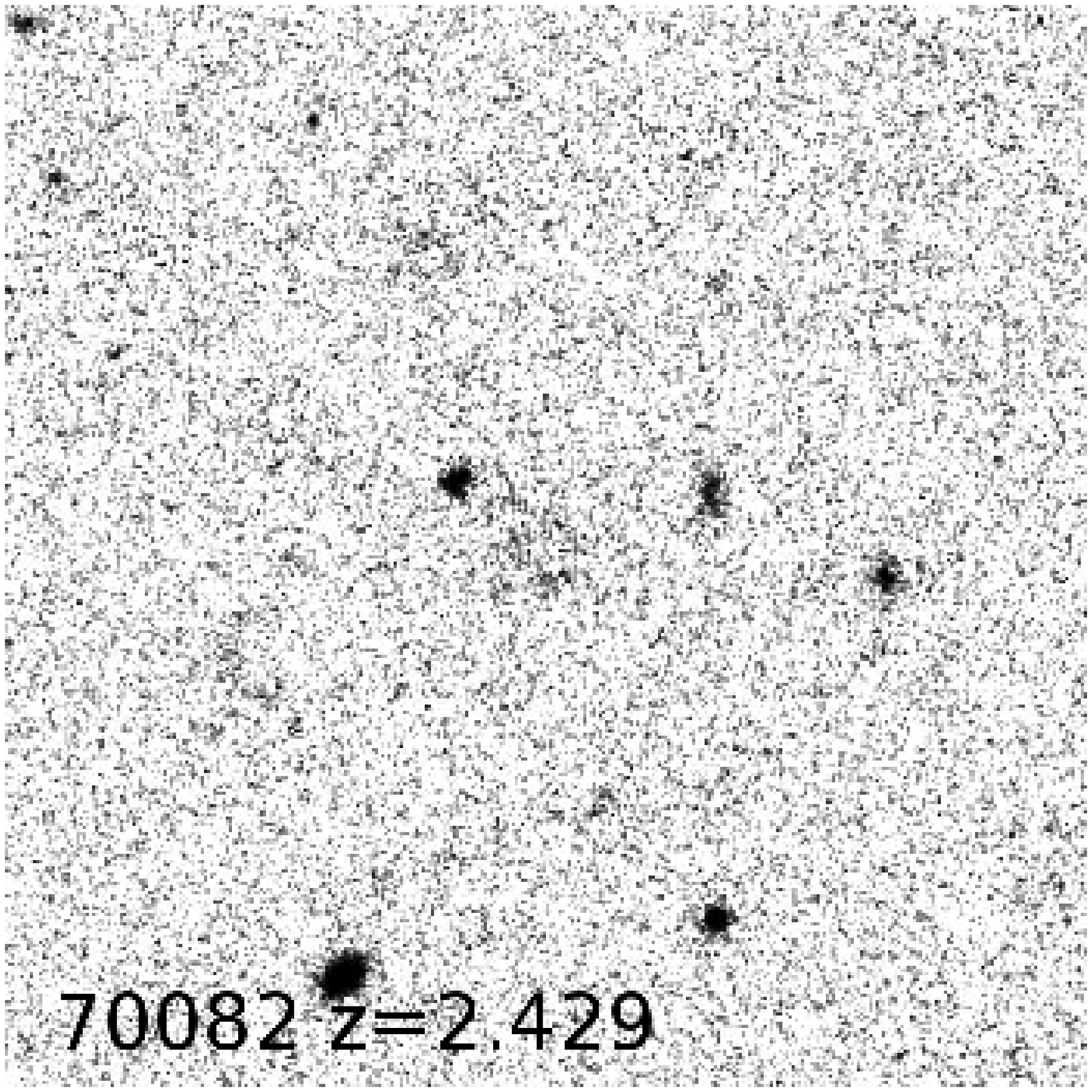}
      
      \vspace{0.3cm}
      
      \includegraphics[width=5cm,height=5cm]{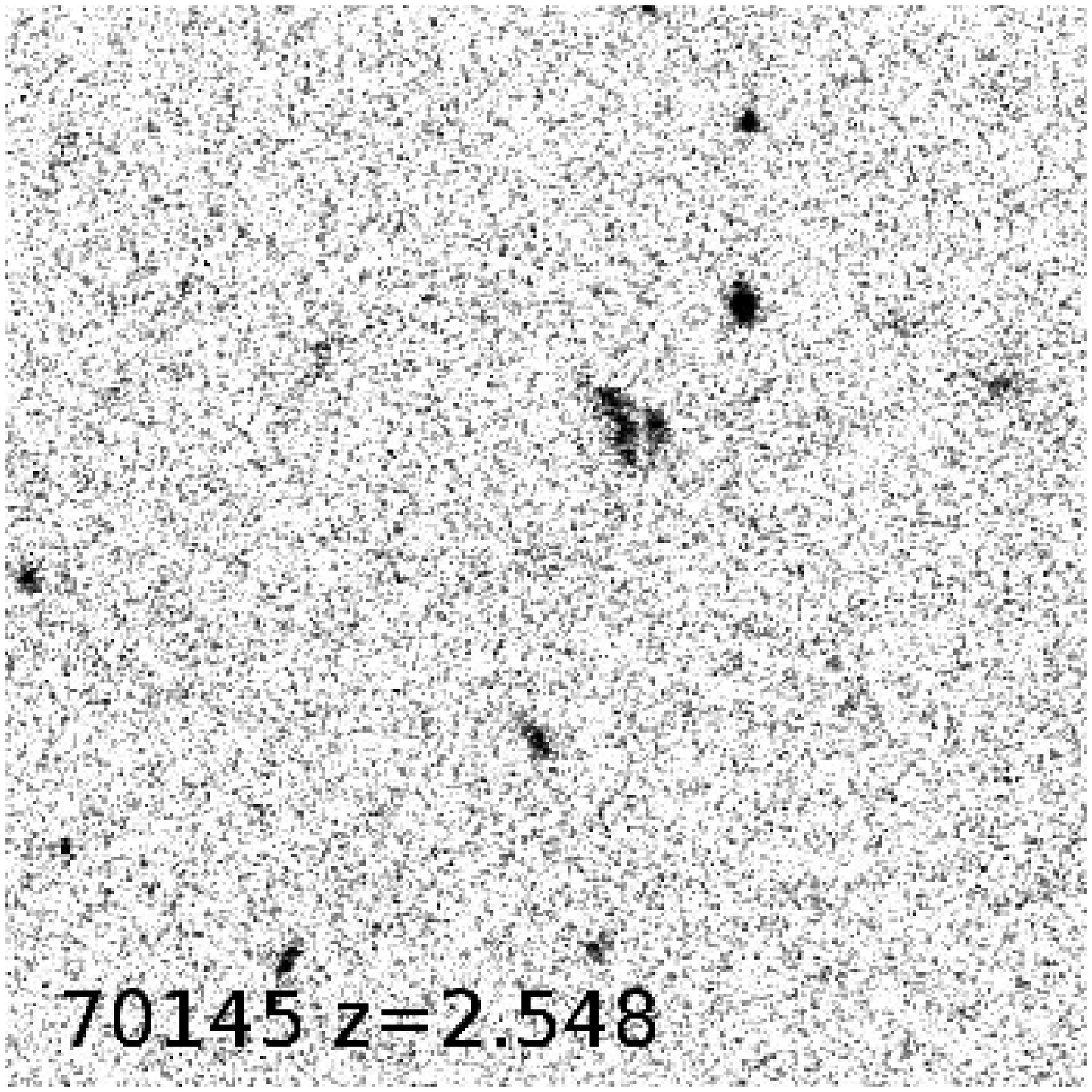}
   \caption{HST-ACS I band $15\arcsec\times15\arcsec$ cut-outs of the COSMOS mosaic (Koekemoer et al. 2007), for the 10 CTK$_f$ sources, ordered by increasing redshift.}
   \label{morph}
   \end{figure*}

\subsection{Morphology}

The key asset of the COSMOS field are the deep (down to I$_{AB}=26$), high resolution (0.09\arcsec\ FWHM and 0.03\arcsec\ pixels) 
Hubble-ACS observations, covering the full 1.8 deg$^2$ (Koekemoer et al. 2007, Scoville et al. 2007) for a total of 590 orbits in Cycles 12–13.
These mosaics allowed the study of the morphology for several hundred thousand galaxies (Leauthaud et al. 2007; Scarlata et al. 2007, Zamojski et al. 2007). 
We exploited this unprecedented dataset with the aim of studying the morphological properties of our small sample of highly obscured sources.
Fig.~\ref{morph} shows $15\arcsec\times15\arcsec$ ACS I band cut-outs for the 10 CTK$_f$ sources, in increasing redshift order.
Apart from XID 2608, our best CT candidate at very low redshift (z=0.125), the majority of CTK$_f$ sources appear as small but resolved galaxies up to z$\sim2$.

We have secure morphological information available for 6 our of 10 CTK$_f$ AGN obtained from the updated Zurich Estimator of Structural Types
(ZEST; Scarlata et al. 2007), known as ZEST+ (Carollo et al., in preparation).
For two source at $z>1$, not present in the ZEST+ catalog, we visually derived a basic classification.
Source XID 70082 at z$=2.429$ is not detected in the ACS imaging while it is detected as a low surface brightness object in CFHT and Subaru optical bands.
Finally source XID 70145 at $z=2.548$ is not detected in the optical, while it becomes visible at longer wavelengths, from K band,
to Spitzer-IRAC MIR channels, and up to $24 \mu m$. 
However, the much poorer resolution of ground-based optical, or IR observations does not allow us to draw any conclusion about the morphology of these two high redshift sources,
and they are therefore excluded from the following discussion.

Interestingly, between 3 and 5 of the 8 sources that we can resolve up to $z\sim2$, show signs of merger/interaction: 
sources XID 60152 and XID 60361 clearly show post-merger morphologies,
source XID 54514 has a close companion at the same spectroscopic redshift, with some structure possibly connecting the two.
For the neighbor of sources XID 54490 there is a spectroscopic redshift (from zCOSMOS) of $z=0.527$ but with very low quality flag,
while for the neighbors of source XID 60342 there is no redshift information, even if a hint of a disturbed morphology can be seen in both sources.
All these morphological indications would translate into a merger/disturbed morphology fraction of $\sim35$\% for our CTK$_f$ sources up to $z\sim2$, that could be even higher, depending on the interpretation for sources XID 54490 and XID 60342. 
We extended this study to the remaining 29 CTN$_f$ sources (17 of them have an entry in the ZEST+ catalog, the remaining were visually inspected), 
and found that a similar fraction of sources have merging/disturbed morphologies, indicative of past or ongoing merging:
excluding the 8 sources that are not detected or resolved in the ACS images (all at $z\simgt2$), we found 8 sources out of 21 ($38\pm12\%$ of the sample)
with merging/disturbed morphologies.

These results can be compared with an optical galaxy merger fraction between 1 and 10\% 
(typically found to increase from z=0 to z=2, Lotz et al. 2008, Bundy et al. 2009, Conselice et al. 2009)
and the average merger fraction of X-ray selected AGN:
Villforth et al. (2014) found a merger fraction between 10 and 20\% for a sample of 76 low luminosity (Log(\lum)$\simlt44$) \xray\ selected AGN at $0.5<z<0.8$, using CANDELS data on the CDFS. 
Similar results are reported in Kocevski et al. (2012), for a sample of 72 \xray\ selected AGN of similar luminosities but higher redshift ($1.5<z<2.5$)  in the same field.
Cisternas et al. (2011) found a merger fraction of $\sim15\%$ in a sample 
of 140 \xray\ selected AGN in the COSMOS field (see also Brusa et al. 2009), and no differences with a control sample of normal galaxies.
Georgakakis et al. (2009) estimated an ``interacting'' fraction of $20\pm3\%$ for 266 sources \xray\ selected AGN at z$=0.5-1.3$. 
Finally, using the same ZEST+ catalog in COSMOS for consistency in the morphological classification,
we derived a fraction of merging/disturbed morphologies of $\sim15\%$, for a sample of 238 type-2 AGN
with average redshift $z\sim0.77$ (with $1\sigma$ dispersion of 0.44) and average Log(L$_{bol}$)=44.6 ($1\sigma$ dispersion of 0.69 dex).

We stress that all these studies explored a luminosity range of Log(L$_{bol}$)$\sim43-45.5$,
while our highly obscured sources are in a higher L$_{bol}$ range, and the merger fraction is thought to increase with L$_{bol}$ (Schawinski et al. 2012; Treister et al. 2012).
Indeed, the majority of them have Log(L$_{bol}$)$>45.5$, the luminosity above which mergers are thought to dominate AGN triggering (see e.g. Hopkins et al. 2013).
A similar L$_{bol}$ range was explored in Mainieri et al. (2011), where they found a merger fraction of $\sim20\%$
for a sample of 142 obscured QSOs, 95\% of them having $22<Log($\nh$)<23.5$.
Therefore, the merger fraction observed in our sample of highly obscured sources (35-45\%) is significantly higher,
even with respect to the merger fraction of high luminosity, \xray\ selected, type-2 AGN,
and the high obscuration seems to play a major role in this difference.
While the studies mentioned above showed that merger has a limited importance in triggering/fueling AGN accretion and SF in general (i.e. in a large range of \nh), 
by selecting CT-AGN specifically, we are picking up sources in which highly obscured AGN activity may indeed be triggered by merger events.

Finally, a number of recent works have demonstrated that looking for merger signatures 
in optical images of high redshift ($\simgt1$) sources, can lead to significantly under-estimated merger fraction, 
because of image quality degradation and because large-scale tidal features and other disturbed structures fade
away or become less prominent at (rest frame) shorter wavelengths (Cameron et al. 2011; Petty et al. 2009; Hung et al. 2014). 
Therefore the merger fraction computed in our sample of highly obscured AGN, derived from I band HST images,
can be slightly under-estimated: we have 8 sources at $1<z<2$ in the full sample, for which the morphological classification can be affected by these effects.

\section{Discussion}

\subsection{Implication for evolutionary models}

Most hierarchical models are based on co-evolution through major mergers (e.g. Di Matteo et al. 2005, Hopkins et al. 2008), 
or steady accretion from disk instabilities (e.g Hopkins \& Hernquist 2006; Bournaud et al. 2011; Gabor \& Bournaud 2013)
in which gas inflow triggers both intense SF and fast BH accretion, in a highly obscured environment. 
This is followed by a blow-out phase, in which supernova and AGN feedback (through radiation pressure and/or hot winds/jets) 
clear the environment from gas and dust, leading to starvation of both the BH accretion and the circumnuclear SF.
All these processes are short lived with respect to the total lifetime of the system ($\sim10^7$ yr, Fiore et al. 2012), 
and are expected to be deeply buried by gas and dust, making them difficult to observe.
The BH accretion rate of these short lived episodes can be very high, even above Eddington (Fabian et al. 2009).

From our results there is strong indication that highly obscured sources tend to have smaller SMBH, accreting at higher 
rates, with respect to type-1 AGN. The significance of these differences is {\it at least} $\sim2.6\sigma$, 
in case of the most conservative assumption of $f_{bulge}=1$ for the  highly obscured sources.
Furthermore there is evidence in favor of a stronger role for interactions/mergers, in
comparison with what found in previous studies for moderately obscured QSOs and moderate/low luminosity AGN.
However, we observe that this increased merger fraction does not translate into a large fraction of extreme starbursts,
of the kind that powers ULIRG and Sub-mm galaxies at low redshift (e.g. Veilleux et al. 2002)
and possibly also at high redshift (Kartaltepe et al. 2010; Hung et al. 2013; but see also Melbourne et al. 2008).
All the highly obscured sources selected in this work, sit in, or are consistent with, the MS of star-forming galaxies.

As already pointed out, even if statistically significant at face value, 
the reliability of all these results is limited by the number of steps needed to go from 
the observables to the physical properties of these sources, and the big uncertainties associated with each of these steps.
Therefore larger sample of sources are needed, with at least comparable high quality multi-wavelength data of the ones provided by COSMOS, 
in order to mitigate these uncertainties and give more reliable results.
We plan to apply the selection method for CT AGN tested here, 
to the full catalog of $\sim4000$ sources expected in the full \chandra\ coverage of the COSMOS field, 
going one order of magnitude fainter in flux limit, 
with the aim of collecting a sample of hundreds of reliable CT AGN candidates.

Finally, for our CT AGN there is indication from the optical spectra, 
(at least for the 4 sources with a detected [OIII] line)
that strong dust reddening may be present.
This should occur on galaxy structures at $\sim0.1-1$ kpc scales, in order to be able to obscure the narrow line region.
On the other hand, given the low quality X-ray spectra available here for our CT AGN, 
it is not possible to distinguish between a small scale obscuring torus 
and a host galaxy scale obscuration. There is therefore the possibility that also the CT obscuration  
that we see in X-rays (or some part of it) is produced by gas in the host.
We stress, however, that a host galaxy scale structure with CT column density is hard to imagine.
Even if there are large uncertainties involved in the determination of the gas mass of such structure 
(i.e size, geometry, filling factor), a very rough upper limit can be obtained assuming e.g. 
a sphere of radius 1 kpc and constant density, resulting in a gas mass of $\sim10^{11} M_{\odot}$. 
In this scenario the gas mass obtained for that density would be larger than the stellar mass of the host galaxy.
On the other hand, this dense gas could be concentrated in a circumnuclear, star forming ring (e.g. Alonso-Herrero et al. 2012).
This scenario would be also in agreement with the increased merger fraction observed for these galaxies,
since galaxy mergers are very efficient in driving the gas to the center (Kauffmann \& Haehnelt 2000).
Recent results (e.g. Goulding et al. 2012) showed, on the other hand, that not all CT AGN show dust extinction in the mid-IR, 
possibly due to different host properties (edge-on and mergers vs. face on).
These results may imply that the gas producing obscuration in X-ray 
and the dust producing extinction in the mid-IR are not always or necessarily related.

\subsection{Implication for CXB models}
Cosmic X-ray Background (CXB) synthesis models have been powerful tools in predicting the existence of a vast population of unseen, highly obscured sources,
able to reproduce the hard shape of the CXB, and its peak at $\sim30$ keV 
(Comastri et al. 1995, Gilli et al. 2007\footnote{available on line at http://www.bo.astro.it/~gilli/counts.html}, 
but see also Akylas et al. 2012).
However, the fraction of CT sources in \xray\ surveys, especially at faint fluxes, is still largely unknown, 
given the already mentioned difficulties in unambiguously identifying CT AGN in low signal to noise spectra.
All current models predict a fraction of CT sources that strongly decrease with increasing \xray\ flux:
at F$_{2-10}=10^{-16}$ erg s$^{-1}$ cm$^{-2}$, the Akylas et al. (2012) model\footnote{available on line at http://indra.astro.noa.gr/} 
predicts a CT fraction of 7\%, while Ueda et al. (2014) and Gilli et al. (2007) 
predict $\sim20\%$; going to F$_{2-10}=10^{-15}$ erg s$^{-1}$ cm$^{-2}$, the predicted fraction drops to 5, 10 and 4\% respectively.

Our sample of CT sources has a too bright flux limit in order to disentangle between different models: 
all of them predict a fraction $\sim1-2\%$ at F$_{2-10} = 0.9\times10^{-14}$ erg s$^{-1}$ cm$^{-2}$, which is the limiting flux of our CTK$_f$ sample (see Tab. 2).
Indeed the fraction of CT AGN observed is roughly in agreement with these predictions, 
since we have $10\pm2$ CT, out of a total sample of $\sim1000$ sources detected in the XMM-COSMOS catalog above that flux limit.

In order to clarify which model is the closest to data, CT searches in deeper \xray\ surveys are needed.
Brightman et al. (2014) already found 64 CT AGN candidates, of which 28 are {\it highly probable} CT AGN in the original 0.8 deg$^2$ C-COSMOS survey.
The planned extension of the analysis presented here, to the full, 2 deg$^2$ \chandra\ coverage in the COSMOS field,
will give a robust measurement (based on a very large sample of sources, with a well tested selection method), 
down to limiting fluxes of the order of $F_{2-10}\sim10^{-15}$ \cgs.
Given the expected total number of sources ($\sim4000$ AGN) the three models mentioned above predict 200, 400 and 160 CT AGN respectively.
With these numbers we should be able to discriminate between the different models.
Finally, the 7 Msec total \chandra\ exposure time that will be reached in the CDFS (P.I. N. Brandt),
combined with the existing 3 Msec of \xmm\ exposure time (Comastri et al. 2011),
will hopefully clarify the situation at even fainter fluxes ($10^{-16}$ \cgs, see Sec.~2).

\section{Summary}

We have exploited the unique multi-wavelength dataset available in the COSMOS field, with the aim of 
producing a robust catalog of CT AGN, \xray\ selected from the XMM-COSMOS catalog.
Our results can be summarized as follows:

\begin{itemize}
\item  We have shown that specially designed spectral modeling is crucial, to correctly identify these highly obscured sources, 
and to produce reliable \nh, luminosities and \feka\ line EW, given their complex \xray\ spectra. 
Deeper \xray\ data available in the same field, 
allowed us to unambiguously identify 10 {\it bona fide} CT AGN (6 of them also have a detected \feka\ line with EW$\sim1$keV)
out of a sample of 39 highly obscured AGN, and to determine an efficiency of the selection of CT AGN, based on the simple two power law model applied to
\xmm\ data alone,  of the order of $\sim80$\%.\\

\item We compared the bolometric luminosity obtained from the \xray\ with the one computed through SED decomposition techniques. Because
the \xray\ based L$_{bol}$ is very sensitive to the \nh\ estimate (the correction can be as large as 1-1.5 dex), 
the fact that the different L$_{bol}$ agrees extremely well, within $\sim0.3$ dex, 
is a further indication that our \nh\ estimates are reliable.\\

\item We studied the distribution of intrinsic SMBH and host properties, such as the $M_{BH}$, $\lambda_{Edd}$ and $M_*$, for the larger sample of highly obscured sources,
with respect to those of a sample of $\sim 240$ type-1 AGN, matched in luminosity and redshift.
Highly obscured AGN show a distribution of $M_{BH}$ significantly shifted toward smaller masses (at $>2.6\sigma$ significance),
and consequently a distribution of $\lambda_{Edd}$ significantly shifted toward higher accretion rates (at the same significance), with respect to type-1 AGN.
A somewhat less significant shift toward smaller $M_*$ for obscured AGN is observed ($\sim2\sigma$).\\

\item We have shown that the sSFR of the selected sources is typical of MS star forming galaxies al all redshifts, 
and no source is observed with sSFR/sSFR$_{MS}>4$, typical of major merger driven starbursts.\\

\item The available optical spectra of CT sources are typical of highly obscured sources, being dominated by the host stellar component.
Four of them show indication of strong dust reddening occurring on host galaxy scales.\\

\item The broad band SED of all the CT sources show that an obscured torus component is needed for all the sources analyzes.\\

\item High resolution, HST-ACS images show that $35-45$\% of our CT sources show some indication of being in merger/disturbed system,
a fraction significantly higher than the one observed in several X-ray selected AGN samples (around 15\%),
and than the observed merger fraction of high luminosity \xray\ selected type-2 AGN (20\%).\\

\end{itemize}

Our results point toward the possibility that X-ray selected,
highly obscured sources preferentially have a small SMBH, accreting close or above Eddington,
harbored by a small, strongly star-forming host galaxy.
Finally, we verified that the number of CT AGN identified in the XMM-COSMOS catalog is consistent with predictions from several CXB synthesis models, 
i.e. around 1\% for a flux limit of F$_{2-10}\sim1\times10^{-14}$ \cgs. 
The depth of the XMM-COSMOS catalog does not allow us to investigate this fraction at fluxes where the various models start to strongly diverge.
More stringent results, in both cases, will be obtained applying the same methodology to the deeper \chandra\ coverage of the COSMOS 2 deg$^2$.

\begin{acknowledgements}

We thank the anonymous referee for constructive comments that have
helped us to improve the quality of this paper.
The authors thank M. Brightman for useful discussions on the \xray\ selection of CT AGN. 
GL and MB acknowledge support from the FP7 Career Integration Grant eEASy (CIG 321913).
GL acknowledges support from the PRIN 2011/2012.
KI acknowledges support by DGI of the Spanish Ministerio de Econom\'ia y Competitividad (MINECO) under grant AYA2013-47447-C3-2-P.
This work  benefited from the {\sc thales} project 383549 that is jointly funded by the European
Union  and the  Greek Government  in  the framework  of the  programme
``Education and lifelong learning''.
This work is based on observations obtained with XMM–Newton, an ESA science mission with instruments and contributions directly funded by ESA Member States and NASA.
Also based on observations made with Chandra X-ray satellite, founded by NASA. 
This research has made use of data and/or software provided by the High Energy Astrophysics Science Archive Research Center (HEASARC), 
which is a service of the Astrophysics Science Division at NASA/GSFC and the High Energy Astrophysics Division of the Smithsonian Astrophysical Observatory.

 \end{acknowledgements}


\begin{appendix}

\section{Detection limits}

\begin{figure}[h]
   \centering
      \includegraphics[width=8cm,height=8cm]{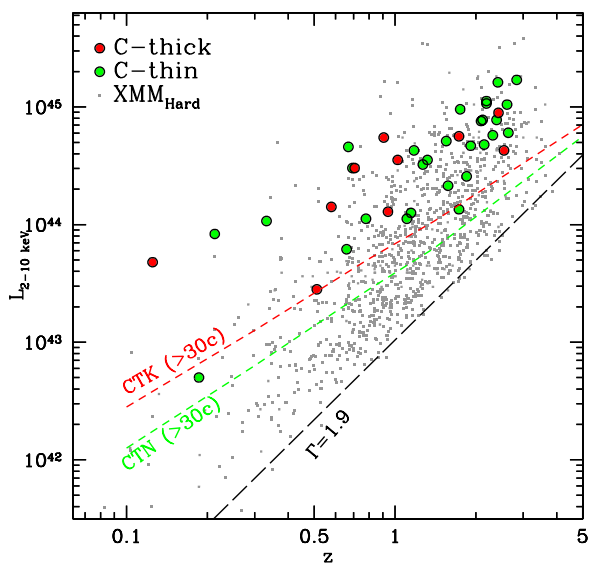}
   \caption{Redshift vs. \lum\ distribution for CTK$_f$ (red) and CTN$_f$ (green) sources. 
   Gray squares represent the 2-10 keV band detected XMM-COSMOS sources. The black dashed line represents the detection limit
   computed for an unabsorbed power-law with $\Gamma=1.9$, while the red (green) dashed line show the detection limit computed for a CTK (CTN) source.}
   \label{zlum}
   \end{figure}

We computed the detection limits in the $z$-\lum\ plane for highly obscured sources, 
assuming two different spectral models, one with \nh$=7\times10^{23}$ for CTN sources and the other with \nh$=1.6\times10^{24}$ cm$^{-2}$ for CTK sources.
We then estimated the maximum redshift at which a CTN or CTK source can be detected, with $> 30$ net counts in full band, for a given
intrinsic luminosity, given the flux limit of the XMM-COSMOS survey (red and green dashed lines in Fig.~\ref{zlum}), 
and compare this with the one computed for a typical unobscured AGN spectrum
i.e. assuming as model an unabsorbed power-law with $\gamma=1.9$ (black dashed line).
Red (green) circles represent CTK$_f$ (CTN$_f$) sources. Gray points represent the 1073 hard (2-10 keV) band detected XMM-COSMOS sources 
(hard band undetected sources suffer from larger uncertainties in the \lum, because an extrapolation from the soft band have to be made). 
We stress that the shape of the X-ray spectrum of CT AGN is responsible for a positive K-correction, 
which shift the X-ray flux of the very hard ($>7-10$ keV) unobscured part of the spectrum in the observing band and favors 
the detection of high z CT with respect to low redshift (relatively to the flux limit of the survey). 
This can be clearly seen in Fig.~\ref{zlum}: the red (CT) detection limit line
is flatter than the black one for the full hard detected sources.
Using these curves we  computed the weight to be applied to each source (CTK or CTN) of a given \lum, 
defined as the ratio between the maximum volume sampled for an unobscured source of the same intrinsic \lum\ and the maximum volume sampled for a CTK or CTN source.
From the plot is clear that the weights are larger for low luminosity sources, and slightly larger for CTK sources with respect to CTN.

\section{Model comparison}

   \begin{figure*}[!t]
   \centering
      \includegraphics[width=8cm,height=8cm]{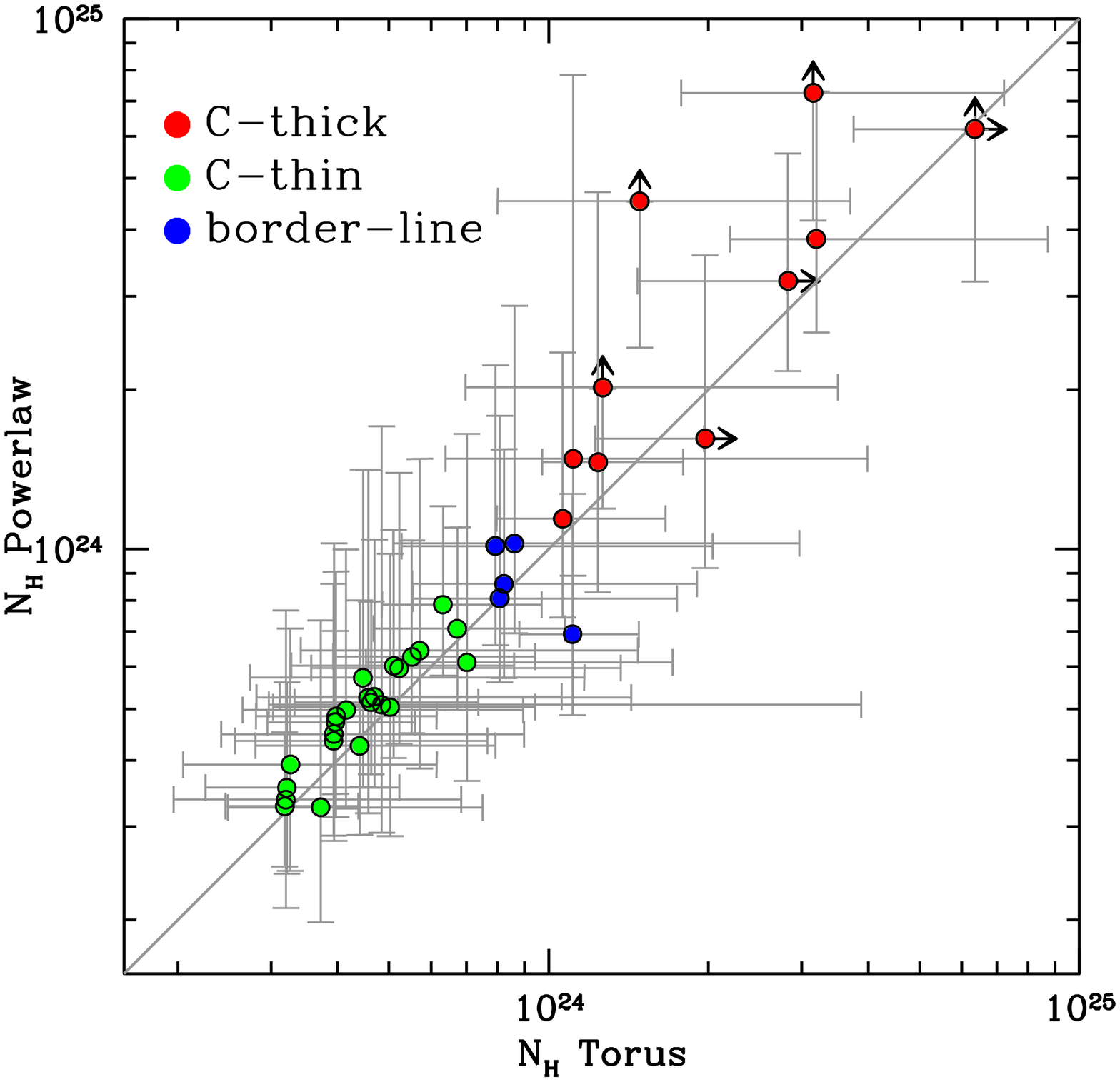}\hspace{0.5cm}
      \includegraphics[width=8cm,height=8cm]{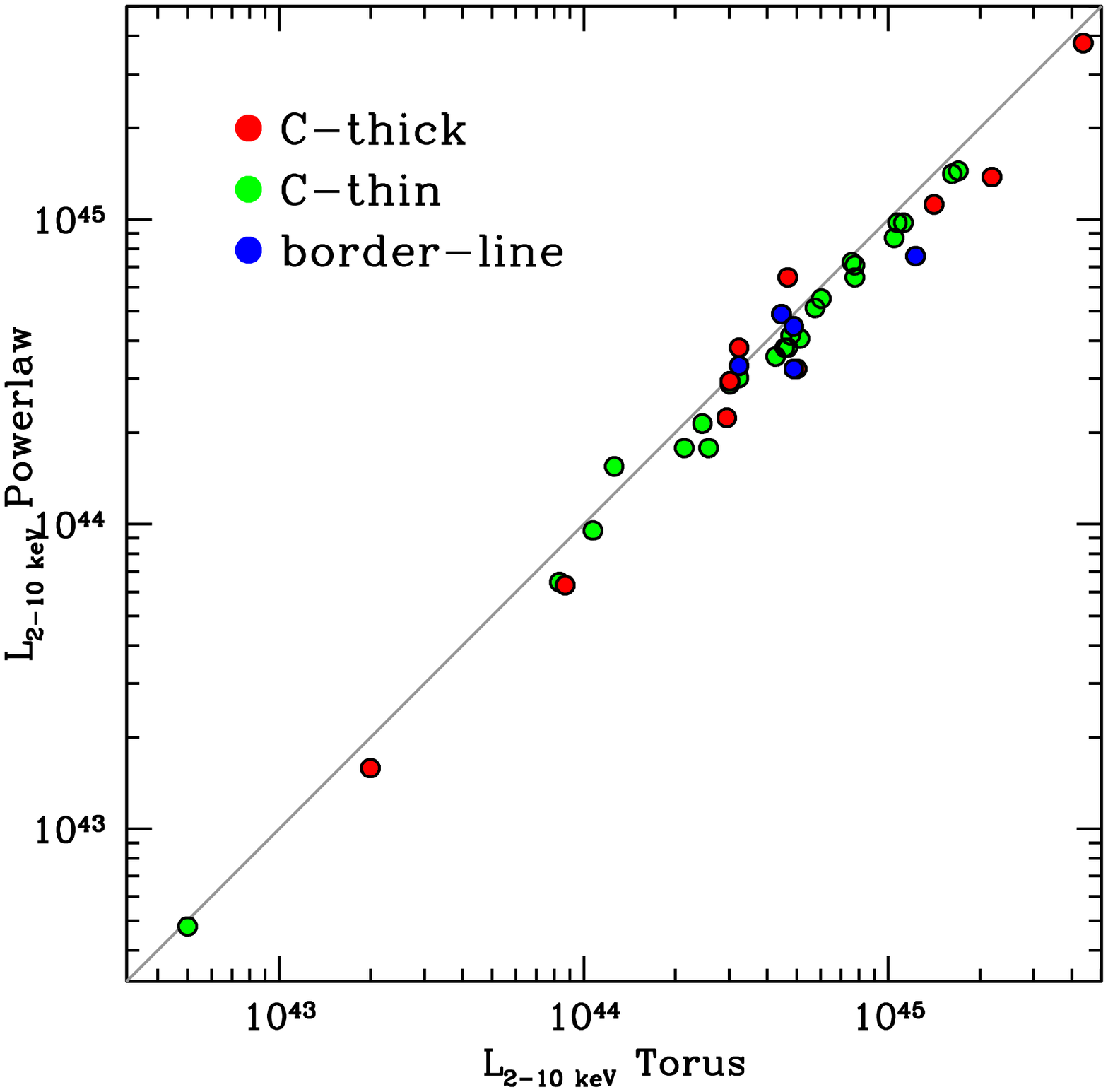}
   \caption{{\it Left panel:} Comparison of the \nh\ best fit obtained from the \plcabs\ (y-axis) and \tor\ (x-axis) models.
   {\it Right panel:} Comparison of the 2-10 keV, absorption corrected luminosity, obtained from the same two models.
   In both panels we show CTN$_i$ sources in green, CTK$_i$ sources in red and border-line sources in blue, and the solid line represents the 1:1 relation.}
   \label{comparison}
   \end{figure*}

As described in Sec.~3, we used two different models to reproduce the observed spectra of highly obscured sources in XMM-COSMOS:
one makes use of the TORUS table developed in Brightman \& Nandra (2011), which was specifically developed to model CT sources,
while the second is built in order to mimic as much as possible the first one, with the advantage of having as output the Fe K$\alpha$ line EW.
Therefore, we are interested in testing how the two models agree in determining the two main parameters involved in our analysis, namely
the \nh\ and the absorption corrected 2-10 keV luminosity.
Fig.~\ref{comparison} (left) shows a comparison of the best fit values obtained for \nh\ from the \tor\ and \plcabs\ models, respectively.
In red are shown the 10 CTK$_i$ sources, in green the 29 CTN$_i$ sources. In blue we highlighted 5 sources that are border-line:
they are either seen as CT from one model but not the other, or they are very close to the dividing line (\nh$=10^{24}$ cm$^{-2}$)
and with large error bars, that makes them fully consistent with being CT (see table 1).
Several CTK$_i$ sources are shown with their lower limit in \nh, because their \nh\ is consistent with the model upper boundaries, 
which is \nh$=10^{25}$ cm$^{-2}$ for the \tor\ model
and \nh$=10^{26}$ cm$^{-2}$ for the \plcabs\ model. 
The \plcabs\ model typically slightly overestimates the \nh\  with respect to the \tor\ model, in both Compton thin and thick regimes. 
However, all the measurements are perfectly consistent within the large error-bars.
Fig.~\ref{comparison} (right) shows a comparison of the 2-10 keV, absorption corrected luminosity, obtained from the 
\plcabs\ and \tor\ model.
The intrinsic luminosity obtained from the \tor\ model is systematically slightly higher ($\sim0.1-0.2$ dex) than the one obtained from the power-law model.
The small differences in \nh\ and \lum\ between the two models are probably related to the different reflection modelization, that slightly underestimate the primary power-law normalization
in the \plcabs\ case with respect to \tor, to reproduce the same data points (see the different levels of the continuum above 10 keV in Fig. 3 left and right).

\section{Unfolded spectrum plus model}

\begin{figure}[h]
   \centering
      \includegraphics[width=8cm,height=8cm]{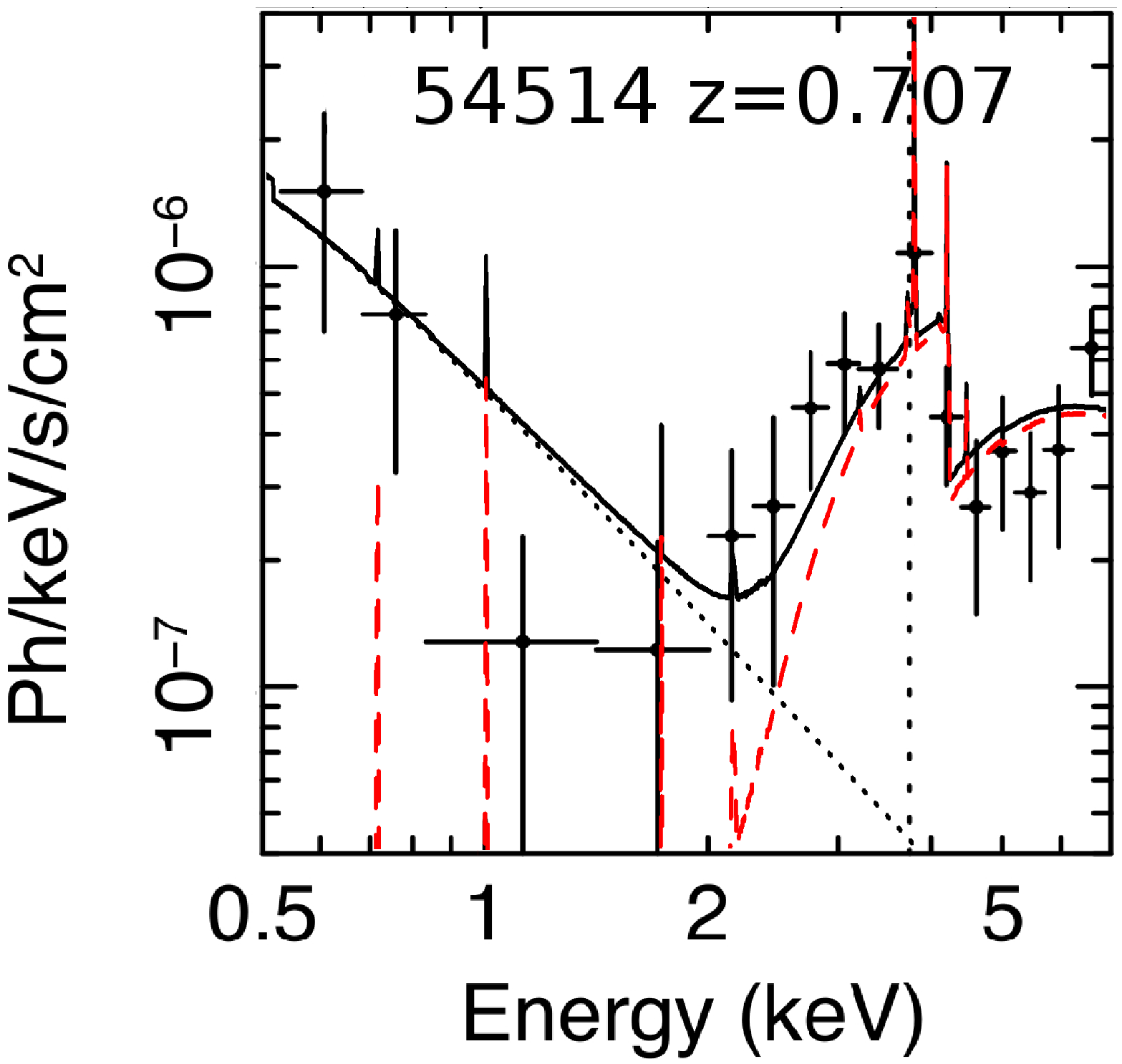}
   \caption{Unfolded, merged pn+MOS+\chandra\ spectrum of one of the sources in the CTK$_f$ sample, shown as an example.
   The best fit model is shown with the black solid curve. The TORUS component is shown with the red dashed curve, while the scattered component
   is shown with the black dotted curve. The XMM source ID and redshift are labeled. The dashed line marks the expected location of the 6.4 keV Fe K$\alpha$ line.}
   \label{54514_model}
   \end{figure}

Fig.~\ref{54514_model} shows the unfoled, co-added spectrum of one of the sources in the CTK$_f$ sample (namely XID 54514), togheter with the 
best fit model obtained from the \tor\ model. The presence of the best fit model helps in interpreting the spectral features observed in this 
as in all the CTK spectra shown in Fig. 4: the sharp flux drop below $7-10$ keV due to absorption, the soft emission 
arising below $1-2$ keV, the presence of the \feka\ line, and the strong feature at $\sim7$ keV due to the absorption edge.

However, we believe that, given the limited quality of these spectra, adding the best fit model in these plots 
would be a very strong guide for the eye, while we do not want to impose a bias to the reader. 
Furthermore it would be misleading, since we are not fitting our models to these co-added spectra, 
but rather to the single \xmm\ pn, MOS and \chandra\ spectra simultaneously. 
Even if the model parameters are the same for all the spectra,
and therefore there is no conceptual difference between fitting the sum of the spectra or simultaneously fit all of them, 
there is a subtle practical difference in the two approach, since there are a number of unavoidable approximation 
that have to be made to produce the co-added spectrum, and therefore the simultaneous fit is always preferable. 
Given all these considerations, we decided to show in Fig. 4 only the co-added spectra, and to show only here, as an example,
one unfolded spectrum plus best fit model.

\section{Comparison with Brightman et al. 2014}

Brightman et al. (2014, B14) reported a search for CT sources in the \chandra\ data of the COSMOS, CDFS and AEGIS fields,
using a similar approach to the one presented here: they used a simple model comprising a Torus component and a scattered component, with
the Cstat statistic applied to lightly grouped spectra (1 counts per bin).
The main differences are in the set of models adopted:
three different models are built with the Torus component
having 60$^{\circ}$, 30$^{\circ}$ and 0$^{\circ}$ half opening angle respectively (models A,B and C, with the C having no 
scattered component),  plus a model with a simple power-law without absorption (model D).
The photon index is fixed to 1.7 for sources with less than 600 counts,
but for each model, if the fit with a free $\Gamma$ is significantly better,
then this parameter is left free to vary even for sources with less than 600 counts.
No cut in minimum number of counts is used.

We compared our results on \nh\ distribution in the entire \xmm\ catalog with the ones published in B14, for the sources in common.
There are 644 \chandra\ counterparts from B14 of our \xmm\ catalog sources.
We excluded from the comparison 111 sources that are not detected in the hard band in the \xmm\ catalog:
the \chandra\ data are much deeper in both soft and hard bands (see Sec.~2.1), and the lack of hard band detection in the \xmm\ catalog make it 
impossible to correctly estimate the amount of obscuration for highly obscured sources (the source is detected in \xmm\ only in the soft band
thanks to the scattered light). 

Because we are interested in comparing the \nh\ distribution, 
we further excluded 70 sources that have a best fit photon index outside the range 1.5-2.5 that was used in our analysis:
these very steep (up to $\Gamma=3$) or flat (down to $\Gamma=0$) spectra may indicate interesting sources, e.g. reflection dominated CT candidates,
but are not useful for the comparison on \nh\ results.
Indeed 23 sources that are obscured in our analysis, are best fitted in B14 with an unabsorbed power-law with $\Gamma<1$. 
Finally we excluded 30 sources with less than 30 counts in the \chandra\ spectrum: the error bars associated are too large,
and below this threshold, the method used to determine which model is preferred, is shown not to work properly in B14.

\begin{figure}[h]
   \centering
      \includegraphics[width=8cm,height=8cm]{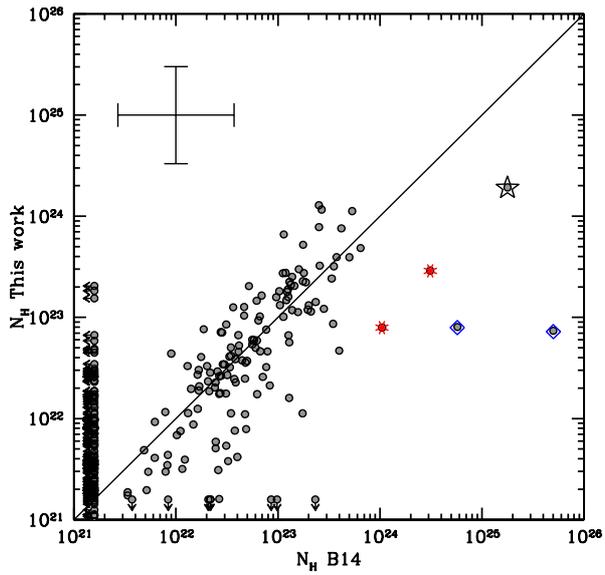}
   \caption{Comparison between of the best fit \nh\ from B14 and this work, for the 433 sources in common.
   The star show the location of source XID 2608.
With cyan diamonds are marked two sources for which the nature of CT sources from \chandra\ data
is not confirmed with better \xmm\ spectra.
In red are shown two sources for which the fit of the \xmm\ spectrum shows a secondary minimum at CT values.
The average error-bar is shown in the top left corner.}
   \label{nhnh}
   \end{figure}

We are left with 433 sources for the comparison. The distribution of \nh\ vs. \nh\ is shown in Fig.~\ref{nhnh}.
There is a global good agreement between the two measurements, within the large uncertainties (the average error-bar is shown in the top left corner).
However only 1 out of 4 CT candidates from our analysis is found to be CT also in B14.
The remaining 3 show a slightly lower \nh, below the CT limit, and there is a general trend of having lower \nh\ values in B14 for sources above $10^{23}$ cm$^{-2}$.
This is due to the different fixed $\Gamma$ adopted (1.7 in B14 instead of 1.9): we tested that the use of $\Gamma=1.7$ gives results consistent with B14 for these sources.

There are two sources (namely XID 2210 and 272, shown with blue diamonds) that are found to be heavily CT in B14 (\nh$>6\times10^{24}$ cm$^{-2}$).
The \xmm\ spectra of both this sources have a factor 2-3 more counts with respect to the \chandra\ ones, and no indication of strong absorption 
can be found. We conclude that these two sources are misclassified in the \chandra\ spectral analysis.
On the other hand, in red are shown two sources for which the fit of the \xmm\ spectra shows a primary minimum
of the probability distribution of the \nh\ at Compton thin values, while a strong secondary minimum is found at CT values,
in agreement with what is found from the \chandra\ spectra.
In this case is not possible to exclude that these are indeed CT sources,
and a simultaneous fit of \chandra\ and \xmm\ data would give a more constrained result.

A similar case is for XID 2608, the best CT candidate in the sample, and the only one being CT both from our analysis and from B14 (marked with a star in Fig.~\ref{nhnh}):
the fit of the \xmm\ data alone gives as best fit an \nh\ of only $1.06\times10^{24}$ cm$^{-2}$ (see Tab. 1), while the fit of \chandra\ data
alone gives an \nh$>10^{25}$ cm$^{-2}$ from both our analysis and B14. 
The joint fit of all the data (including also MOS) give an intermediate and well constrained value (see Tab. 2). 

There is finally a large population of sources for which we measure a moderate obscuration,
while B14 found that an obscured model is not required for these sources,
and the best fit is a simple power law (possibly with free $\Gamma$).
The much higher effective area of \xmm\ below 1 keV, extending down to 0.3 keV (instead of 0.5 keV as in \chandra) clearly plays a role in determining what is
the minimum \nh\ that can be constrained for each spectrum 
(depending on the number of counts and the source redshift).

\end{appendix}

\end{document}